\documentclass[10pt,journal,compsoc]{IEEEtran}
\ifCLASSOPTIONcompsoc
  \usepackage[nocompress]{cite}
\else
  \usepackage{cite}
\fi
\usepackage[pdftex]{graphicx}
\usepackage{amsmath}
\usepackage{amsfonts}
\usepackage{amsthm}
\usepackage[utf8]{inputenc}
\usepackage[english]{babel}
\usepackage[usenames, dvipsnames]{color}

\usepackage{subfig}
\usepackage{tabularx}
\usepackage{array}
\usepackage{multirow}
\theoremstyle{definition}

\ifCLASSINFOpdf
\else
\fi
\begin{document}
\title{Robust and High Fidelity Mesh Denoising}
\author{Sunil~Kumar~Yadav,
	Ulrich~Reitebuch,
	and~Konrad~Polthier
	\IEEEcompsocitemizethanks{\IEEEcompsocthanksitem Sunil~Kumar~Yadav, Ulrich~Reitebuch and Konrad~Polthier are with the Department
		of Mathematics and Computer Science, Freie Universit{\"a}t Berlin.\protect\\
		E-mail: sunil.yadav@fu-berlin.de}
}

\markboth{}%
{Shell \MakeLowercase{\textit{et al.}}: Bare Demo of IEEEtran.cls for Computer Society Journals}

\IEEEtitleabstractindextext{%
\begin{abstract}
This paper presents a simple and effective two-stage mesh denoising algorithm, where in the first stage, the face normal filtering is done by using the bilateral normal filtering in the robust statistics framework. \textit{Tukey's bi-weight function} is used as similarity function in the bilateral weighting, which is a robust estimator and stops the diffusion at sharp edges to retain features and removes noise from flat regions effectively. In the second stage, an edge weighted Laplace operator is introduced to compute a differential coordinate. This differential coordinate helps the algorithm to produce a high-quality mesh without any face normal flips and makes the method robust against high-intensity noise. 
\end{abstract}

}

\maketitle

\IEEEdisplaynontitleabstractindextext

%
\IEEEpeerreviewmaketitle

\IEEEraisesectionheading{\section{Introduction}\label{sec:introduction}}

\IEEEPARstart{M}{esh} denoising is one of the most active and fascinating research areas in geometry processing as digital scanning devices become widespread to capture the 3D points of a surface. In the process of data acquisition, noise is inevitable due to various internal, and external sources. Mesh denoising algorithms are focused to remove this undesired noise and compute a high-quality smooth function on a triangular mesh for further use in rendering, scientific analysis, reverse engineering and medical visualization. In general, both sharp features and noise are high-frequency components on the surface and are ambiguous in nature. During the surface denoising, it is a challenging task to decouple these components, which is essential to compute a smooth surface with sharp features. 

\subsection{Related Work}
In last two decades, a wide variety of smoothing algorithms
have been introduced to remove undesired noise,
while preserving sharp features in the geometry. For a comprehensive review on
mesh denoising, readers are referred to \cite{PMP:2010}, \cite{Centin} and \cite{yadav17}. Here, an overview of current state-of-the-art methods is given, which are based on differential coordinates and bilateral filtering.

The concept of differential coordinates was initially introduced by Alexa \cite{Alexa} as a local shape descriptor of a geometry. Differential coordinates
are able to preserve the fine details on a triangular mesh, which leads to their further applications in mesh processing, for example, Lipman et al. \cite{LaplacianMeshEditingIJSM:2005} used the same concept for mesh editing. Mesh quantization \cite{Sorkine:2003:HQM:882370.882376} and shape approximation \cite{Sorkine:2005:GBS:1042201.1042391} are also done by using the concept of differential coordinates. Sorkine et al. \cite{CGF:CGF999} introduced a general differential coordinate-based framework for mesh processing. Later, Su et al. \cite{diffCoordinate} utilized differential coordinates to remove noise and compute a smooth surface.  

The bilateral filtering was initially proposed by Tomasi et al. \cite{BilateralImage} for image smoothing. The relation between anisotropic diffusion \cite{aniso} and bilateral filtering is explained by Barash \cite{bilAniso}. In continuation, Black et al.\cite{RoAni} and Durand et al.\cite{DurandBilImage} expressed anisotropic diffusion and bilateral filtering in the robust statistics framework\cite{Hampel86a}. The concept of bilateral smoothing was extended for mesh denoising by Fleish et al.\cite{BilFleish} and later, Jones et al.\cite{nonItMesh} explained the same concept in the robust statistics framework. Bilateral filtering was applied to the two-stage surface denoising method by Zheng et al. \cite{BilNorm}, where the weight functions were computed based on the normal differences (similarity measurement) and spatial distances between neighbouring faces. Later, researchers proposed a general framework for bilateral and mean shift filtering in any arbitrary domain \cite{Solomon}. Recently, several bilateral normal-based denoising algorithms have been published. For example, Wei et al.\cite{binormal} utilized both face and vertex normals to produce a smooth surface and Yhang et al.\cite{Guidedmesh} proposed a guided mesh normal filtering based on the joint bilateral filtering \cite{JointBilateralUpsamplingImage}. A tensor voting guided mesh denoising was also introduced by Wei et al.\cite{tvgWei}, where feature classification was done using the normal voting tensor and then MLS fitting was applied to remove noise.  

The compressed sensing framework is also utilized to produce feature preserving mesh denoising algorithms. He et al.\cite{L0Mesh} extended the $L^0$ image smoothing concept to a mesh denoising algorithm where the algorithm maximizes the piecewise flat regions on noisy surfaces to remove noise. The $L^1$ optimization based algorithms were introduced by Wang et al.\cite{L1Mesh} and Wu et al.\cite{ROFL1} where features were preserved from the residual data by the extended ROF (Rudin, Osher and Fatemi) model. In continuation, Lu et al.\cite{robust16} detected features on the noisy surface using quadratic optimization and then removed noise using $L^1$ optimization while preserving features. Later, researchers utilized the $L^1$-median normal filtering along with vertex preprocessing to produce a noise free surface \cite{LU201749}. Recently, Yadav et al.\cite{yadav17} proposed a binary optimization-based mesh denoising algorithm, where noise was removed by assigning a binary values to a face normal-based covariance matrix. Centin et al.\cite{Centin} proposed a normal-diffusion-based mesh denoising algorithm, which removes noise components without tempering the metric quality of a surface. In this paper, a mesh denoising algorithm is introduced to provide a high fidelity mesh structure against different levels of noise without creating any normal flips.        

\subsection{Contribution}
We propose a two-stage mesh denoising method, where the face normal processing is mainly motivated by the robust statistics framework in anisotropic diffusion and bilateral filtering. Basically, the robust statistics framework is focused on developing different estimators which are robust to outliers. In the mesh denoising and robust statistic framework, sharp features are considered as \textit{outliers} and a proper estimator will avoid these \textit{outliers} to preserve sharp features and remove undesired noise components from a noisy surface.

We apply \textit{Tukey's bi-weight function} for the similarity function as a robust estimator which stops the diffusion at sharp features and produces smooth umbilical regions. 
In the vertex update step, we use a differential coordinate-based Laplace operator along with edge-face normal orthogonality constraint. This Laplace operator is computed using the edge length as the weight function to produce a high-quality mesh without face normal flips and it also makes the algorithm more robust against high-intensity noise.

\section{Method}
The proposed method is implemented in two different stages. In the first stage, noisy face normals are processed using the bilateral filtering in the robust statistics framework and in the second stage, a high fidelity mesh is reconstructed using a edge-face normal orthogonality constraint and a edge-weighted differential coordinate. 

\subsection{Robust Bilateral Face Normal Processing}
In the first step of the proposed algorithm, we smooth noisy face normals using the bilateral filter, which is a non-linear filter and it computes output using the weighted  average of the input: 
\begin{equation}
\label{equ:bil}
	\tilde{\mathbf{n}}_i=\frac{1}{ K_{i}} \sum\limits_{j \in \Omega_i} a_jf(\lvert \mathbf{c}_i-\mathbf{c}_j \rvert, \sigma_c)g(\lvert\mathbf{n}_i-\mathbf{n}_j\rvert, \sigma_s) \mathbf{n}_j, 
\end{equation}
where $\mathbf{n}_j$ represents a noisy face normal and belongs to the geometric neighbour disk $\Omega _i$ of the central face normal $\mathbf{n}_i$. The geometric neighborhood is more robust against non-uniform meshes compared to the combinatorial neighborhood \cite{yadav17}. The normalization factor $K_i$ is defined as:
\begin{equation*}
K_i= \sum\limits_{j \in \Omega_i} a_j f(\lvert\mathbf{c}_i-\mathbf{c}_j\rvert,\sigma_c)g(\lvert\mathbf{n}_i-\mathbf{n}_j\rvert,\sigma_s), 
\end{equation*}
 where $a_j$ is the area of the corresponding face. The term $f(\lvert\mathbf{c}_i-\mathbf{c}_j\rvert,\sigma_c)$ and $g(\lvert\mathbf{n}_i-\mathbf{n}_j\rvert,\sigma_s)$ represent the closeness and the similarity functions whose kernel widths are controlled by $\sigma_c$ and $\sigma_s$ respectively and $\mathbf{c}_i$ and $\mathbf{c}_i$ are the centroids of face $i$ and $j$. 
 
 \subsubsection*{Robust Estimation} The problem of the estimation of a smooth surface from  a noisy surface can be written by using the tools of robust statistics. As shown by Black et al.\cite{RoAni}, Durand et al. \cite{DurandBilImage} and Jones et al. \cite{nonItMesh}, to compute a smooth surface, Equation (\ref{equ:bil}) can be written as the following minimization problem:
 \begin{equation}
 \operatorname*{min}_{\mathbf{n}_i}  \sum\limits_{j \in \Omega_i} a_j f(\lvert\mathbf{c}_i-\mathbf{c}_j\rvert, \sigma_c)\rho(\lvert\mathbf{n}_i-\mathbf{n}_j\rvert,\sigma_s), 
 \label{equ:minRob}
 \end{equation}
where $\rho$ is a robust error norm. The minimization of the above function will lead to smooth face normals. To solve Equation (\ref{equ:minRob}), the first derivative of the $\rho$-function should be zero which is a so-called \textit{influence function} $\Psi(x,\sigma_s) = \rho\prime (x,\sigma_s)$, where $x$ can be any variable (in our algorithm $x =(\mathbf{n}_i-\mathbf{n}_j) $). The influence function shows the behaviour of a $\rho$-function. 

For anisotropic diffusion and bilateral filtering, the influence function $\Psi(x,\sigma_s)$ and $\rho(x,\sigma_s)$-function can be written in terms of the similarity function $g(x, \sigma_s)$ \cite{DurandBilImage}:

 \begin{equation*}
\Psi(x, \sigma_s) = xg(x, \sigma_s)\text{,} \quad \rho(x,\sigma_s) = \int xg(x,\sigma_s) dx. 
 \label{equ:influence}
 \end{equation*}
The selection of a $\rho$-function will be crucial to avoid outliers. In our framework, the outliers will be sharp features and an appropriate $\rho$ function will help the algorithm to retain sharp features and remove noise components. 

\begin{figure}[bth]%
	\centering
	\subfloat[$g(x,\sigma_s)$]{{\includegraphics[width=2.8cm]{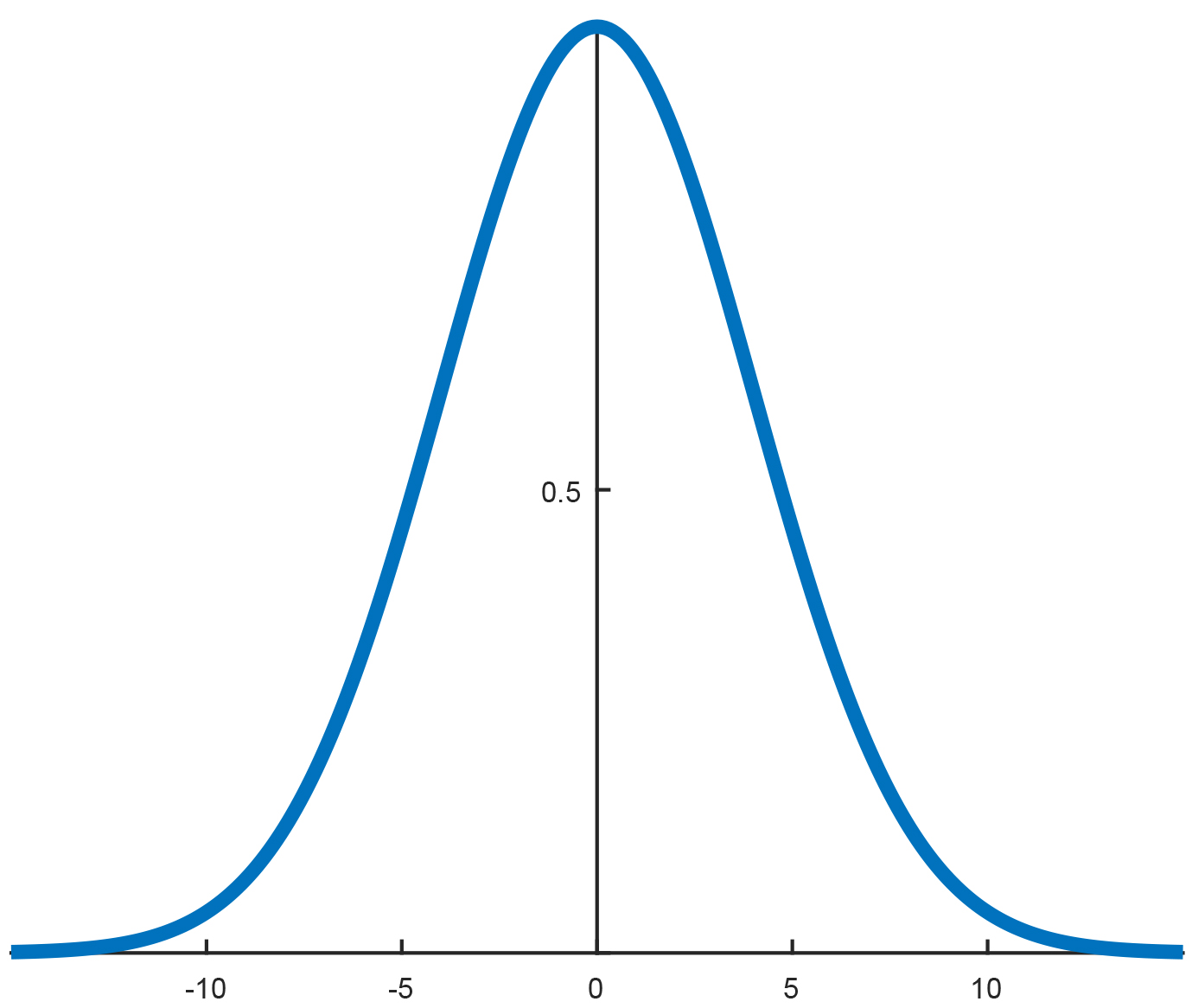} }}%
	\subfloat[$\Psi(x,\sigma_s)=xg(x,\sigma_s)$]{{\includegraphics[width=3.0cm]{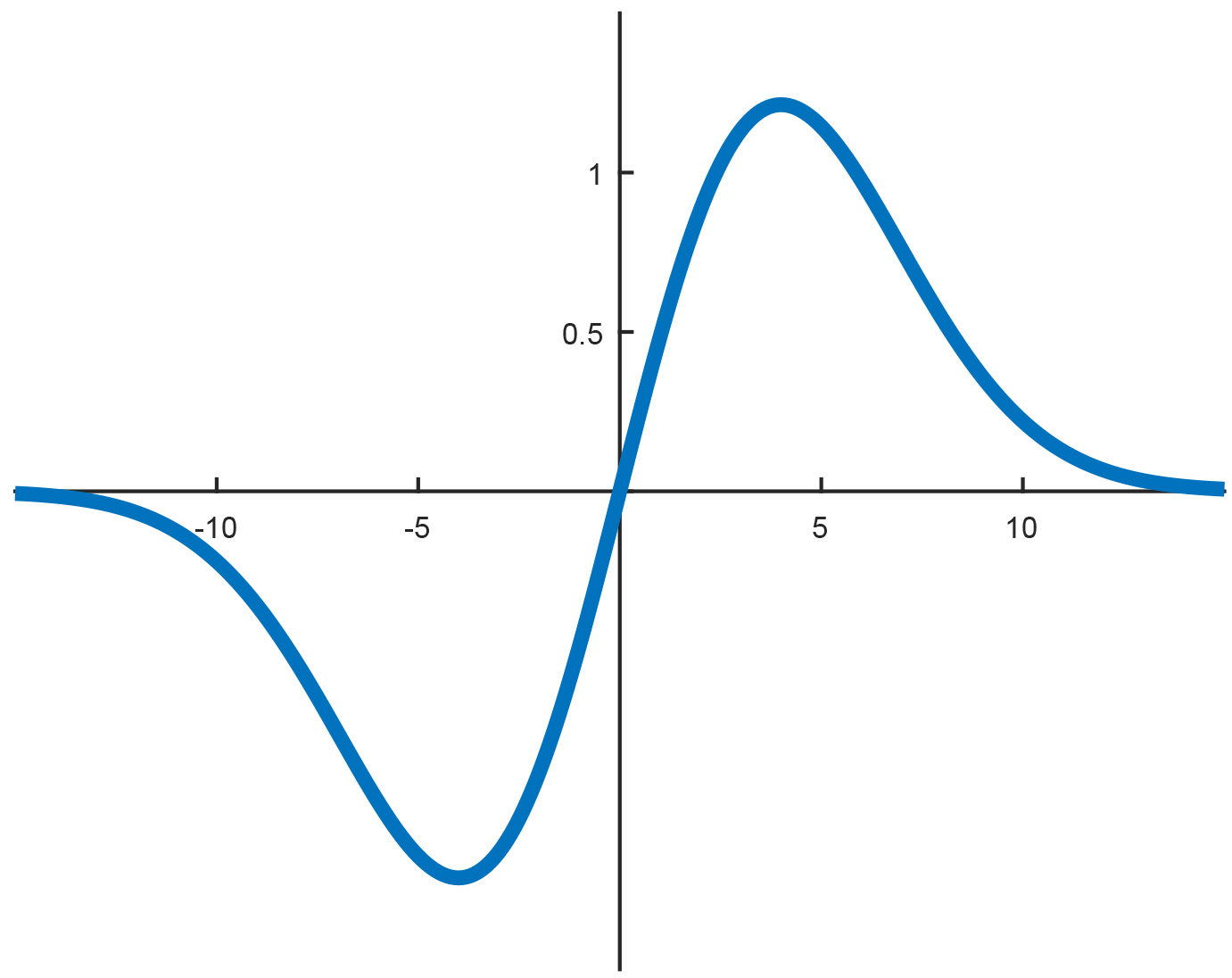} }}%
	\subfloat[$\rho(x,\sigma_s)$]{{\includegraphics[width=2.8cm]{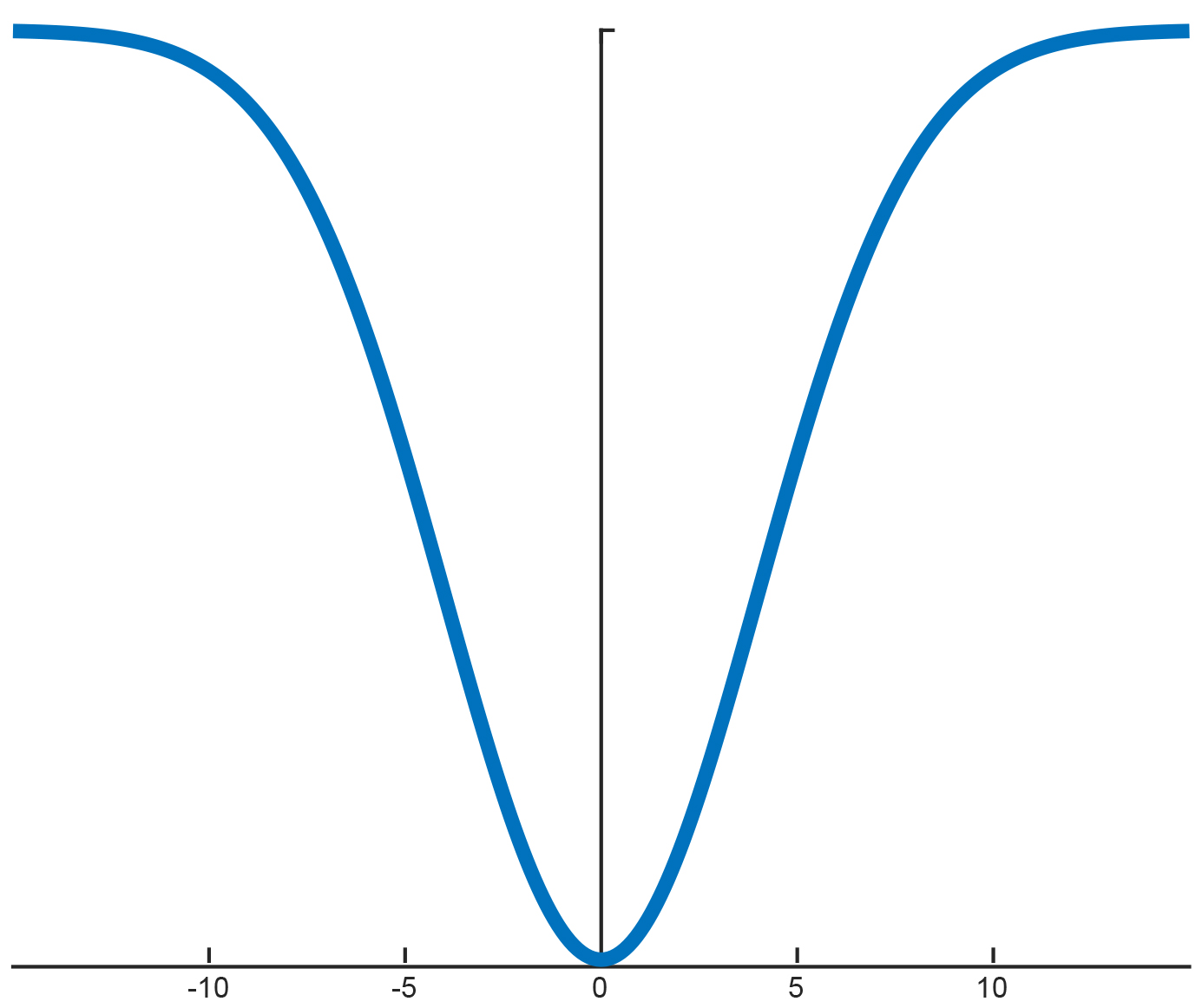} }}%
	\caption{A Gaussian-based similarity function and corresponding influence function and  $\rho$-function \cite{RoAni}.   }%
	\label{fig:RobustGauss}%
\end{figure}

\begin{figure}[bth]%
	\centering
	\subfloat[$g(x,\sigma_s)$]{{\includegraphics[width=2.8cm]{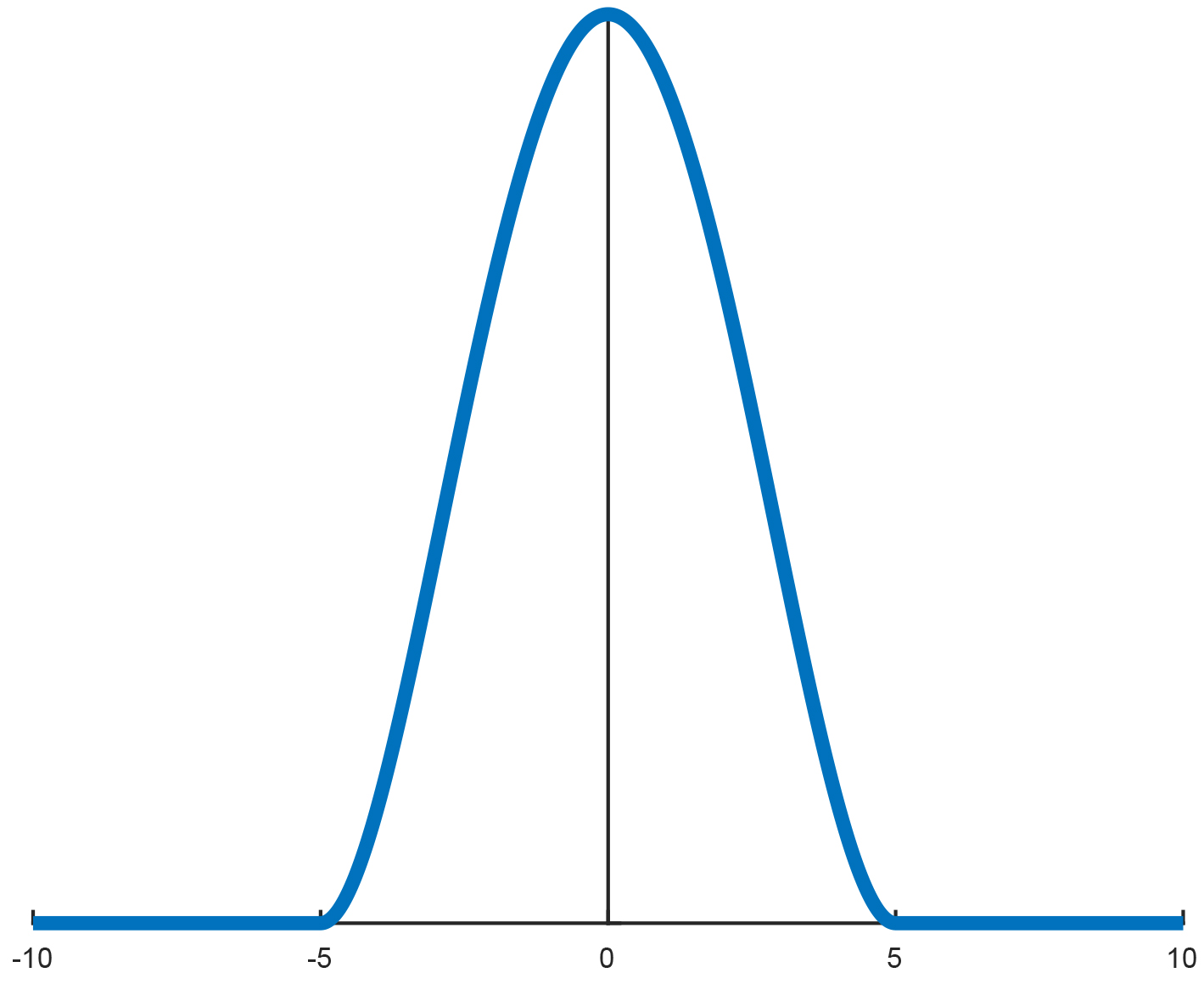} }}%
	\subfloat[$\Psi(x,\sigma_s)=xg(x,\sigma_s)$]{{\includegraphics[width=3.0cm]{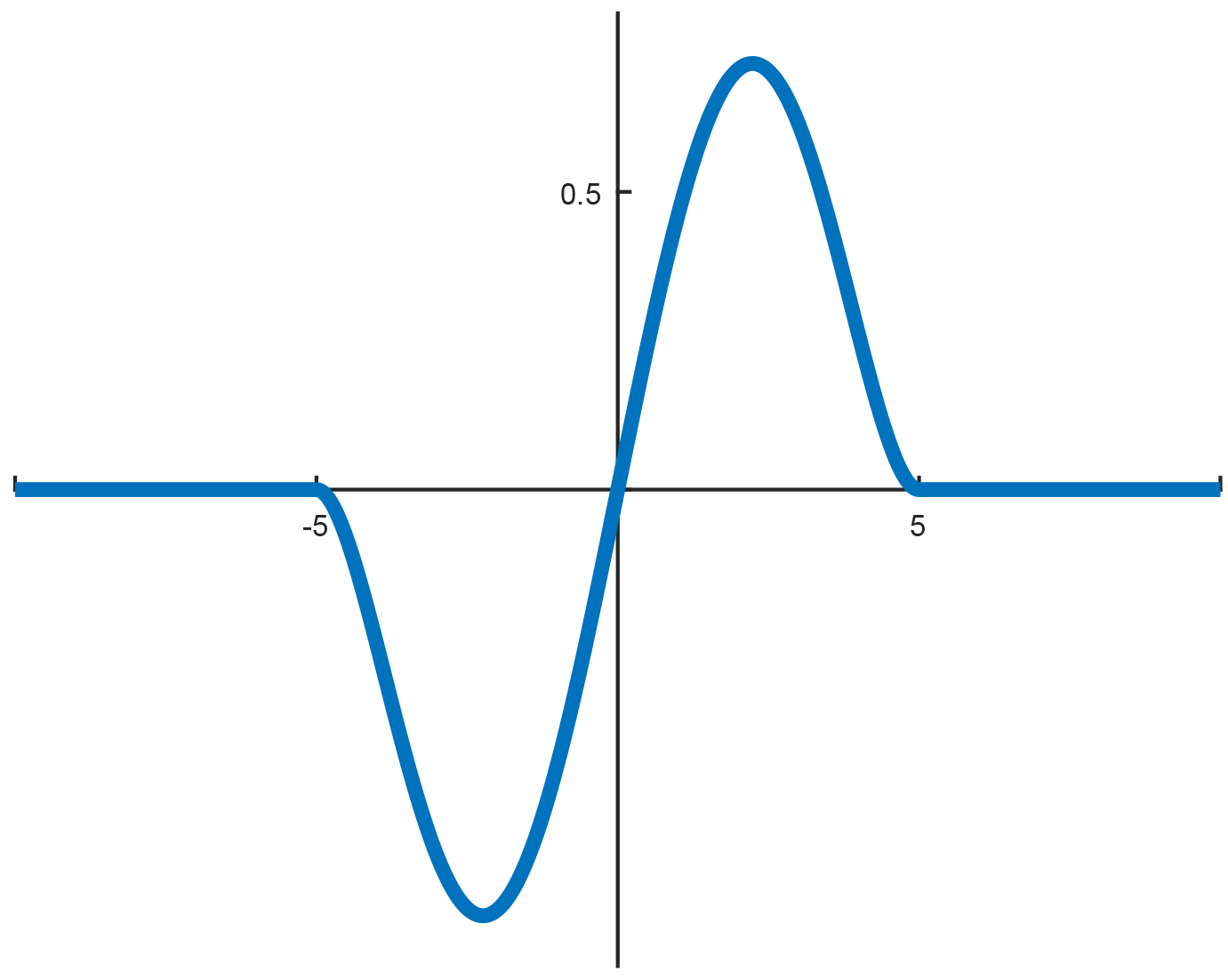} }}%
	\subfloat[$\rho(x,\sigma_s)$]{{\includegraphics[width=2.8cm]{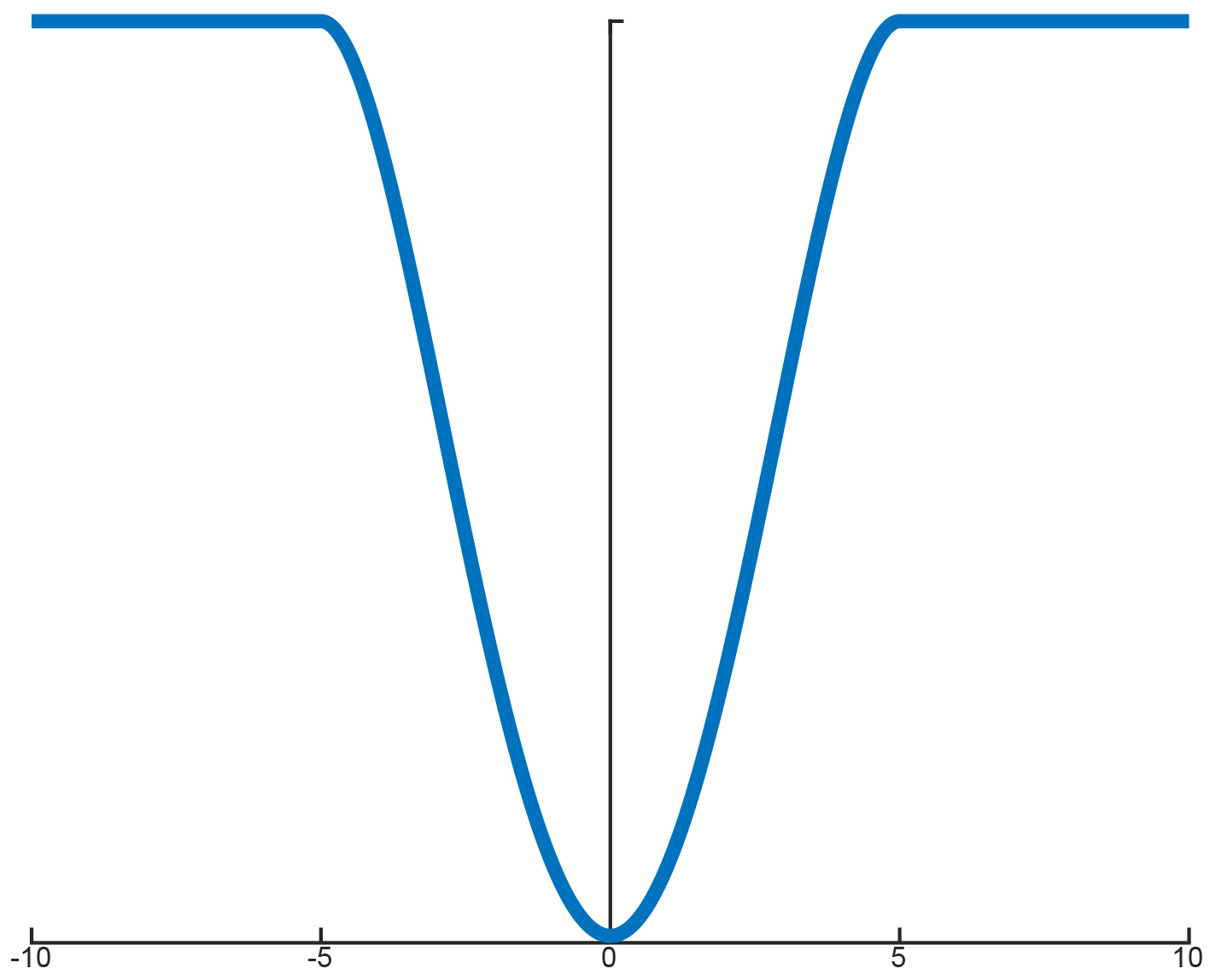} }}%
	\caption{Tukey bi-weight similarity function and corresponding influence function and  $\rho$-function \cite{RoAni}.   }%
	\label{fig:RobustTukey}%
\end{figure}
For example, the least square estimator (a quadratic $\rho$-function) will be the most naive estimator because it will have a linear influence function without any bounds and will be very sensitive to outliers. Figure \ref{fig:RobustGauss} shows the Gaussian-based $\rho$-function which has a bounded influence function $\Psi(x)$ and gives smaller weight to outliers compared to the least square estimators. Figure \ref{fig:RobustTukey} shows that \textit{Tukey's bi-weight function} produces more robust results compared to the least square and Gaussian-based estimators because it completely cuts off the diffusion at sharp features (outliers) and removes noise from non-feature areas. \textit{Tukey's bi-weight function} is defined as\cite{RoAni}:
\begin{equation}
g(x,\sigma_s) = \begin{cases} \frac{1}{2} \{1-(x/\sigma_s)^2\}^2 &\mbox{if } \lvert x\rvert \leq \sigma_s, \\
0 & \mbox{if } \lvert x\rvert > \sigma_s. \end{cases}
\label{equ:LBP} 
\end{equation}
In terms of preserving features, \textit{Tukey's bi-weight function} is more effective compared to the Gaussian function because it does not allow the diffusion across sharp features and removes noise components along these sharp features effectively. In our algorithm, we use \textit{Tukey's bi-weight function} as the similarity function to produce feature preserving smooth face normals. For the closeness function $f(\mathbf{c}_i-\mathbf{c}_j,\sigma_c)$, a Gaussian function is used.   
\subsection{High Fidelity Mesh Reconstruction}
In the second step of the denoising process, we need to impose the effect of processed face normals to the corresponding vertices.
To reconstruct a high fidelity mesh, we minimize the following quadratic energy function:
\begin{equation}
\begin{aligned}
		\operatorname*{min}_{\tilde{\mathbf{v}}_i} \Big\{  {\| \mathbf{v}_i - \tilde{\mathbf{v}}_i \|}^2 + \sum\limits_{j=0}^{N_v(i)-1}\sum\limits_{(i,j)\in \partial F_k}^{} {\| \tilde{\mathbf{n}}_k\cdot (\mathbf{v}_i -\mathbf{v}_j)\|}^2\\ + {\| \mathbf{R}_i\|}^2 \Big \},
	\label{vertEnergy}
\end{aligned}
\end{equation} 
where $\tilde{\mathbf{v}}_i$ and ${\mathbf{v}}_i$ are the newly computed and the noisy vertex positions. The terms $N_v(i)$ and $\partial F_k$ represent the number of vertices and the boundary edge of the vertex star of $\mathbf{v}_i$ respectively. The second term of the energy function synchronizes a vertex position to the corresponding face normal by using the concept of orthogonality between edge vector and face normal\cite{Taubin01linearanisotropic}.  The term $\mathbf{R}_i$ helps algorithm to produce a high fidelity mesh and makes the proposed algorithm robust against high-intensity noise. This term is computed using a differential coordinate and can be written as:
\begin{equation}
 \mathbf{R}_i = \mathbf{D}_i + \mathbf{D}_{t_i} ,
\end{equation} 
\begin{figure}[bth]%
	\centering
	\subfloat[$\mathbf{D}_i = 0$]{{\includegraphics[width=2.5cm]{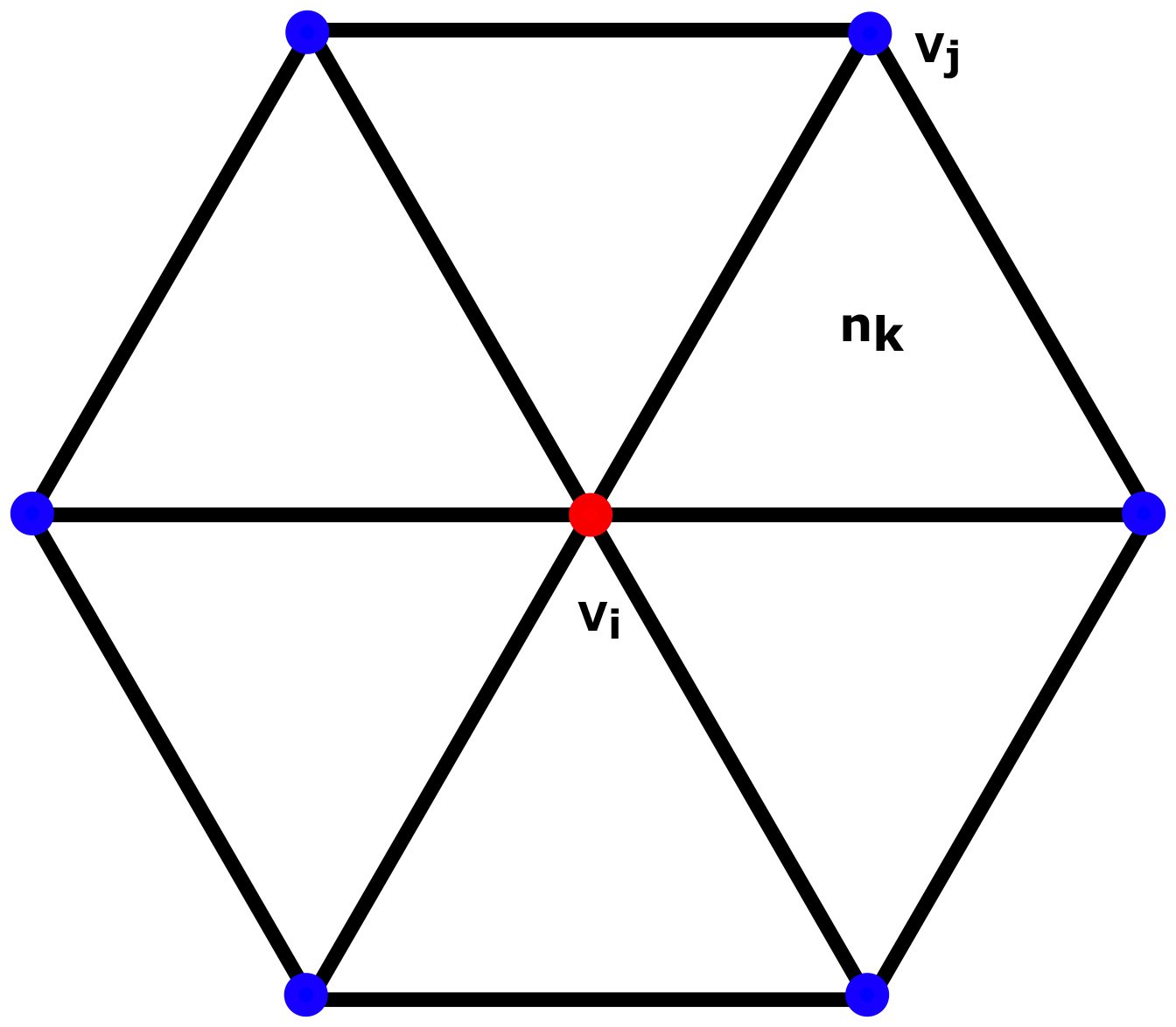} }}%
	\subfloat[$\mathbf{D}_i \neq 0$]{{\includegraphics[width=2.5cm]{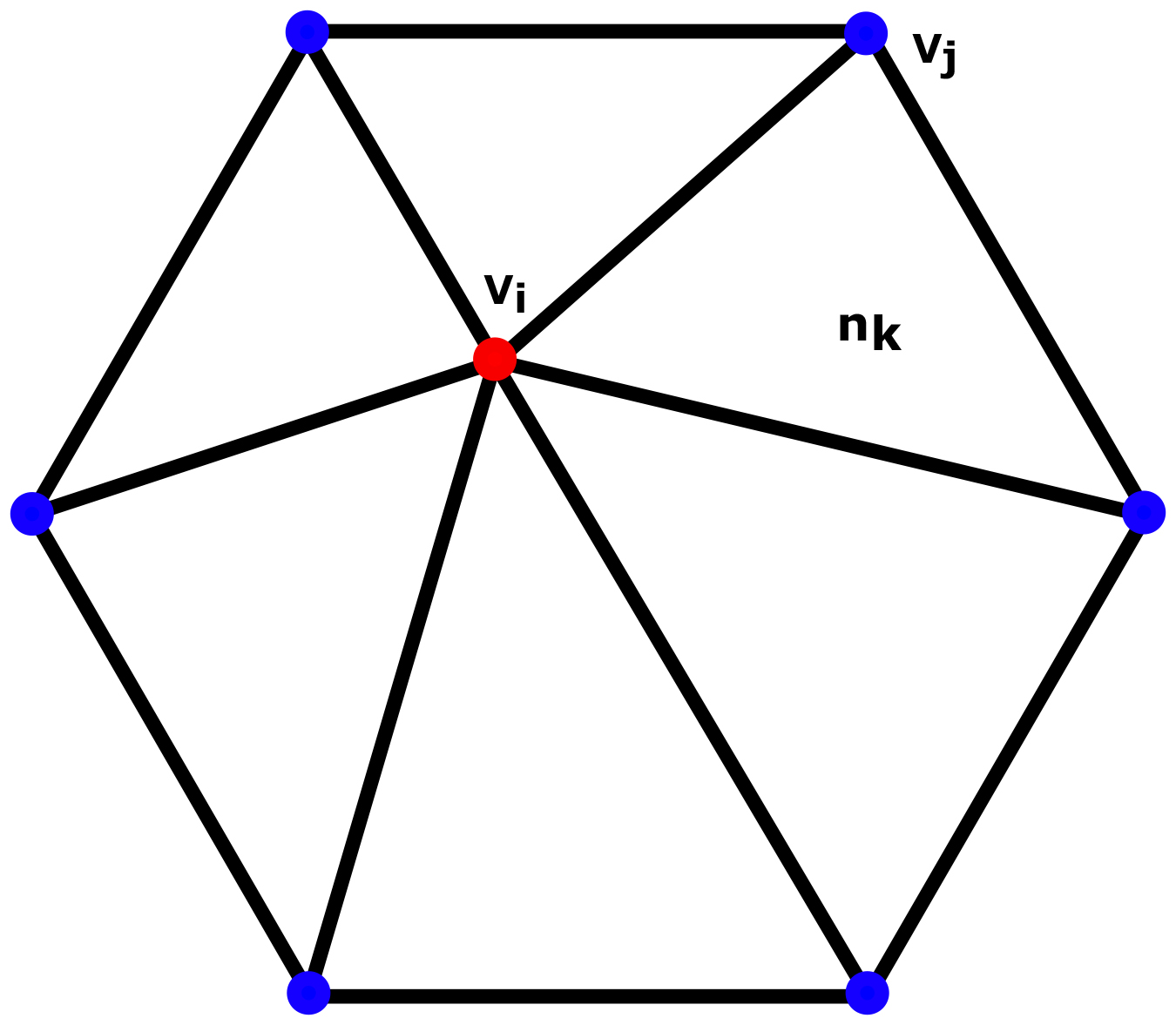} }}%
	\caption{ An example to show the effect of differential coordinate $\mathbf{D}_i$ at the second stage of the proposed algorithm.    }%
	\label{fig:Mesh}%
\end{figure}
where $\mathbf{D}_i$ and $\mathbf{D}_{t_i}$ represent the differential coordinate and its tangential component respectively. The differential coordinate at the vertex $\mathbf{v}_i$ and is defined as: 
\begin{equation}
\mathbf{D}_i=\triangle\mathbf{v} _i-\mathbf{v} _i,  
\end{equation}
where 
\begin{equation}
\triangle\mathbf{v} _i=\frac{ 1 }{\sum\limits_{j\in N(i)}^{}{w}_{ij} } \sum\limits_{j\in N(i)}^{}\\ {w}_{ij}\mathbf{v}_j.
\end{equation}
The weighting term is defined as $w_{ij}=\|\mathbf{v}_i-\mathbf{v}_j\|^2$, which is basically based on the edge length of connected vertices. As shown in Figure \ref{fig:Mesh}, the differential coordinate $\mathbf{D}_i$ computes the deviation vector from the centroid of the 1-ring neighbour.

 \begin{figure}[bth]%
 	\centering
 	\subfloat[Noisy]{{\includegraphics[width=2.8cm]{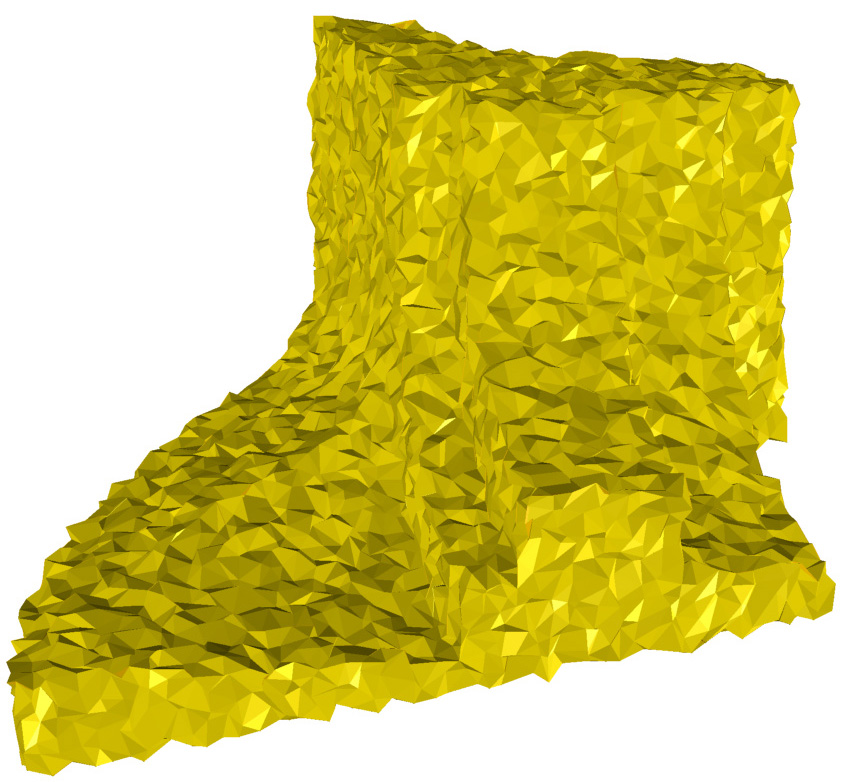} }}%
 	\subfloat[Tangent direction]{{\includegraphics[width=2.8cm]{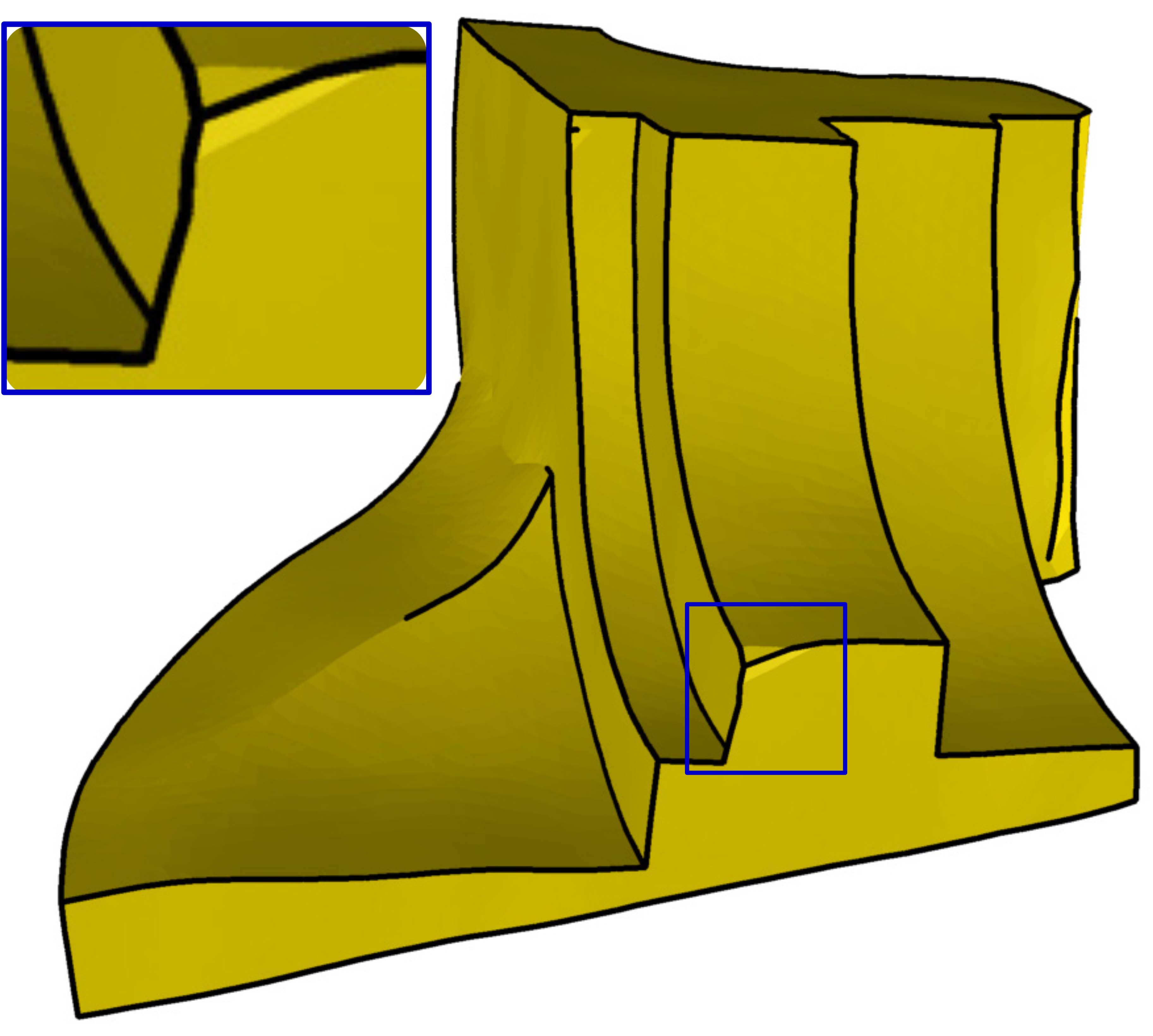} }}%
 	\subfloat[Using Equation (\ref{vertEnergy})]{{\includegraphics[width=2.8cm]{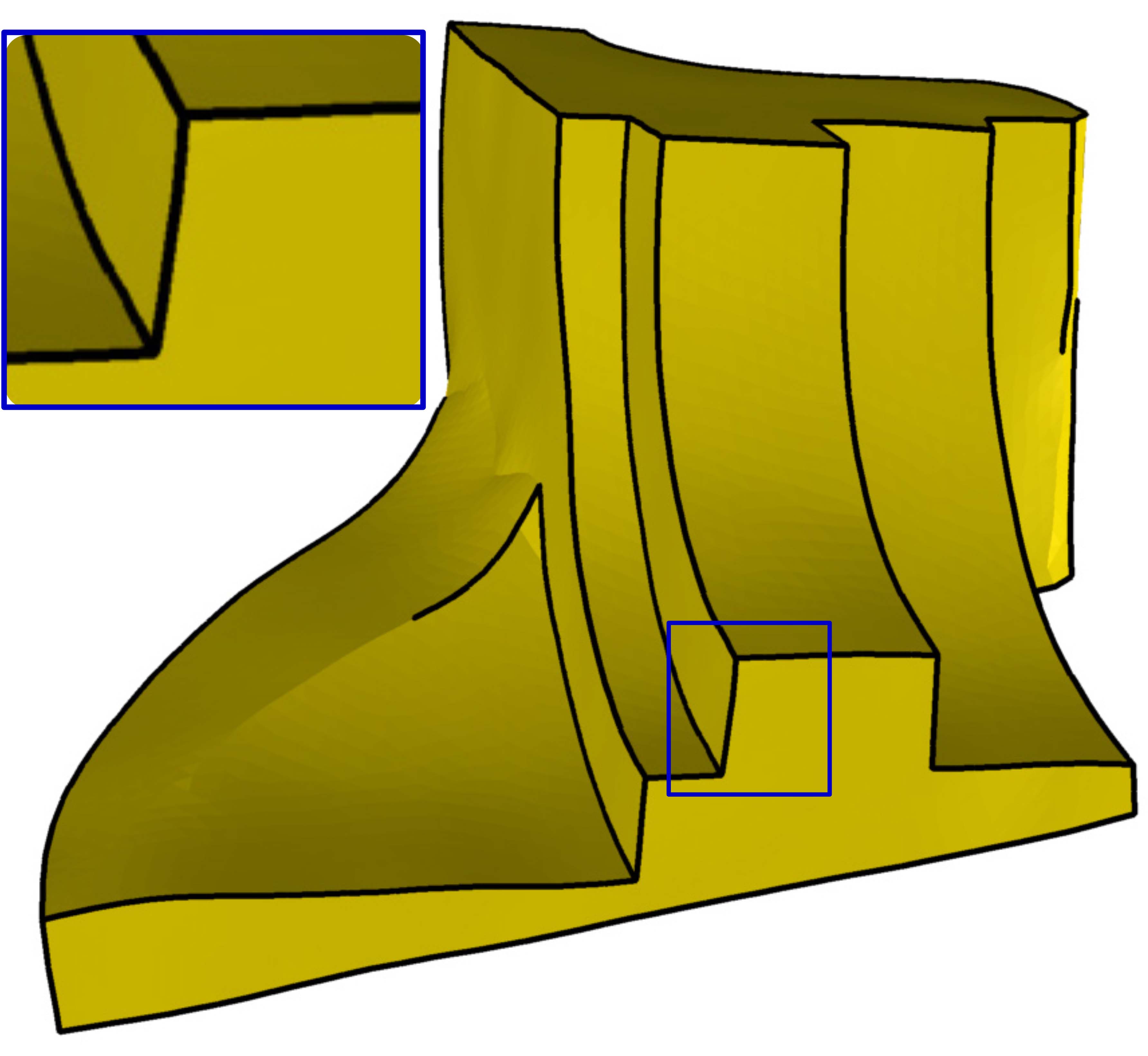} }}%
 	\caption{ The Fandisk model corrupted by a uniform noise ($\sigma_n = 0.65 l_e$) in random direction where $l_e$ is the average edge length. Figure (b) and (c) show effect of the diffusion of $\mathbf{D}_i$ only in the tangent direction and using Equation (\ref{vertEnergy}) respectively. The sharp feature curve (black curve) is computed using the dihedral angle ($\theta = 35^\circ$). The magnified view shows that the diffusion only in tangent direction is not able to reconstruct the sharp corner.}%
 	\label{fig:tanDir}%
 \end{figure}
 
 \begin{figure*}
 		\centering
 			\subfloat[Noisy]{\includegraphics[width=2.6cm]{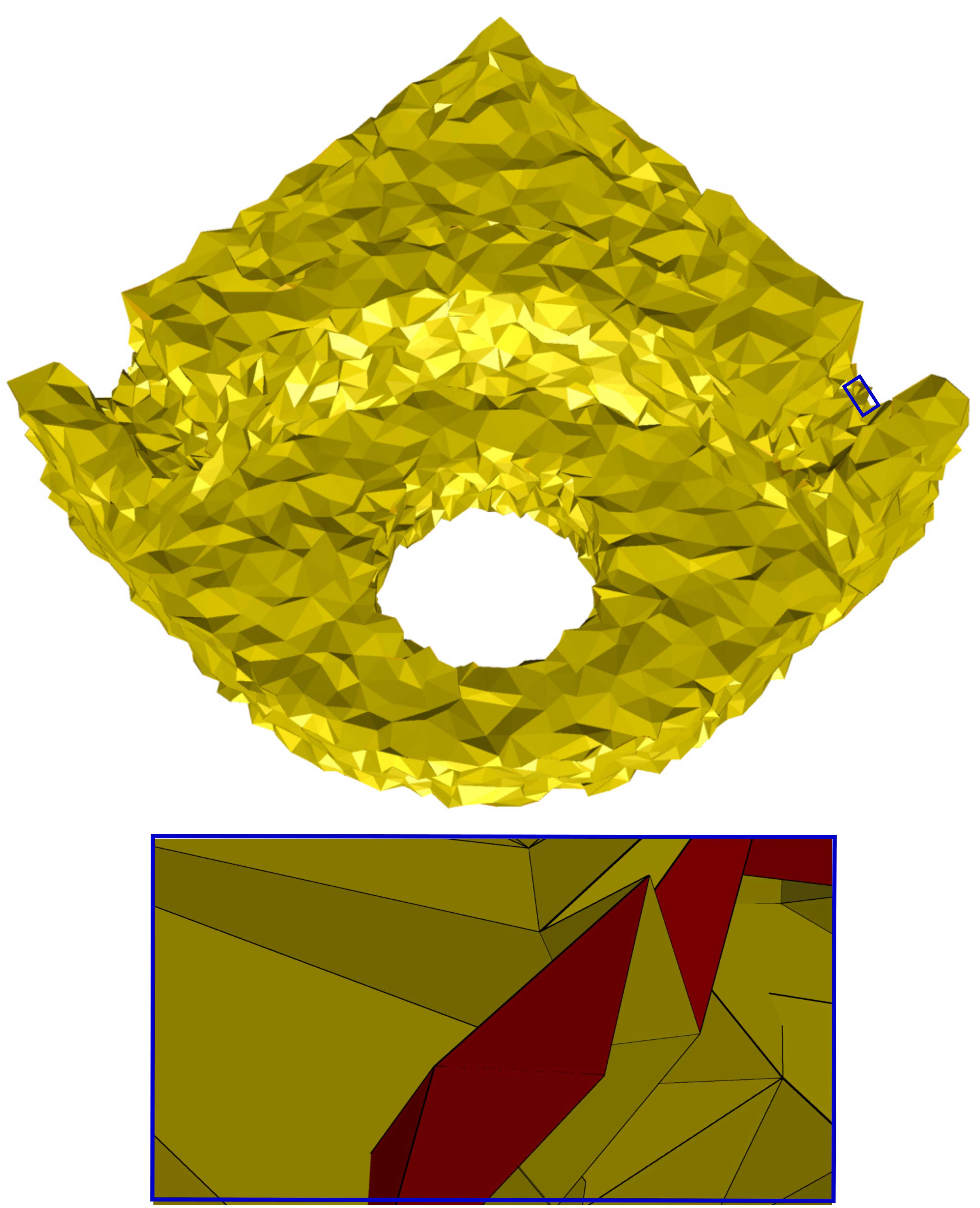}} 
 			\subfloat[$\sigma_s= 0.1$]{\includegraphics[width=2.6cm]{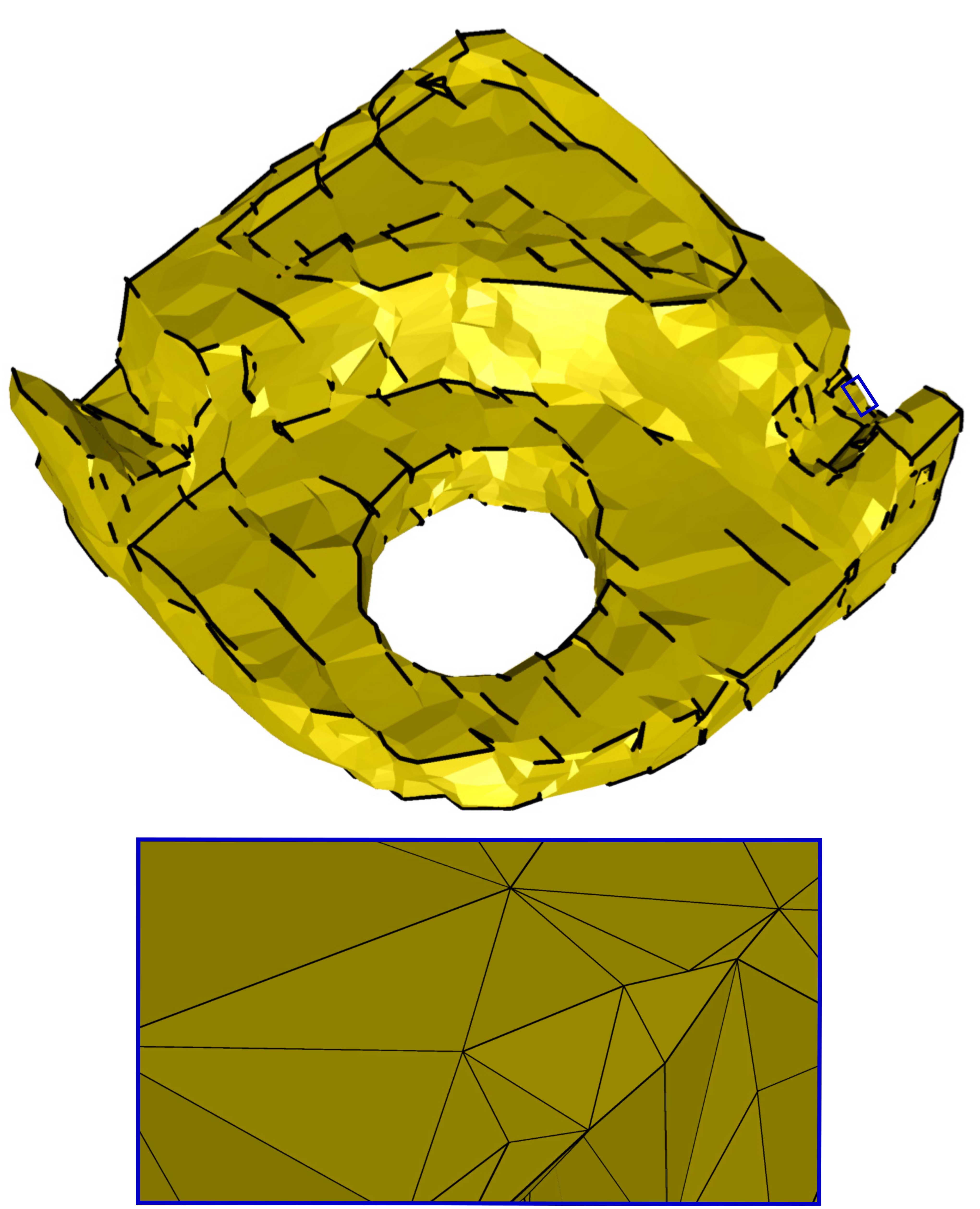}} 
 			\subfloat[$\sigma_s= 0.5$]{\includegraphics[width=2.6cm]{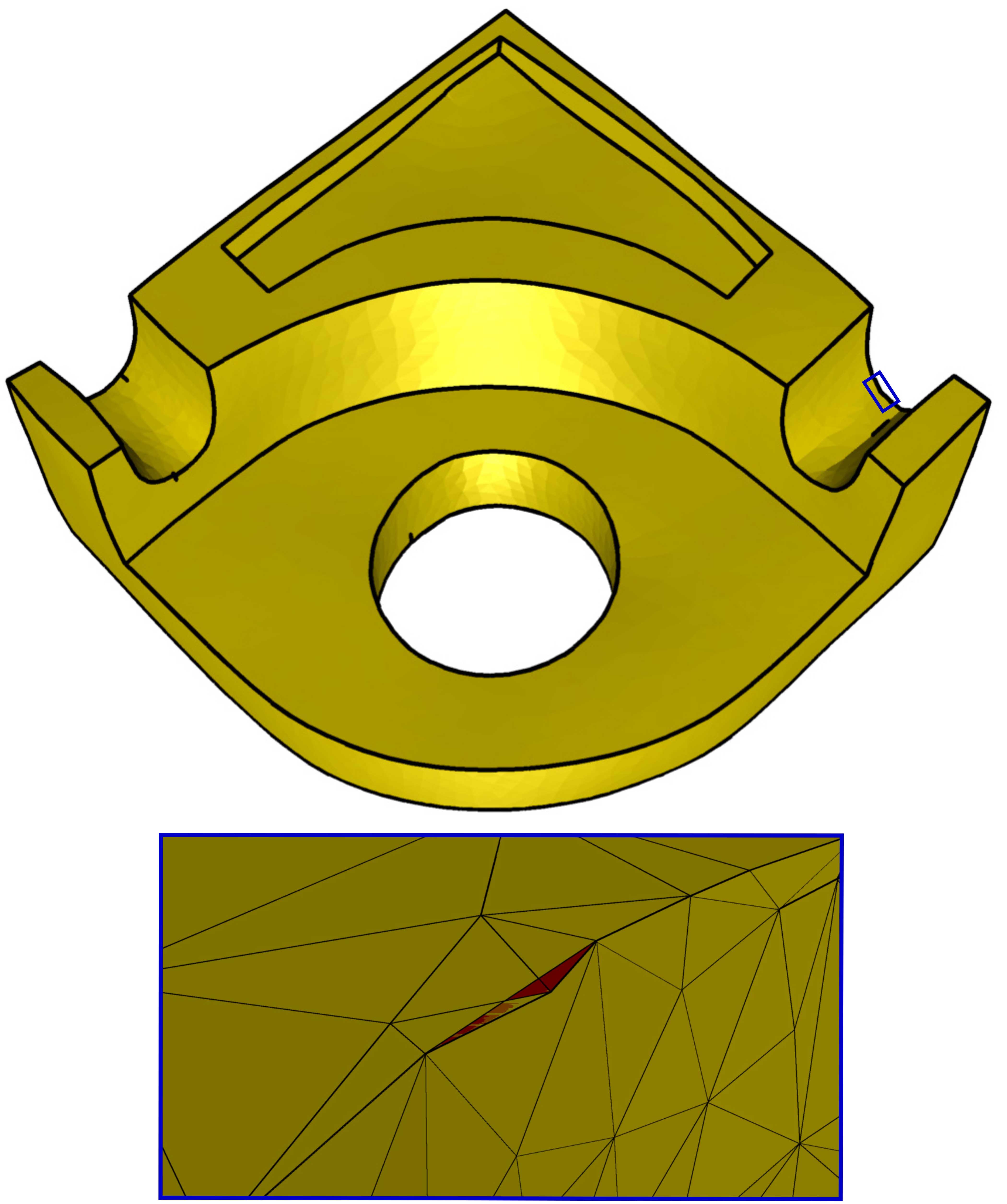}}
 			\subfloat[$\sigma_s= 1.5$]{\includegraphics[width=2.6cm]{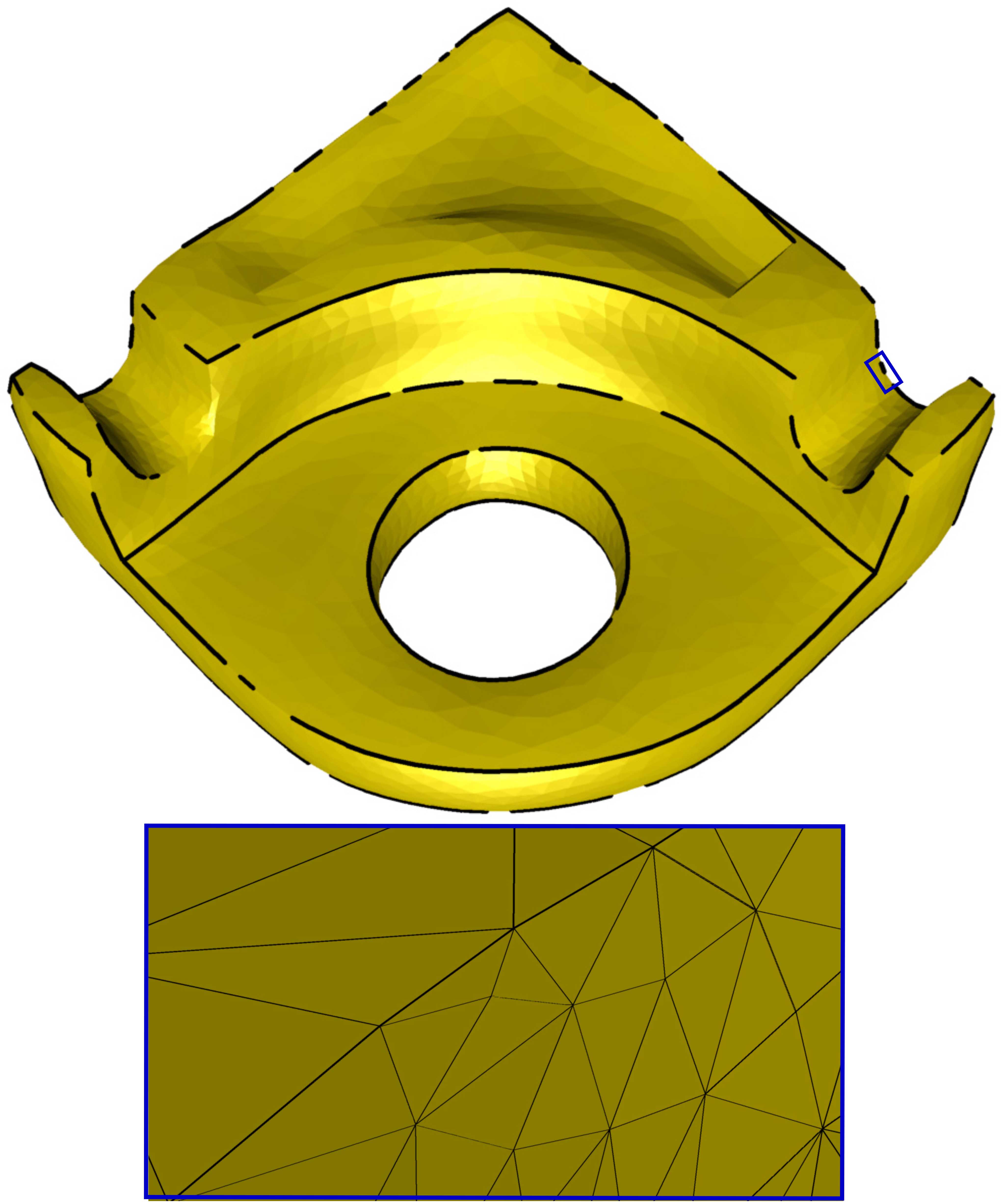}}
 			\subfloat[$\lambda_I= 0.0$]{\includegraphics[width=2.6cm]{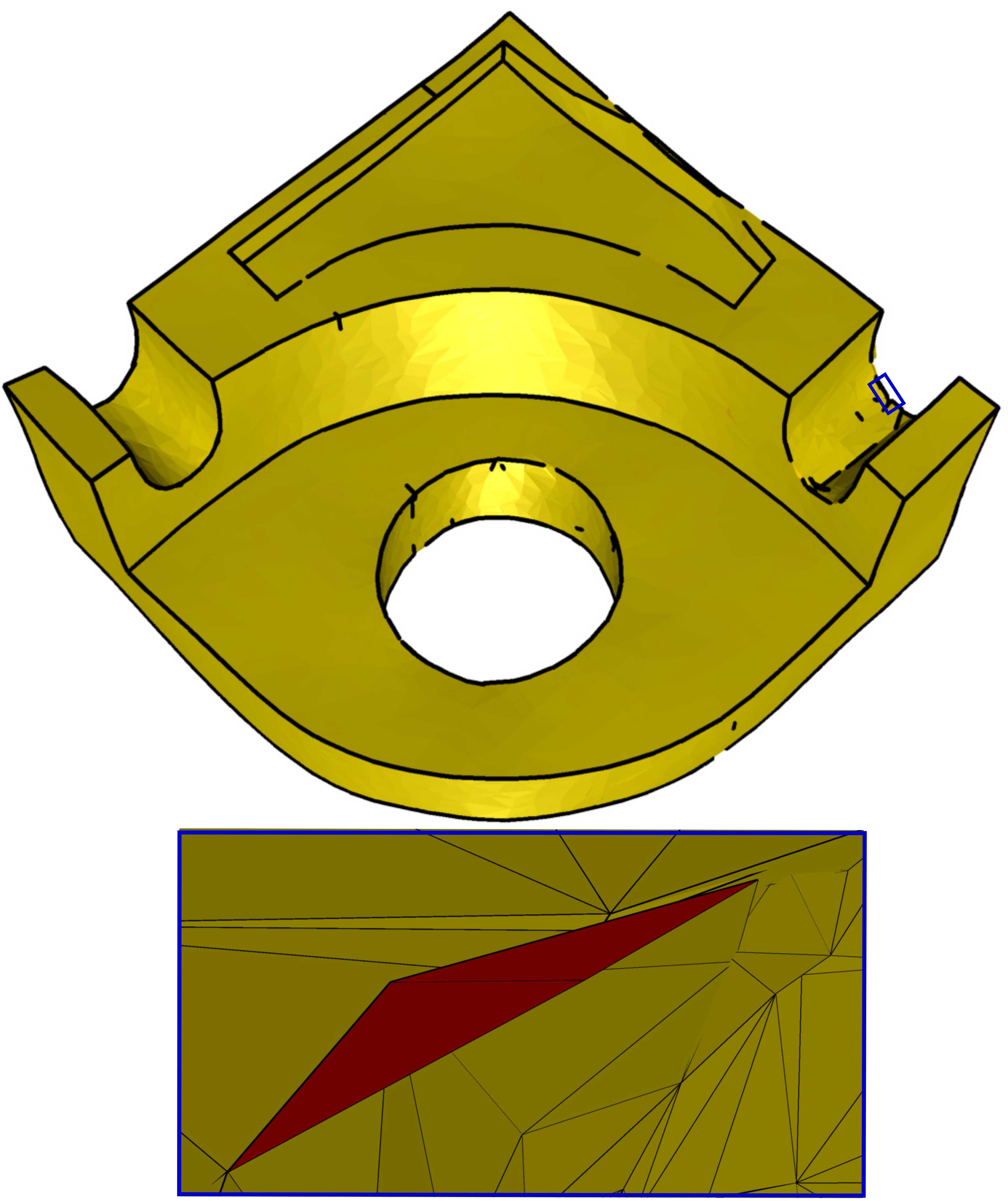}}
 			\subfloat[$\lambda_I= 0.2$]{\includegraphics[width=2.6cm]{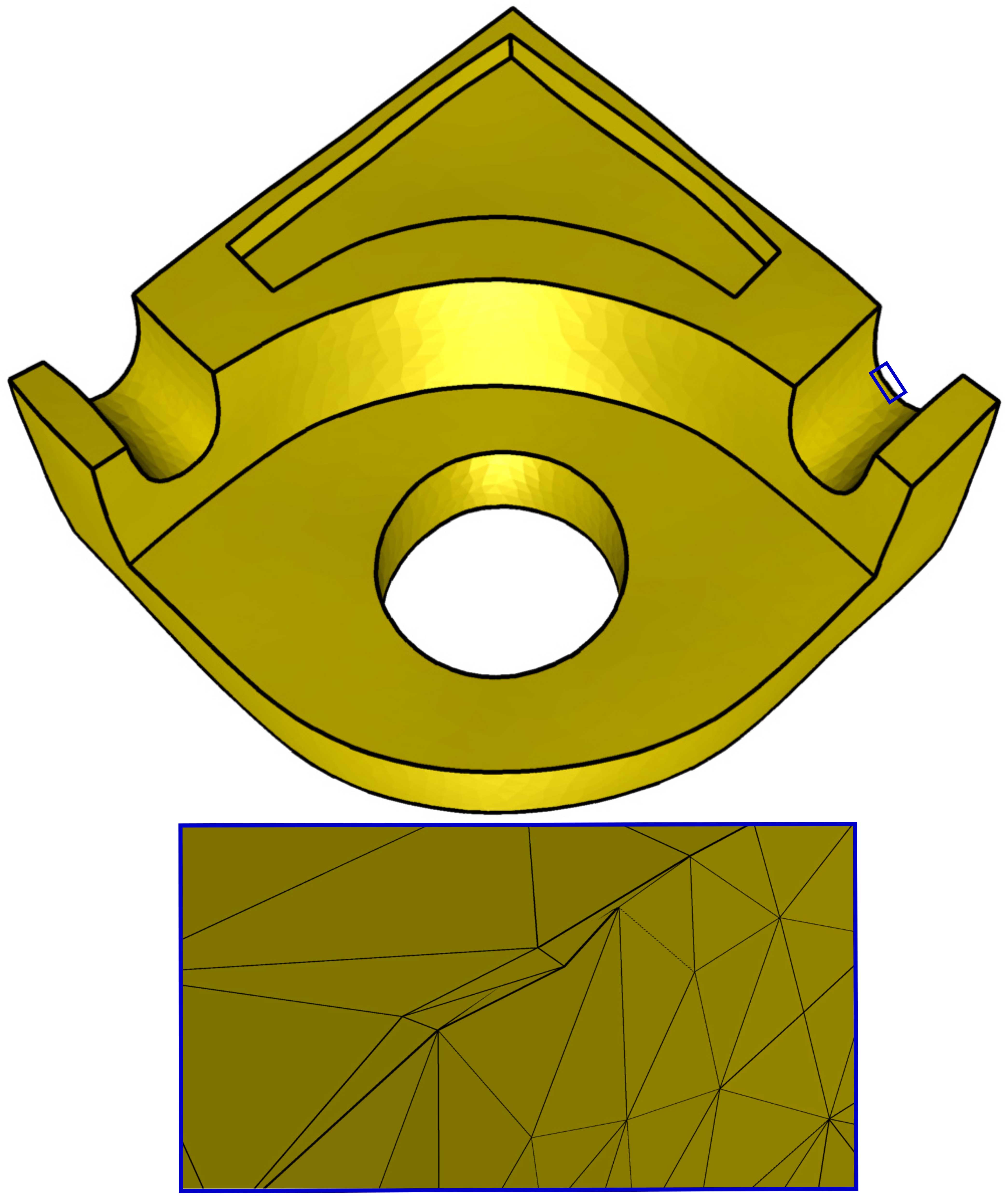}} 
 			\subfloat[$\lambda_I= 0.6$]{\includegraphics[width=2.6cm]{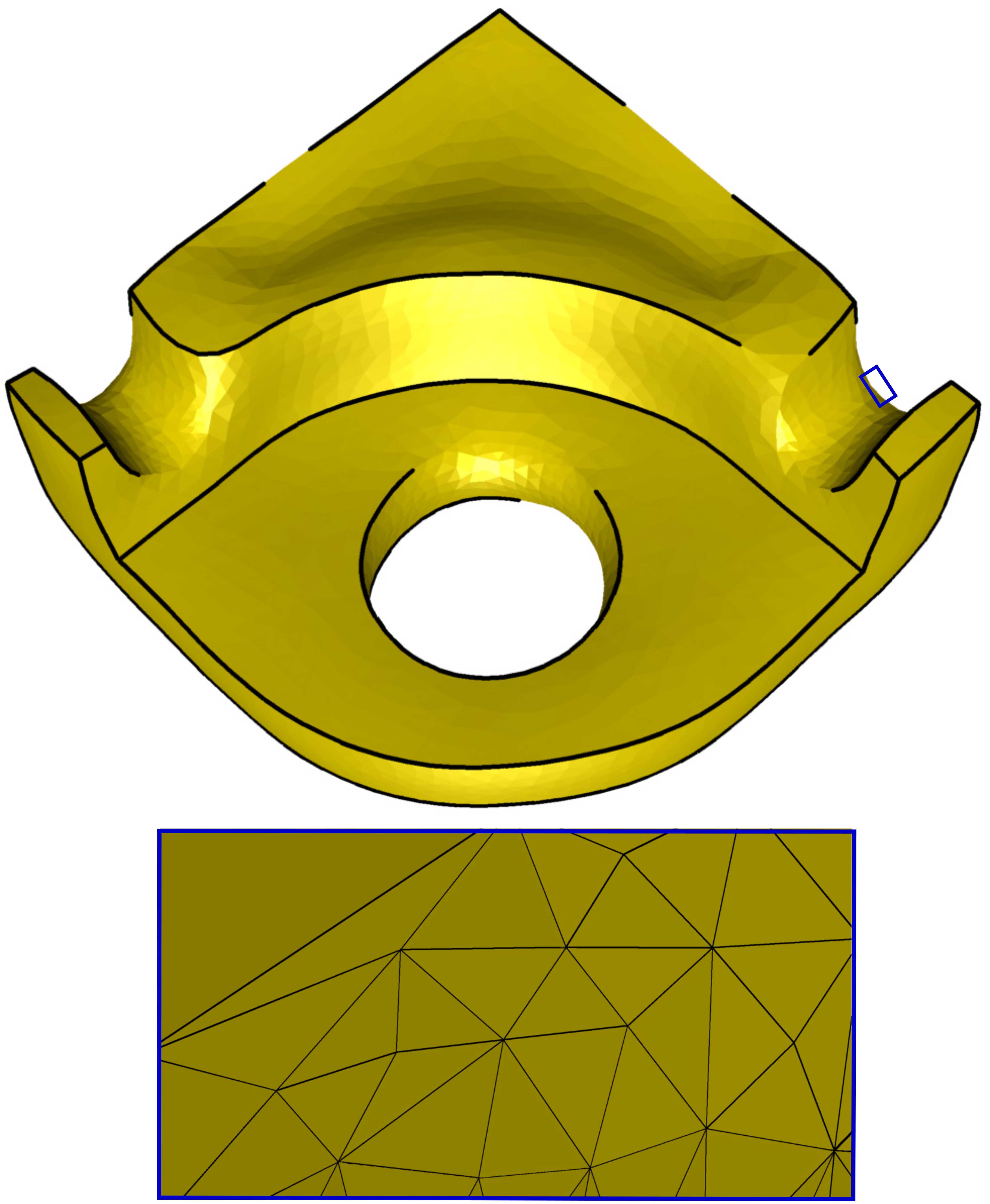}} 
 	\caption{The Bearing model is corrupted with an uniform noise ($\sigma_n = 0.5l_e$) in random direction (a). For Figure (b)-(d), $\lambda_I = 0.02$ and for Figure (e)-(g), $\sigma_s = 0.8$. The black curve shows the sharp edge information in the smooth geometry and is detected using the dihedral angle $\theta = 65^ \circ$. The second row shows the magnified view mesh structure at a sharp edge. The red color faces are with wrong orientation (flipped normals).}
 	\label{fig:params}
 \end{figure*}

 \begin{figure*}
 	\def\tabularxcolumn#1{m{#1}}
 	\begin{tabularx}{\linewidth}{@{}cXX@{}}
 		\centering
 		\begin{tabular}{ccccc}
 			\subfloat[Original]{\includegraphics[width=3.3cm]{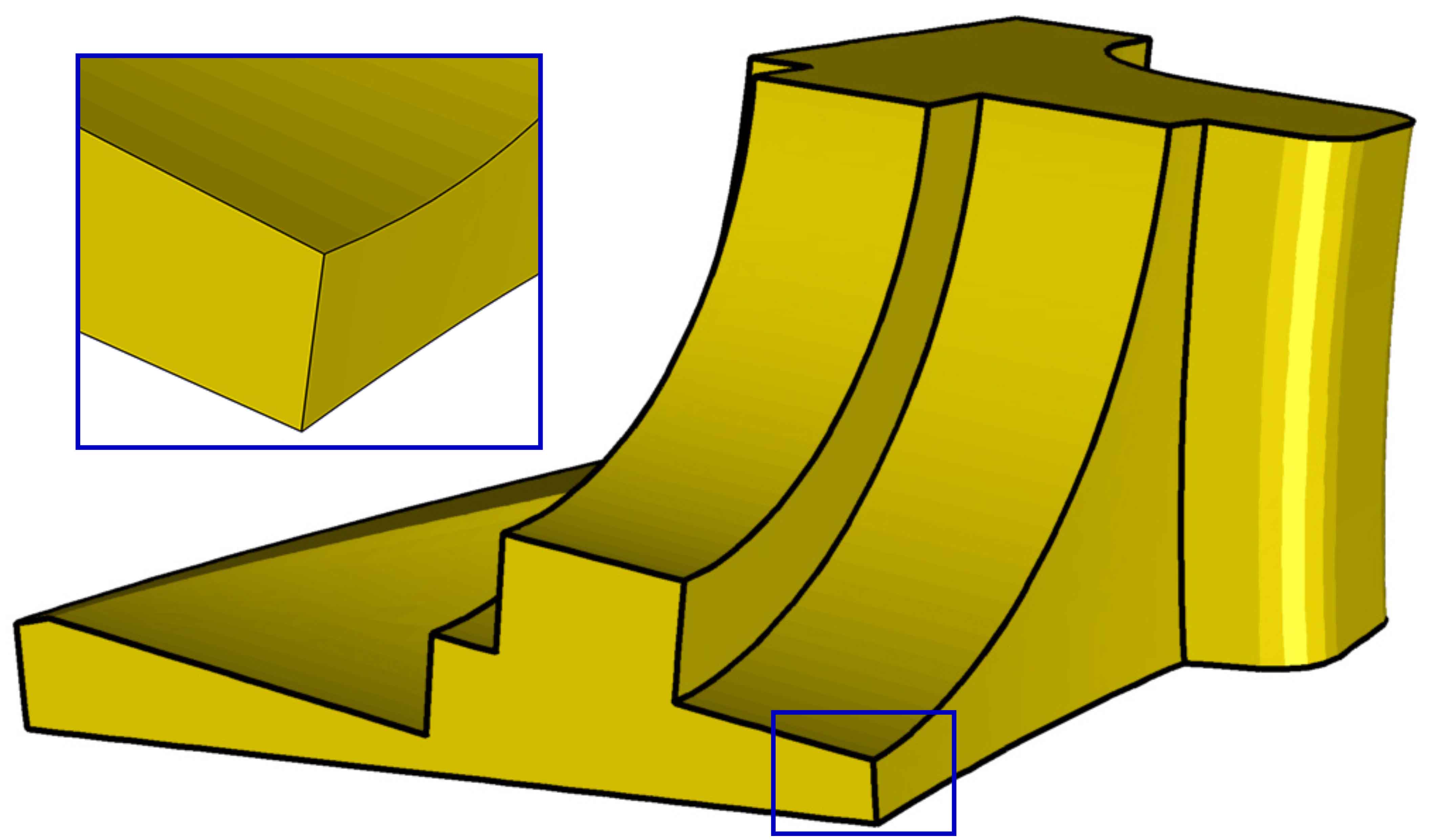}} 
 			& \subfloat[Noisy]{\includegraphics[width=3.3cm]{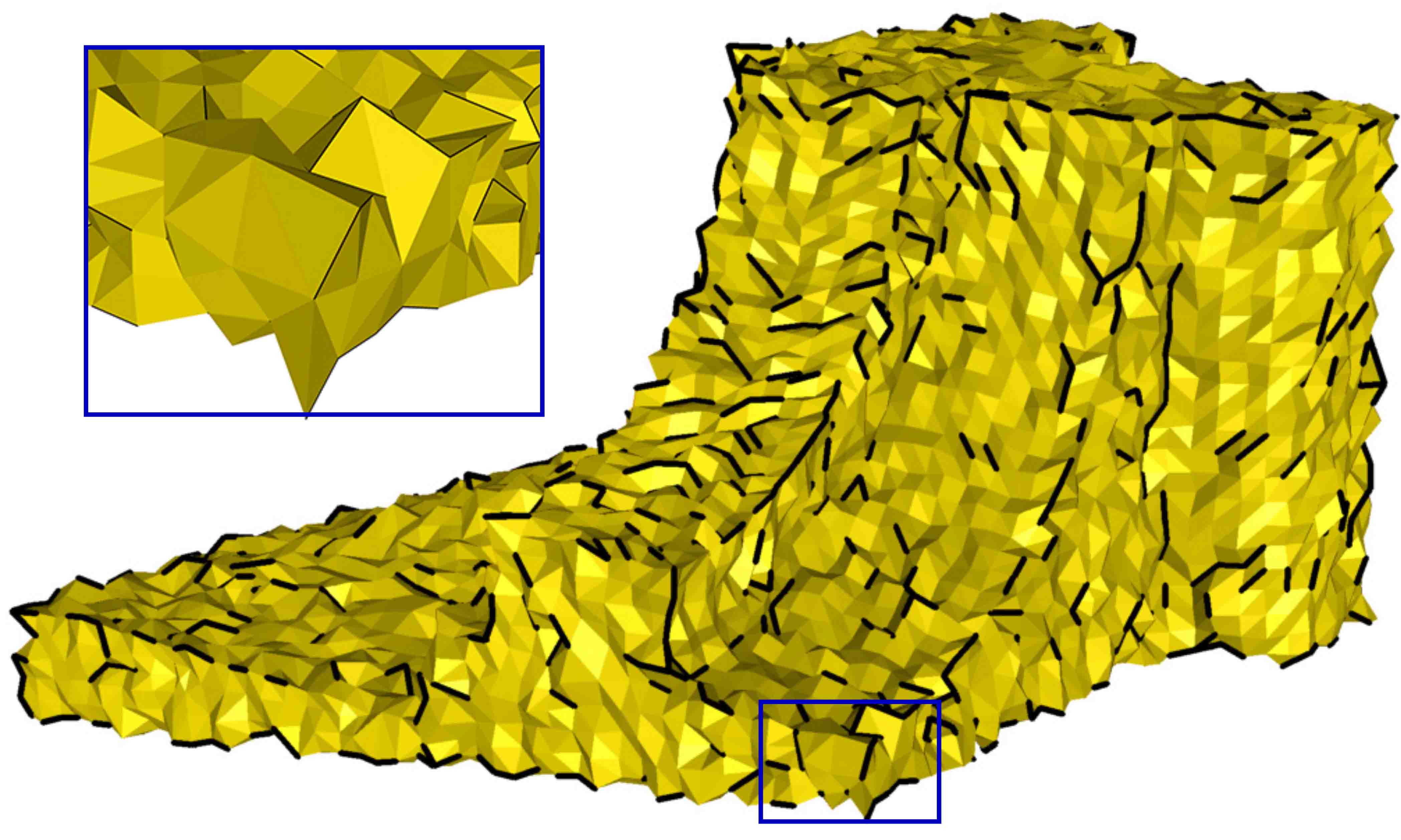}} &
 			 \subfloat[\cite{aniso}]{\includegraphics[width=3.3cm]{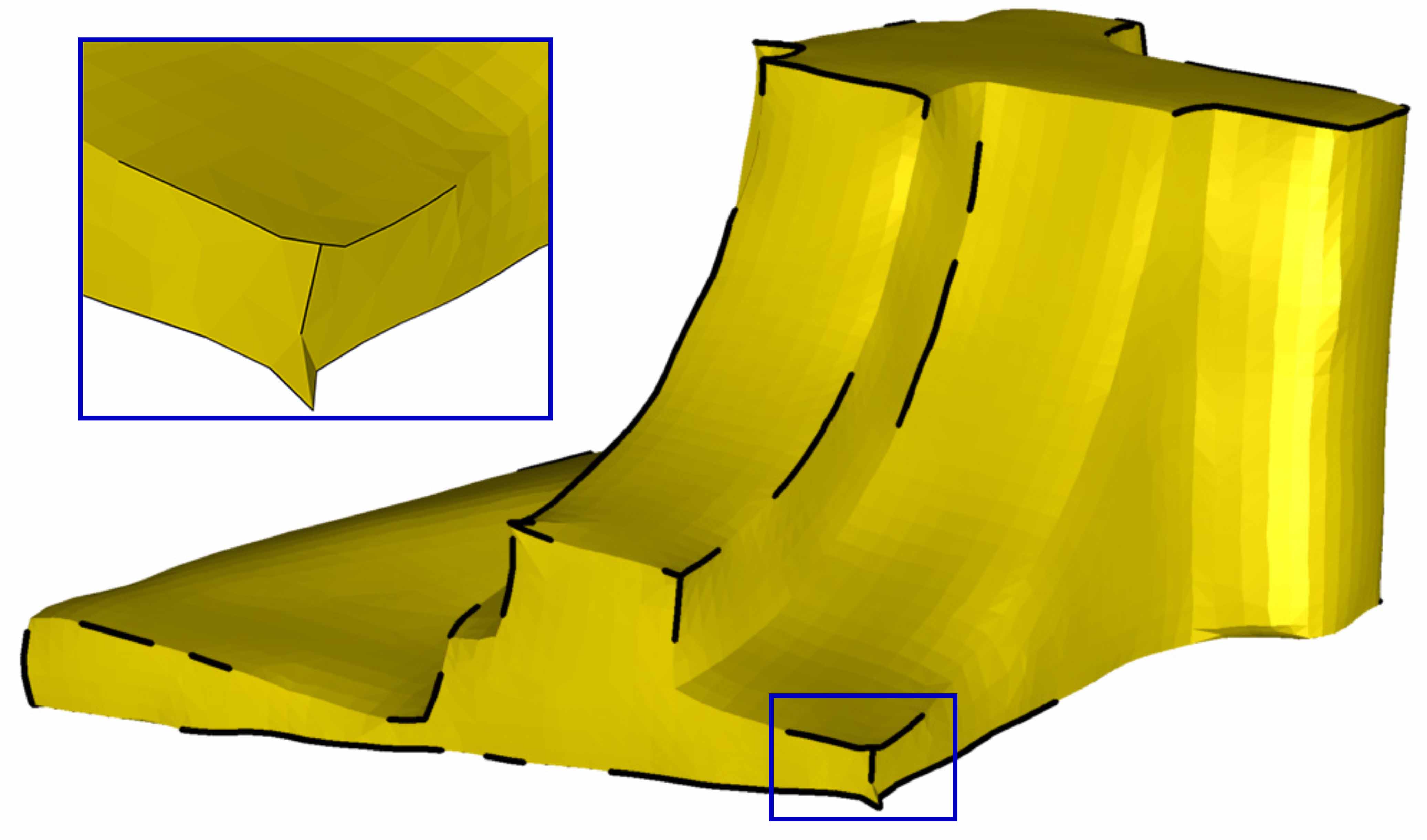}}&
 			  \subfloat[\cite{BilNorm}]{\includegraphics[width=3.3cm]{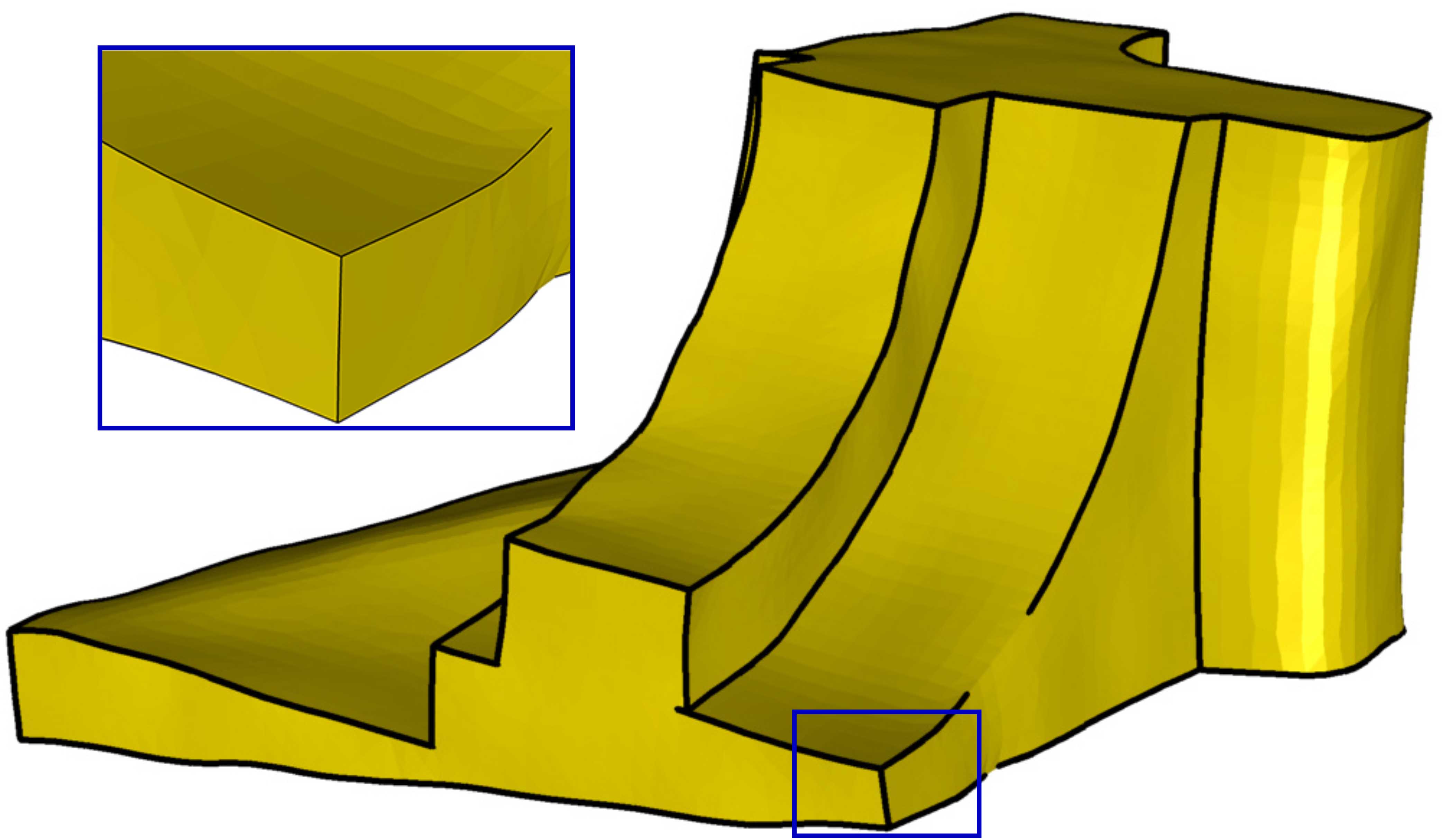}}&
 			\subfloat[\cite{L0Mesh}]{\includegraphics[width=3.3cm]{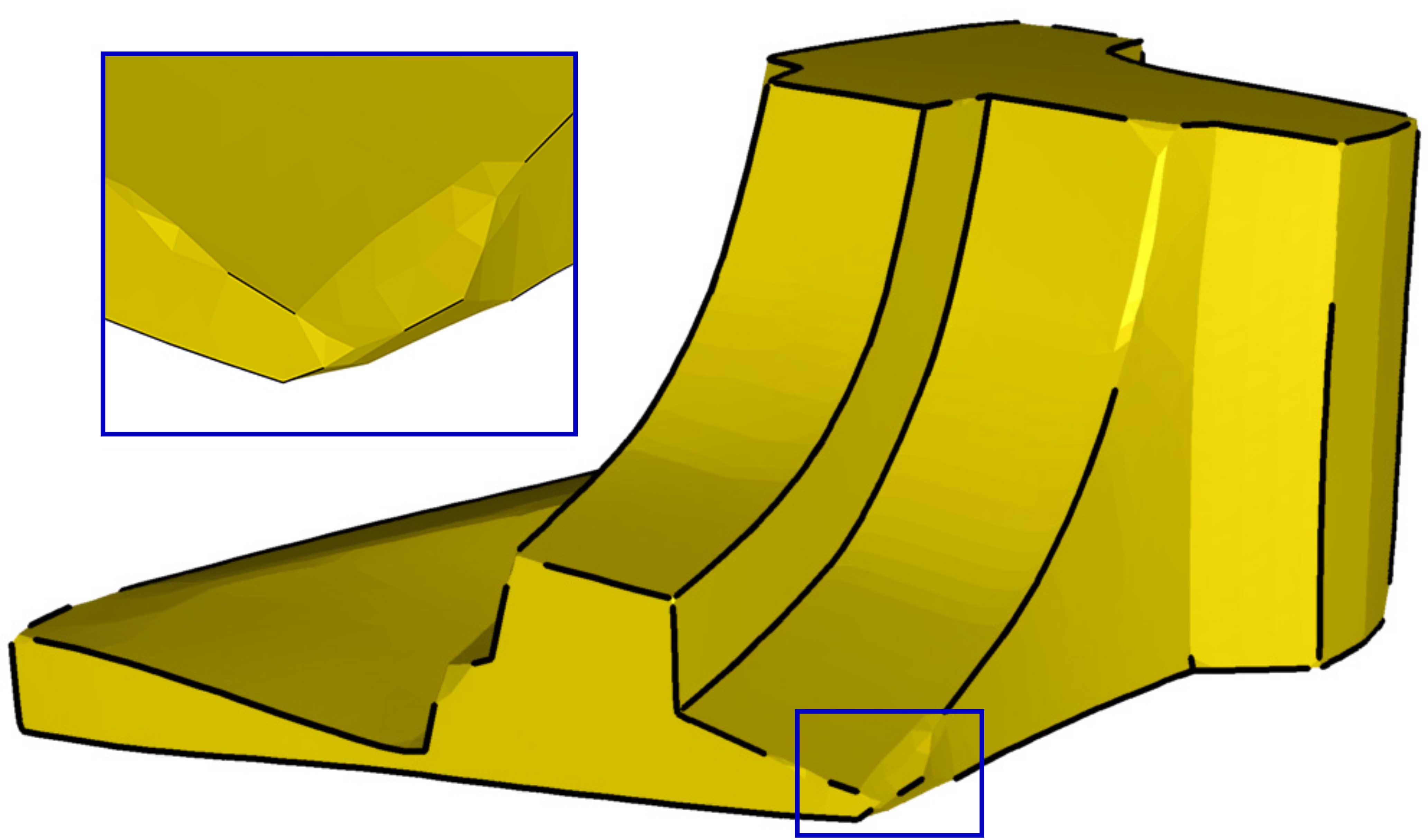}}\\
 			\subfloat[\cite{Guidedmesh}]{\includegraphics[width=3.3cm]{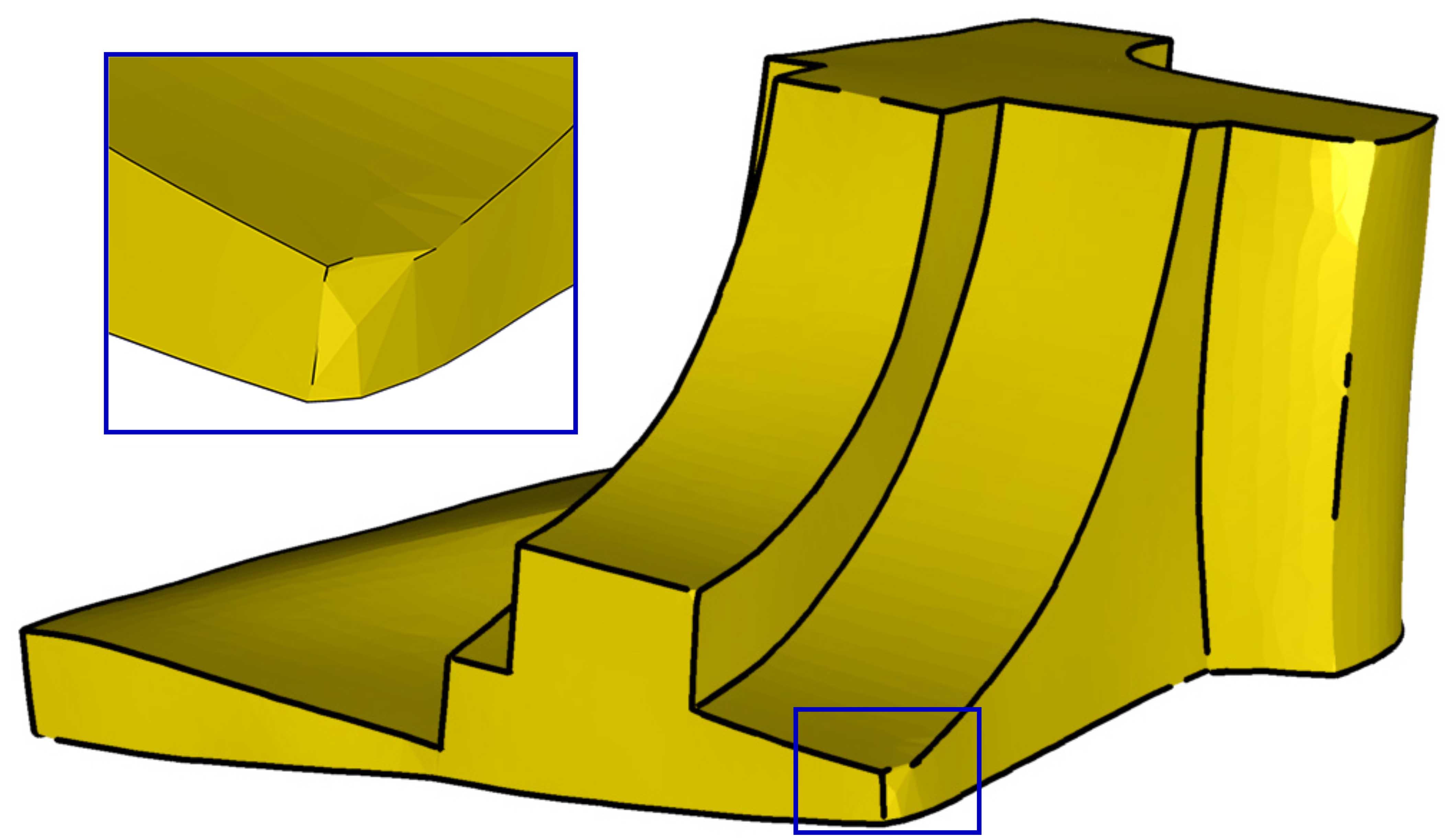}} &
 			\subfloat[\cite{binormal}]{\includegraphics[width=3.3cm]{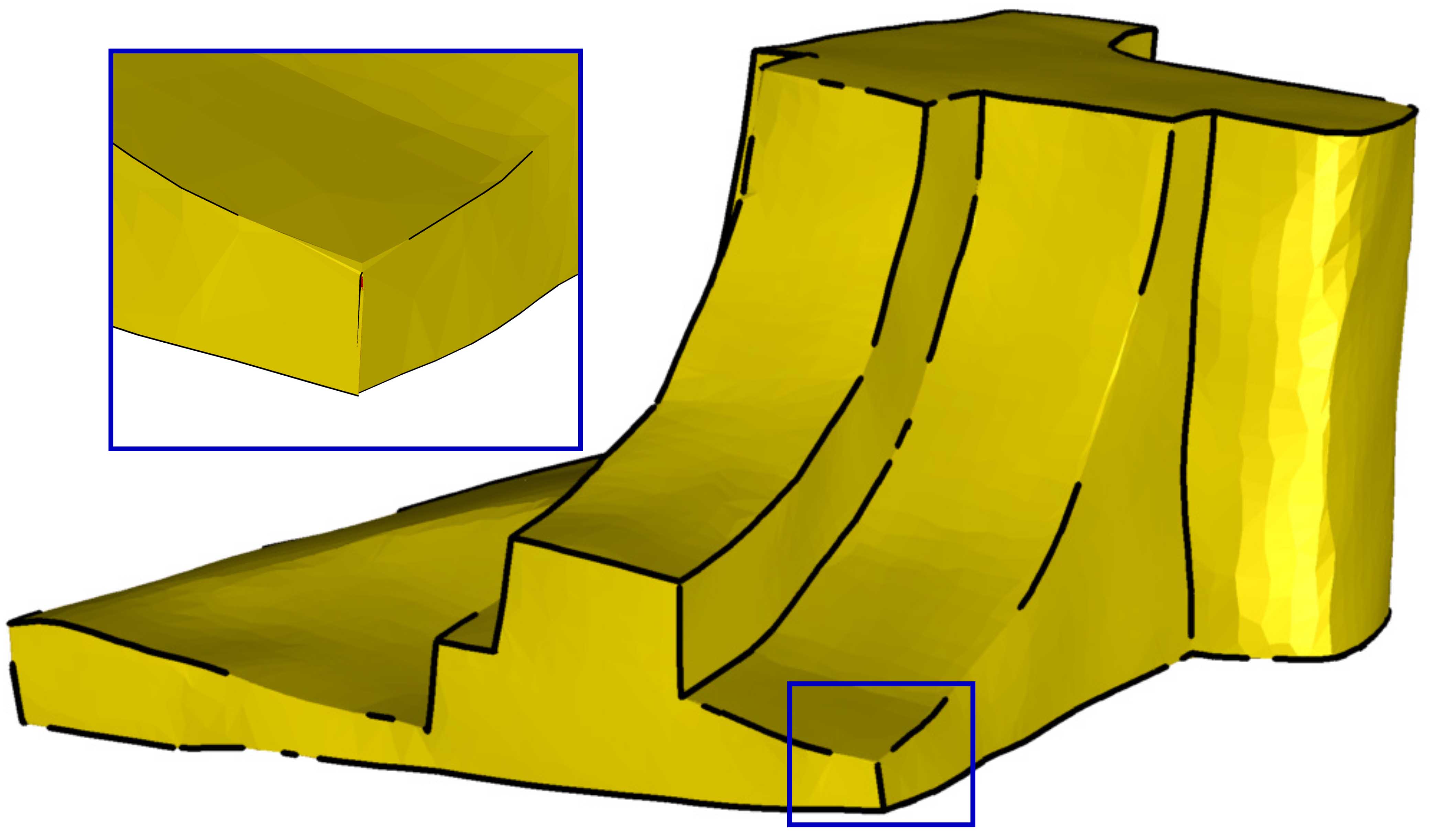}}&
 			 \subfloat[\cite{robust16}]{\includegraphics[width=3.3cm]{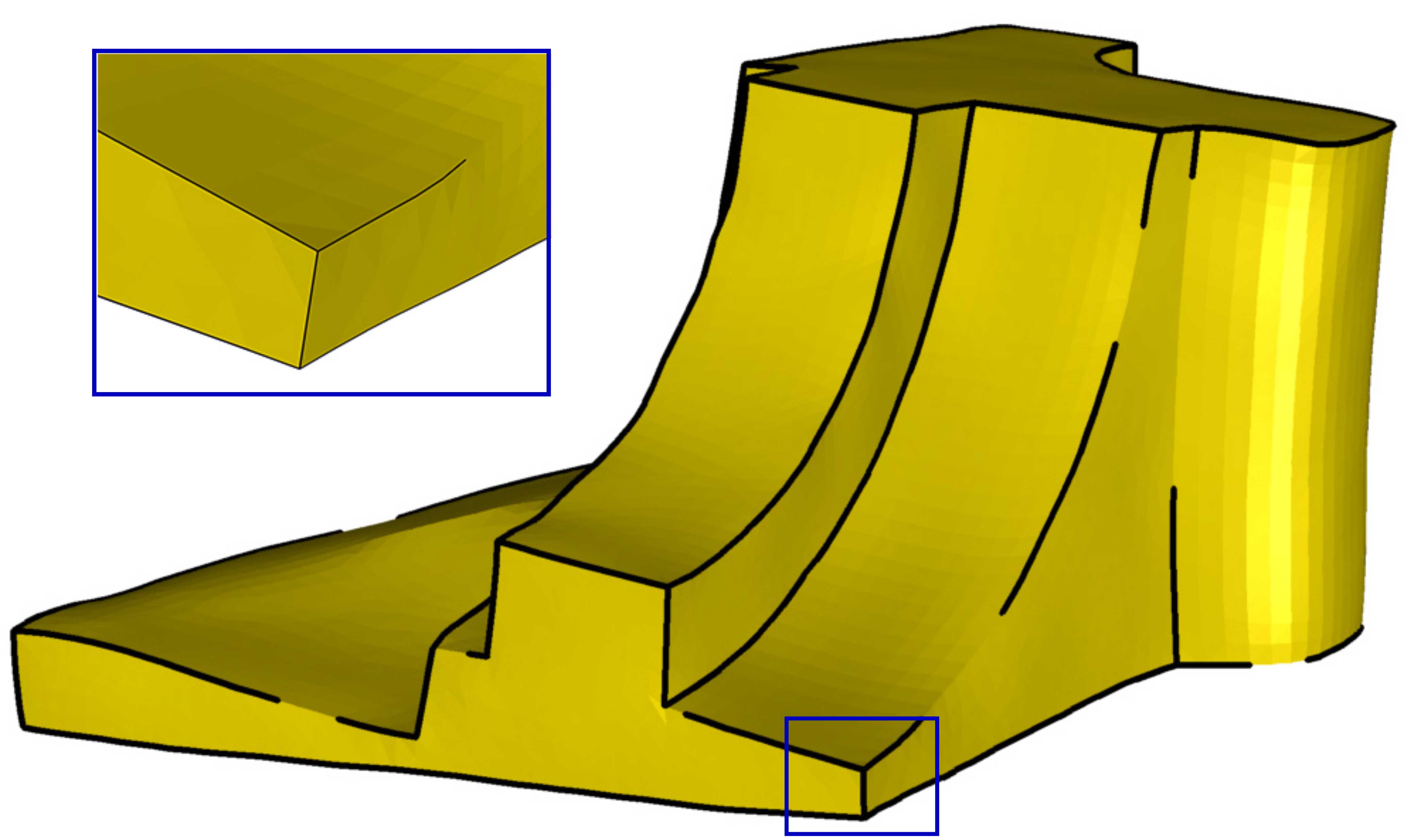}}&
 			  \subfloat[\cite{yadav17}]{\includegraphics[width=3.3cm]{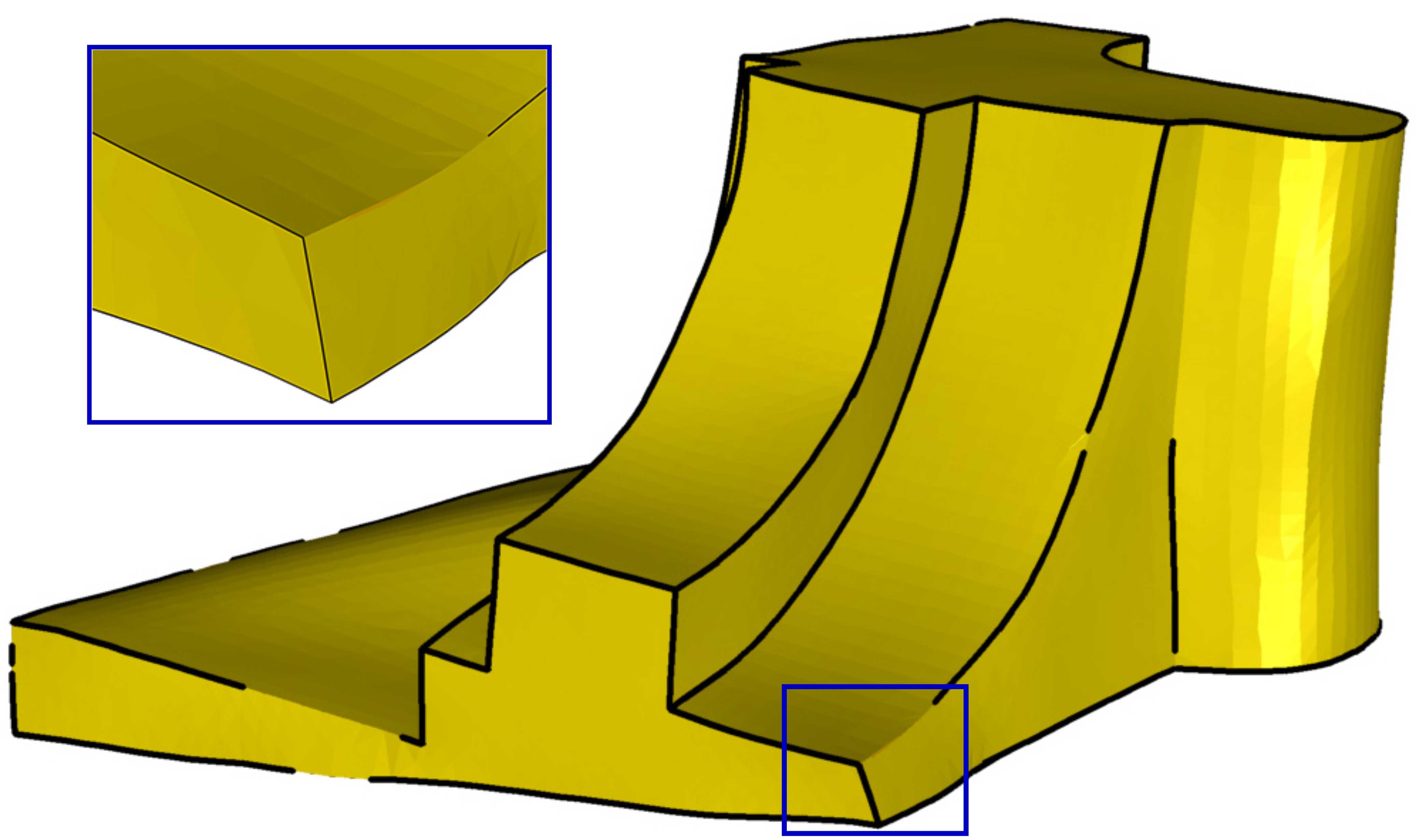}}& 
 			   \subfloat[Ours]{\includegraphics[width=3.3cm]{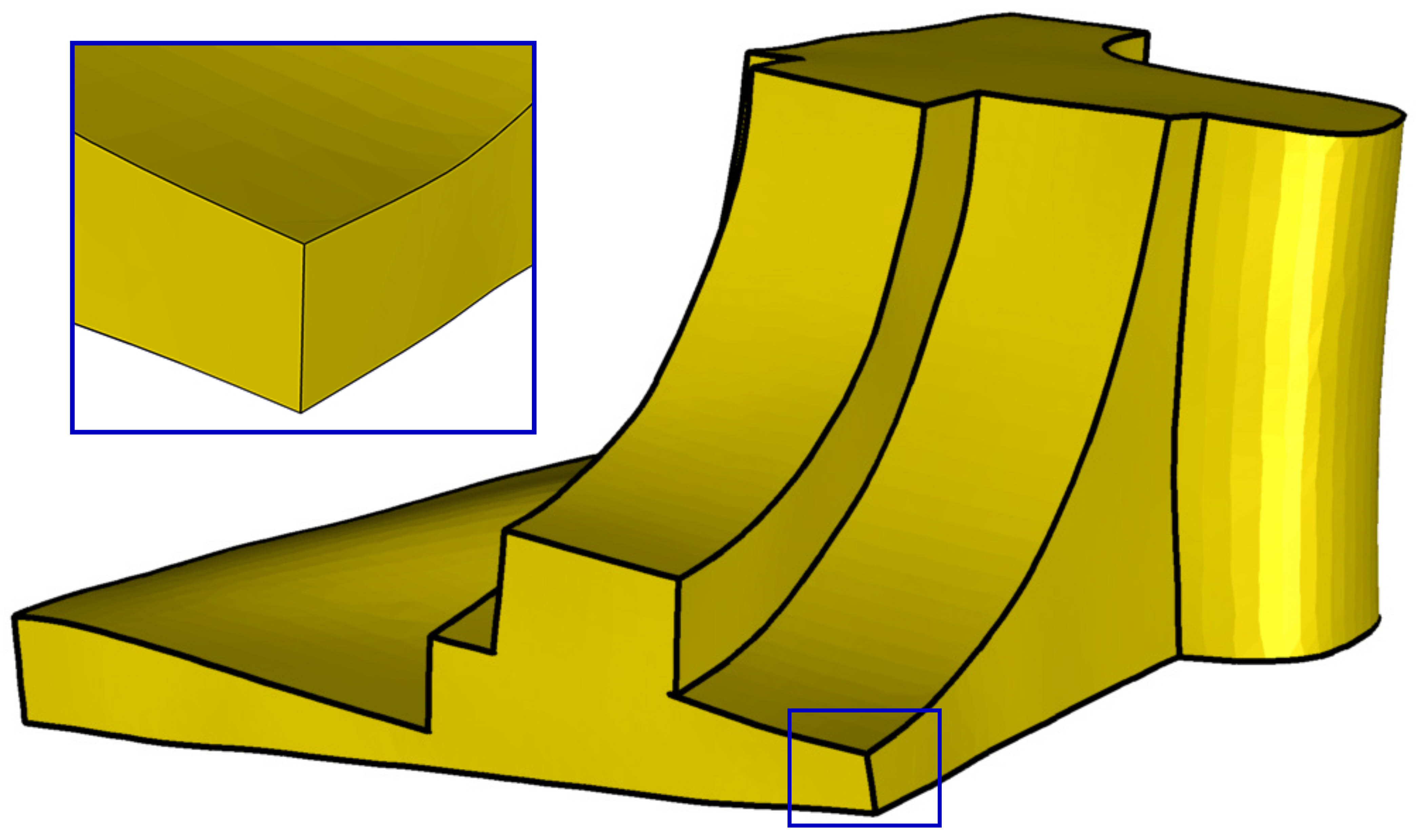}}
 		\end{tabular}
 	\end{tabularx}
 	\caption{Fandisk model corrupted with a Gaussian noise ($\sigma_n = 0.3l_e$) in random direction. The first row shows the results
 		produced by state-of-the-art methods and our proposed method. The
 		black curve shows the sharp edge information in the smooth
 		geometry and is detected using the dihedral angle $\theta = 70^\circ$.  }
 	\label{fig:fandisk}
 \end{figure*}

  \begin{figure*}[h]%
  	\centering
  	\subfloat[Original]{{\includegraphics[width=2.2cm]{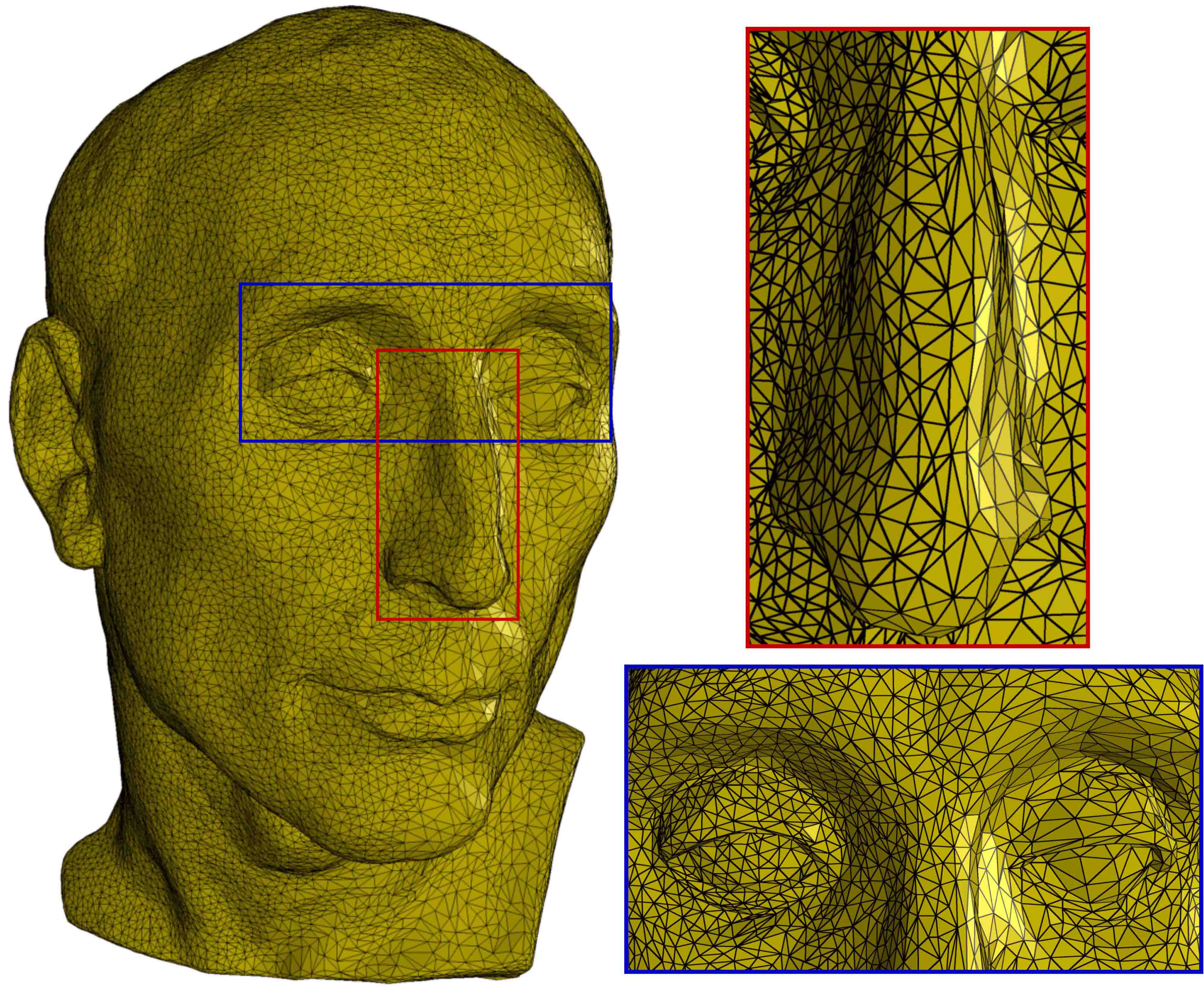} }}%
  	\subfloat[Noisy]{{\includegraphics[width=2.2cm]{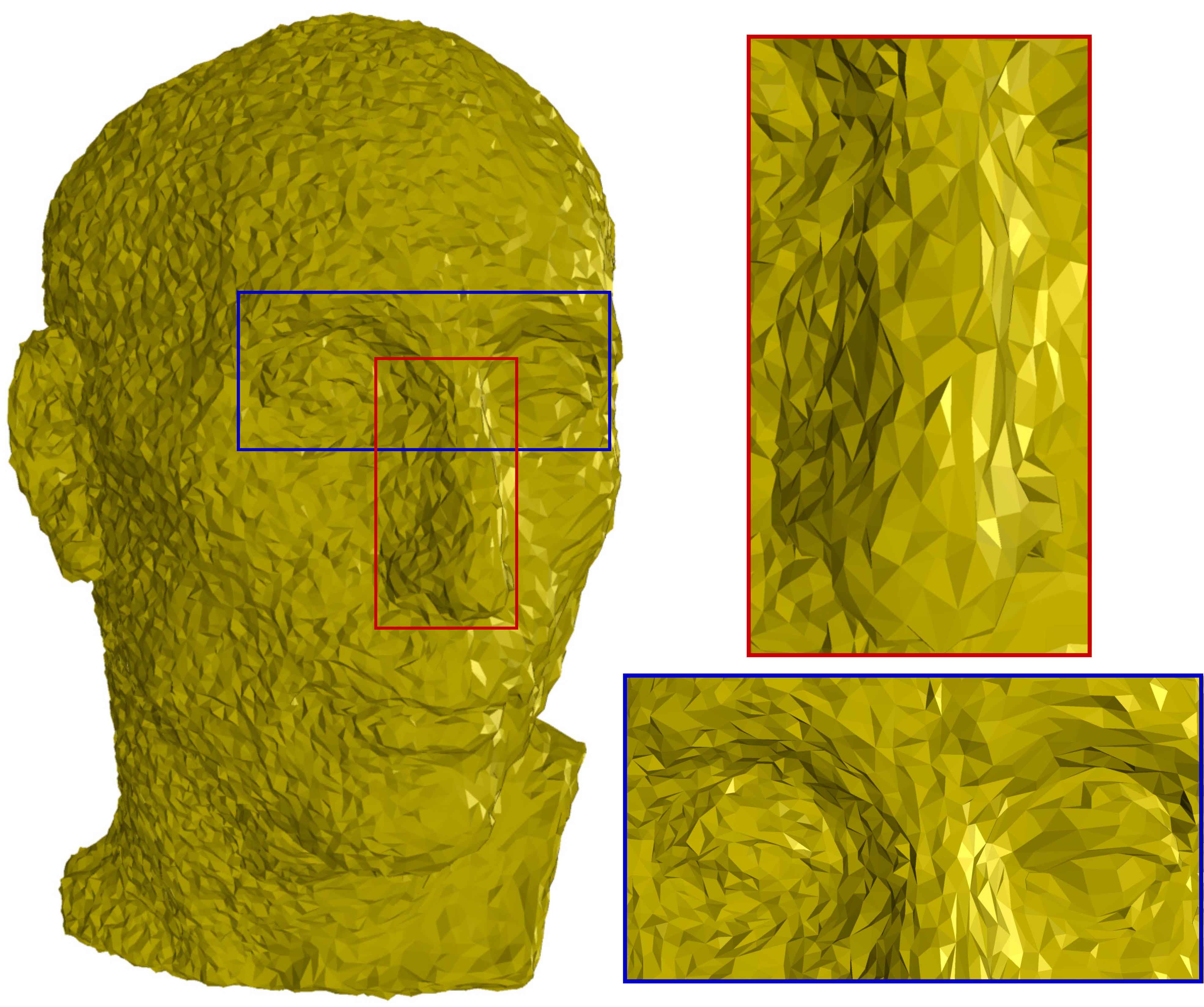} }}%
  	\subfloat[\cite{BilNorm}]{{\includegraphics[width=2.2cm]{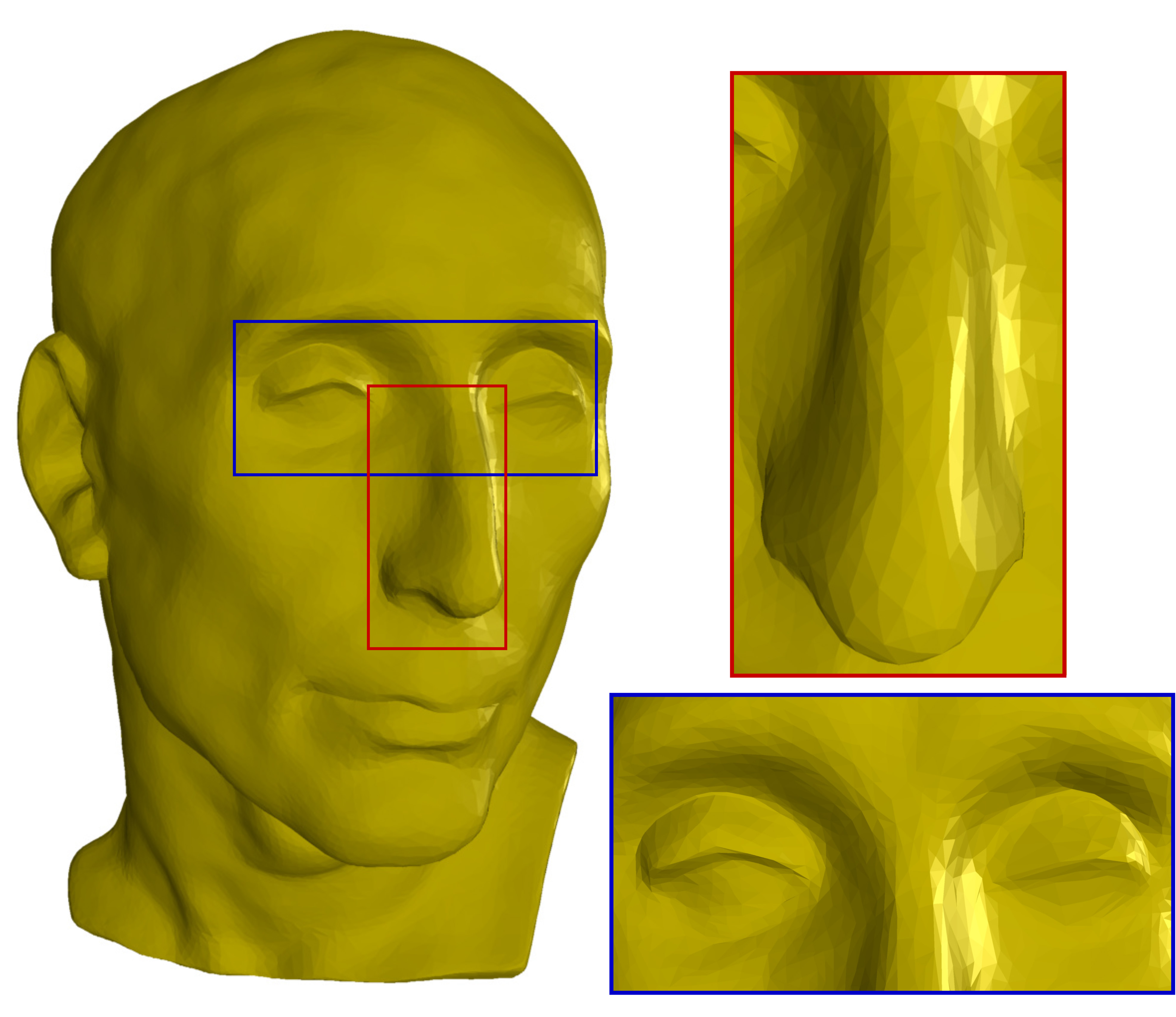} }}%
  	\subfloat[\cite{L0Mesh}]{{\includegraphics[width=2.2cm]{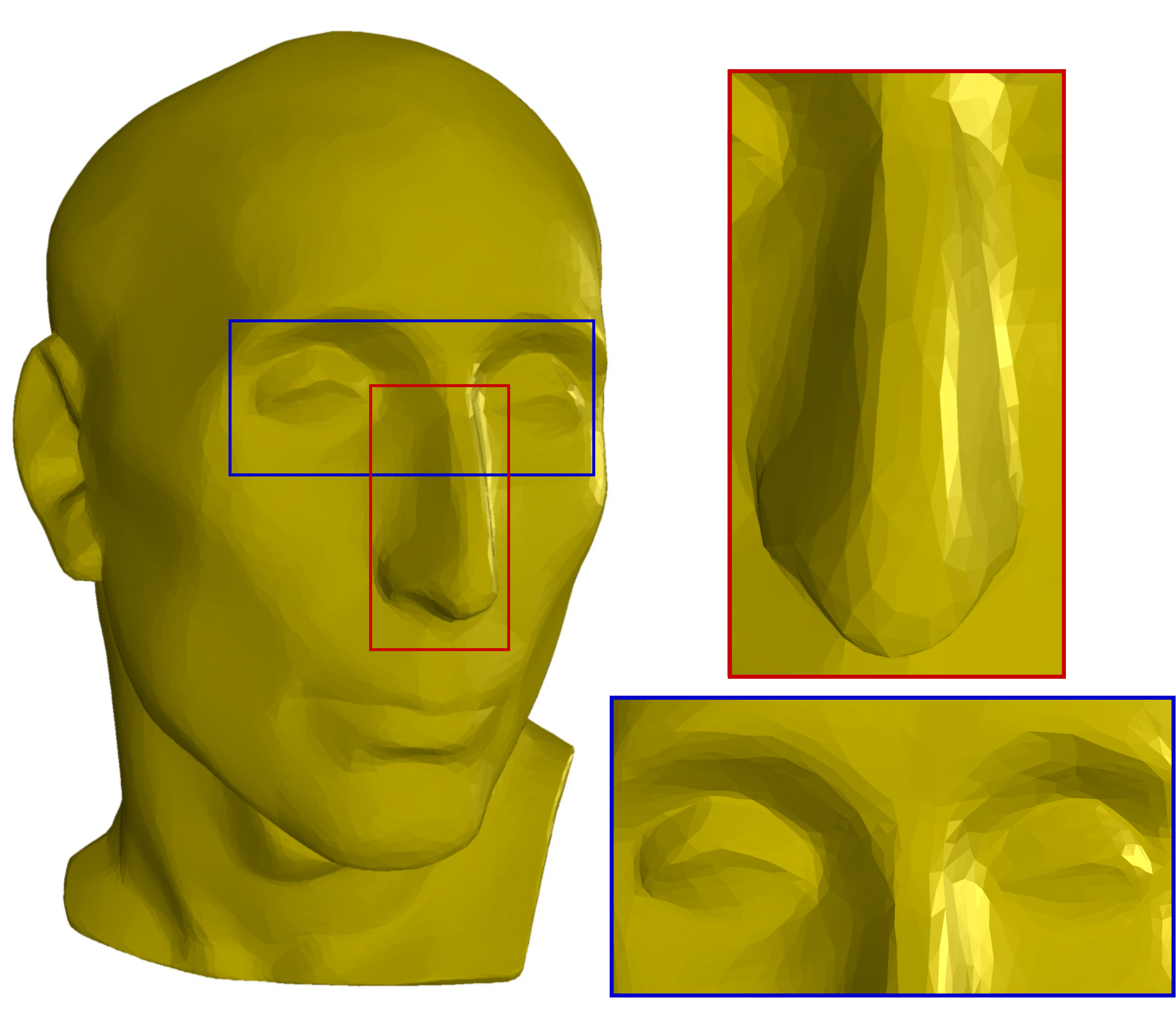} }}%
  	\subfloat[\cite{Guidedmesh}]{{\includegraphics[width=2.2cm]{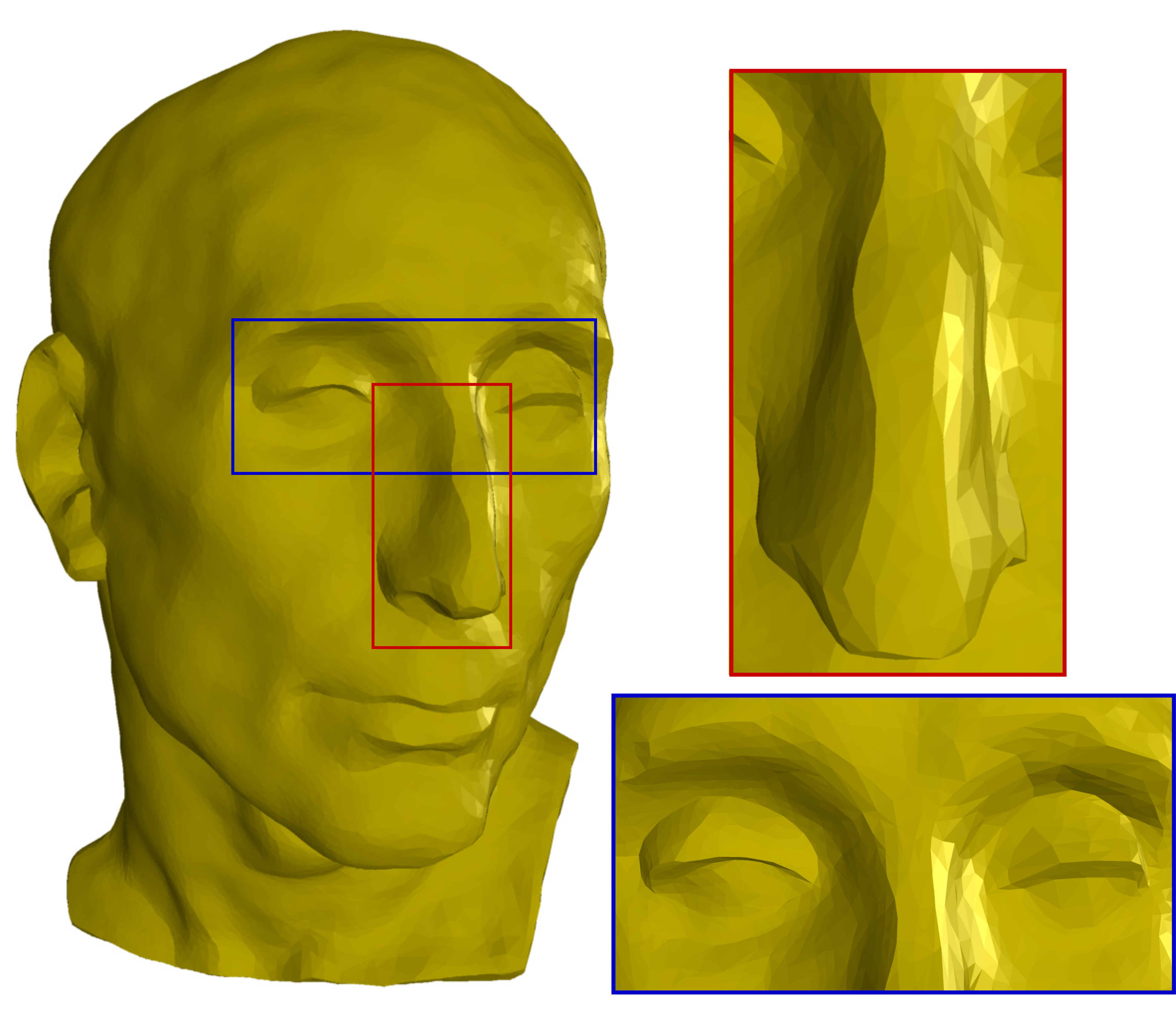} }}%
  	\subfloat[\cite{yadav17}]{{\includegraphics[width=2.2cm]{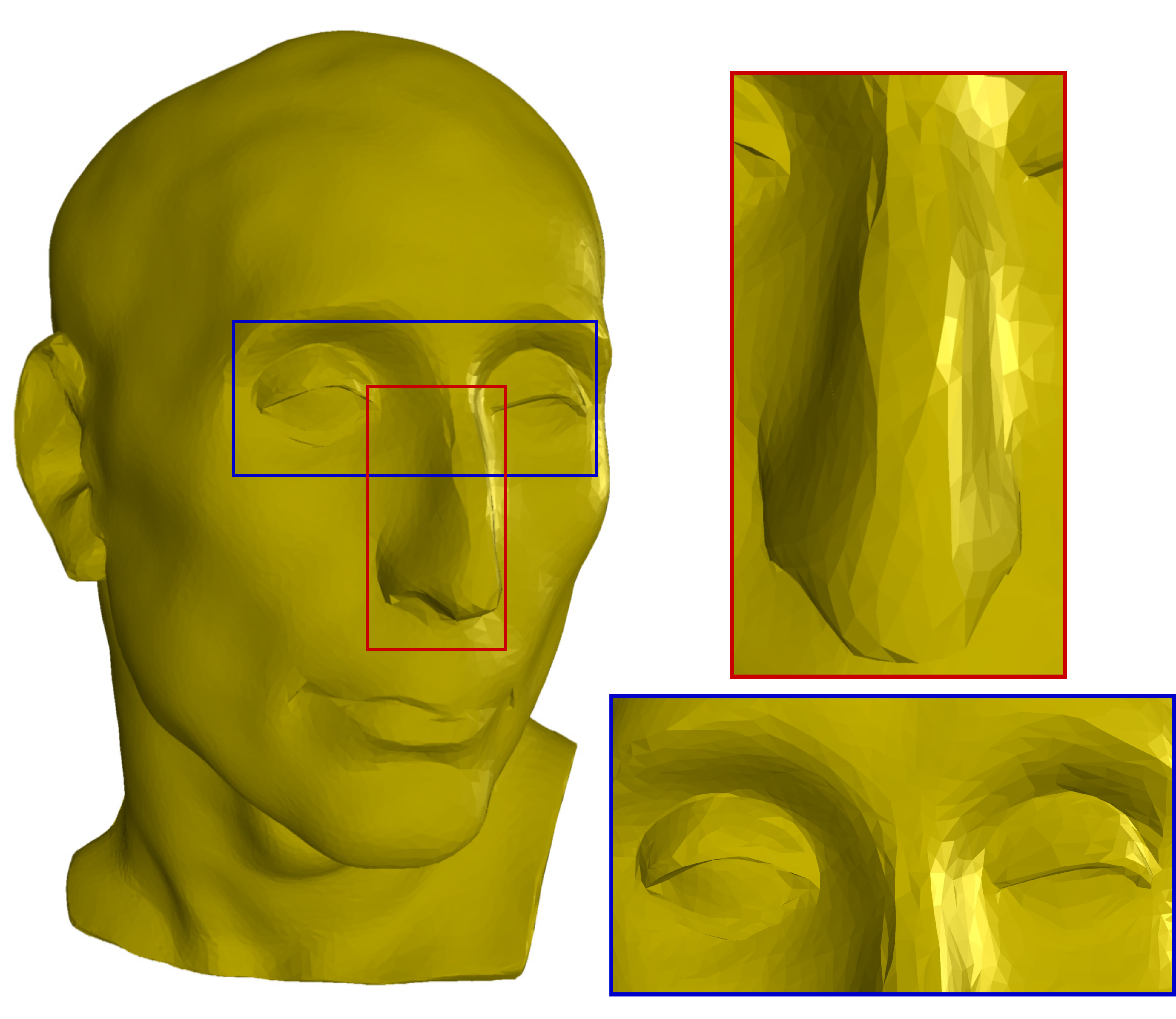} }}%
  		\subfloat[\cite{Centin}]{{\includegraphics[width=2.2cm]{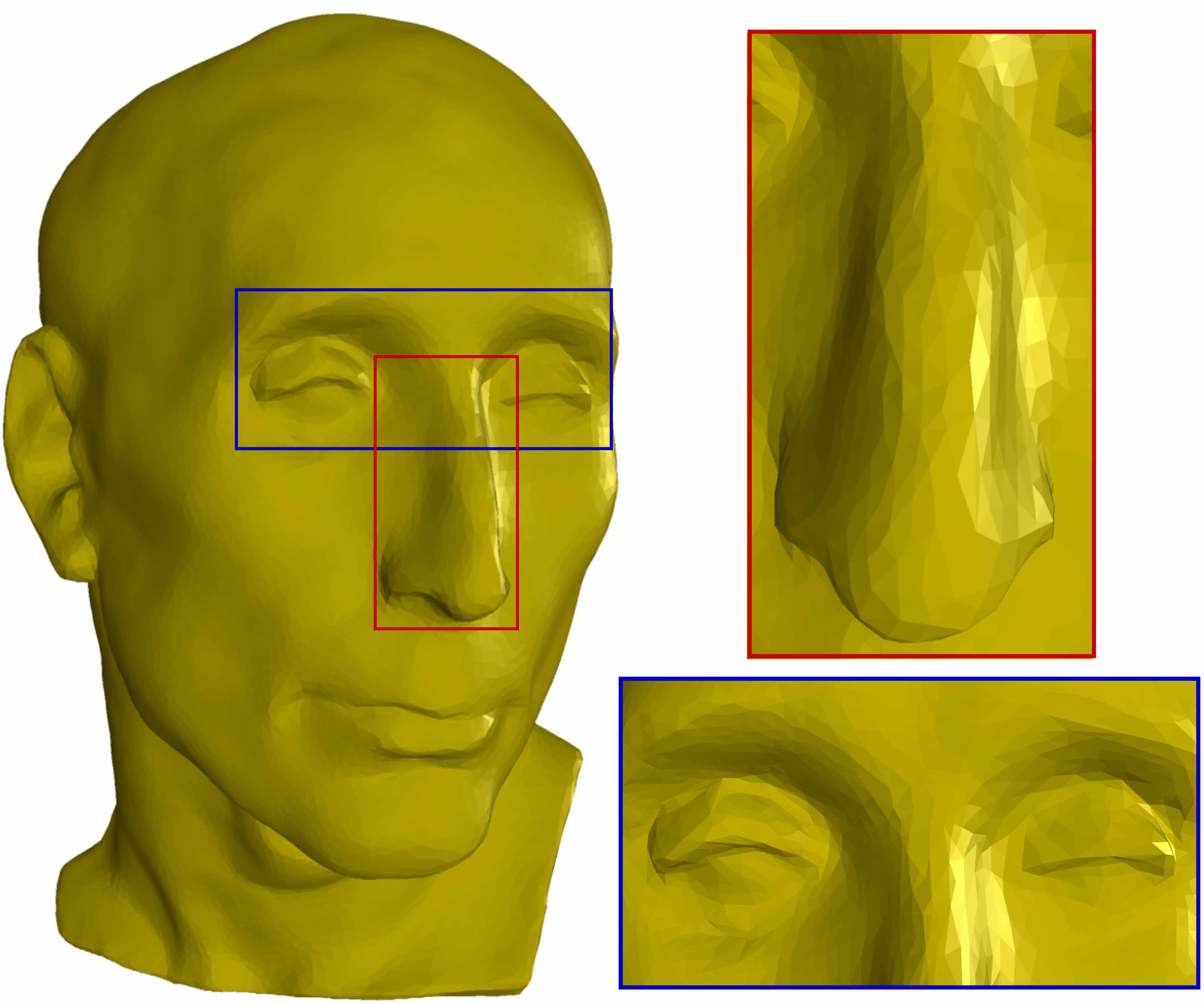} }}%
  	\subfloat[Ours]{{\includegraphics[width=2.2cm]{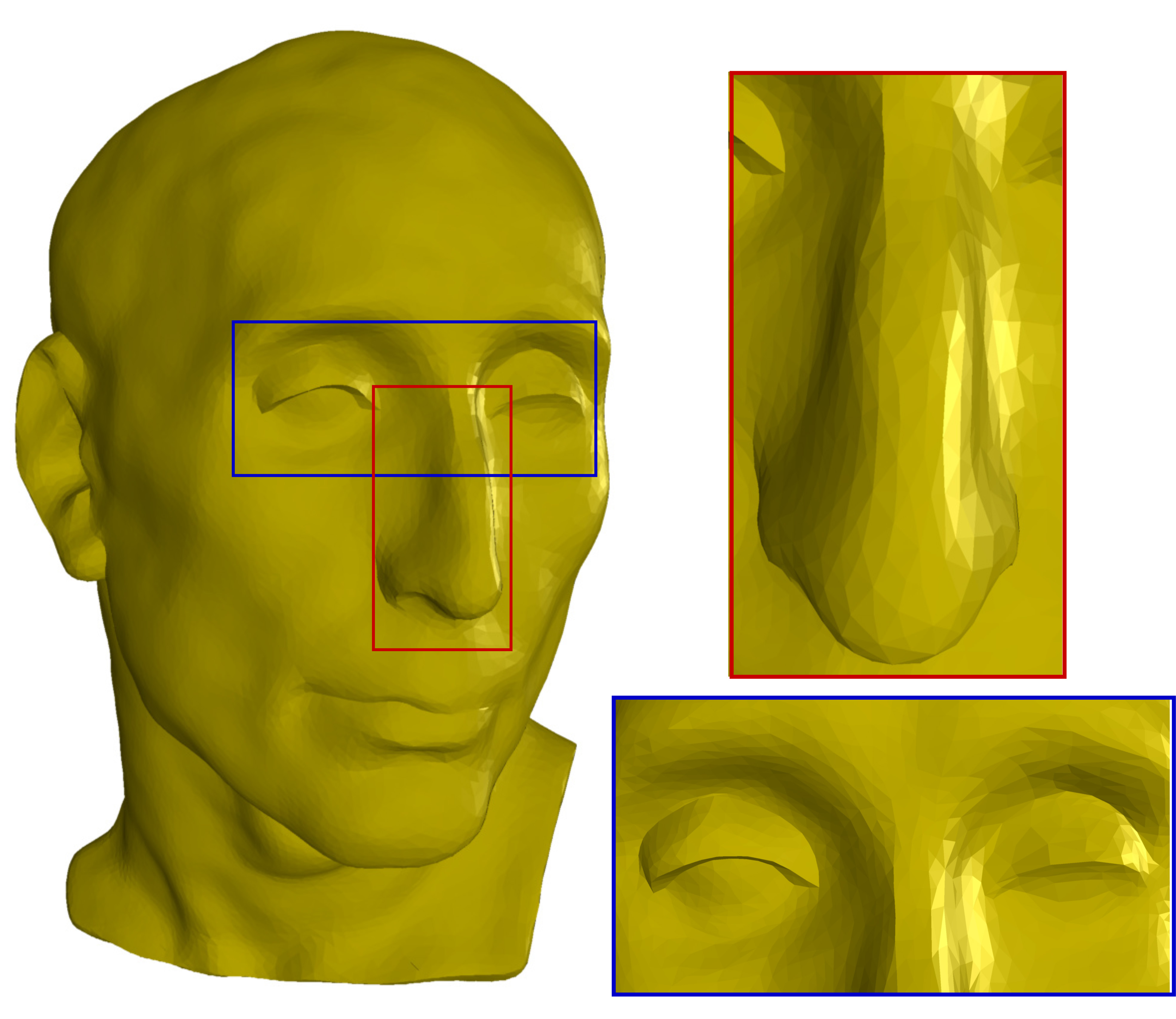} }}%
  	\centering
  	\caption{The Nicola model, a non-uniform triangulated mesh surface corrupted by Gaussian noise in random direction. The magnified view of the eyes and the nose shows that the proposed method preserves better sharp features without creating piecewise flat areas compared to state-of-the-art methods.  }
  	\label{fig:nicola}
  \end{figure*}

 \begin{figure*}[h]%
 	\centering
 	 \subfloat[Original (non-uniform mesh)]{{\includegraphics[width=2.2cm]{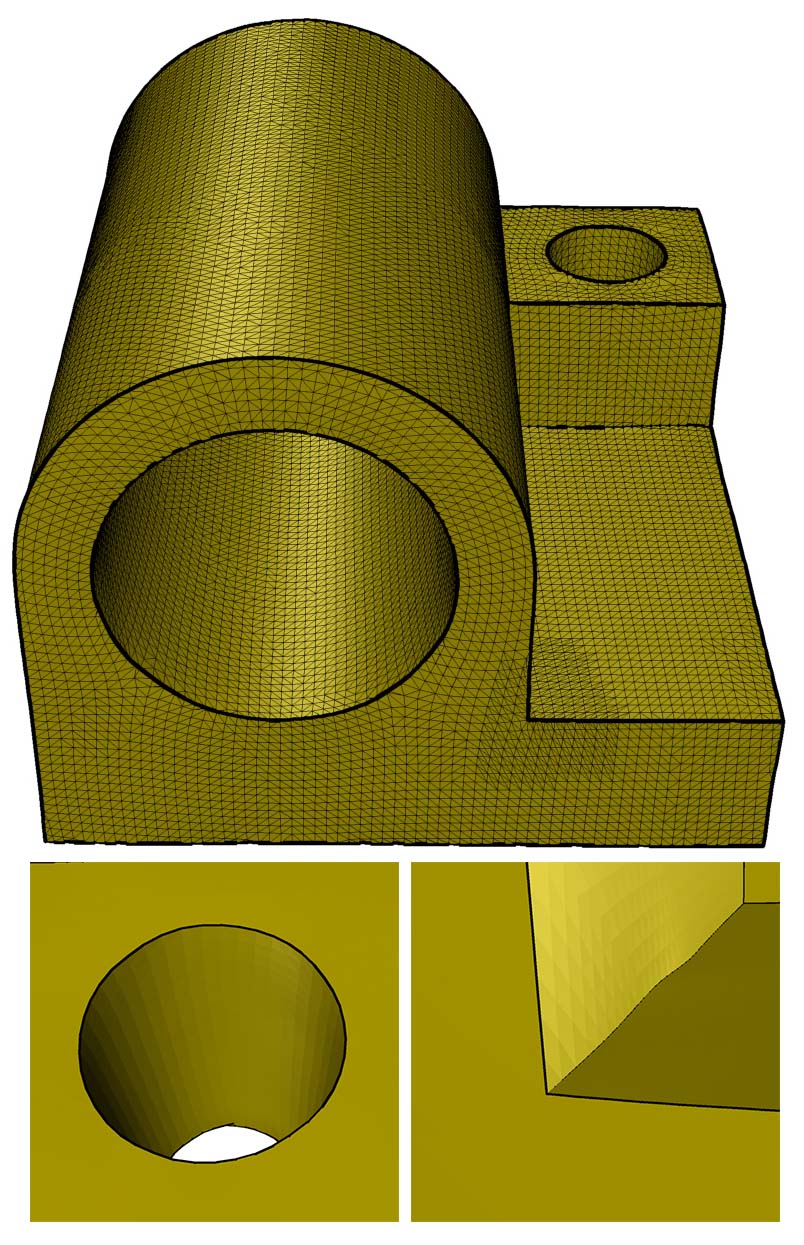} }}%
 	\subfloat[Noisy]{{\includegraphics[width=2.2cm]{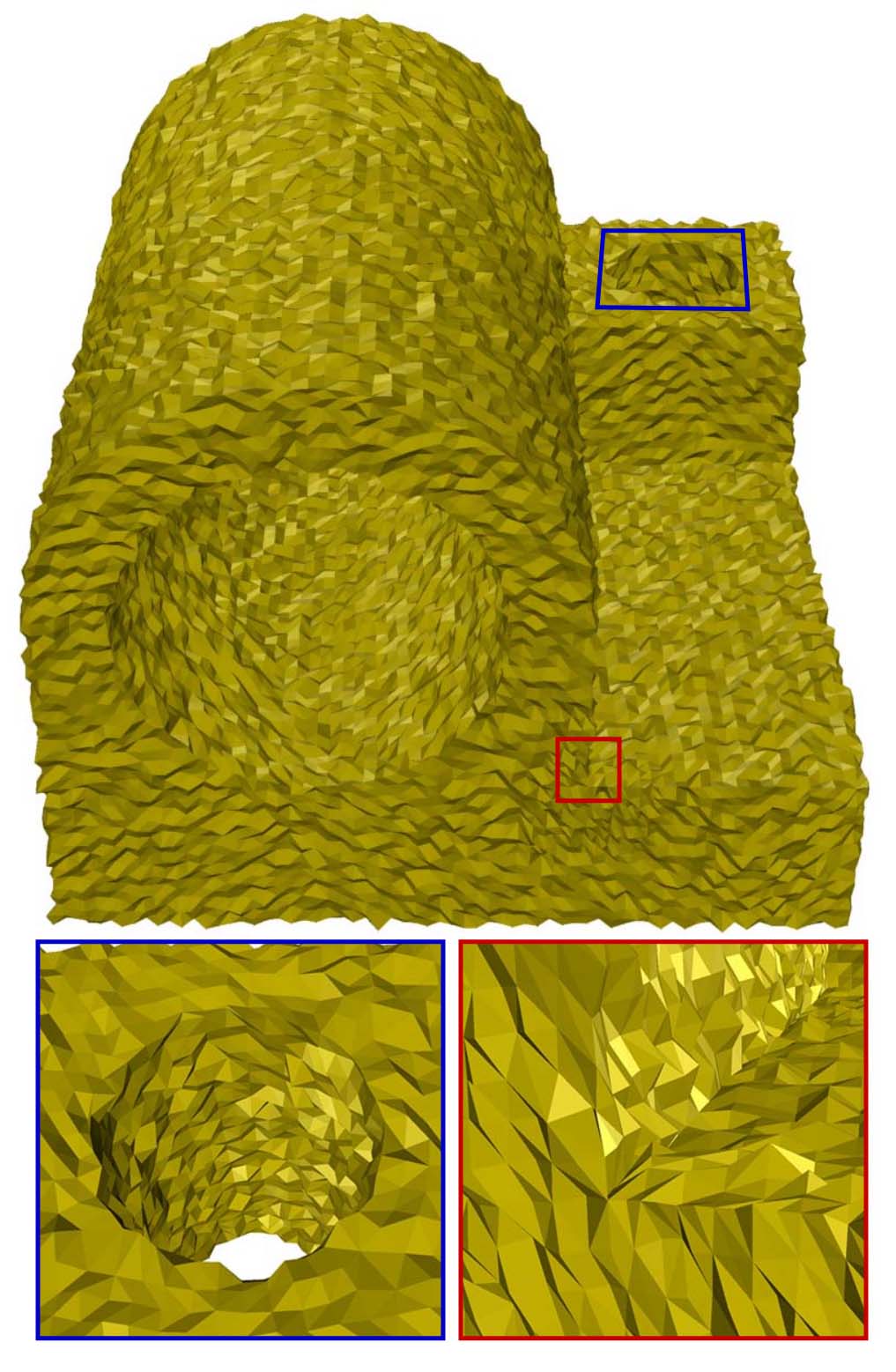} }}%
 	\subfloat[\cite{BilNorm}]{{\includegraphics[width=2.2cm]{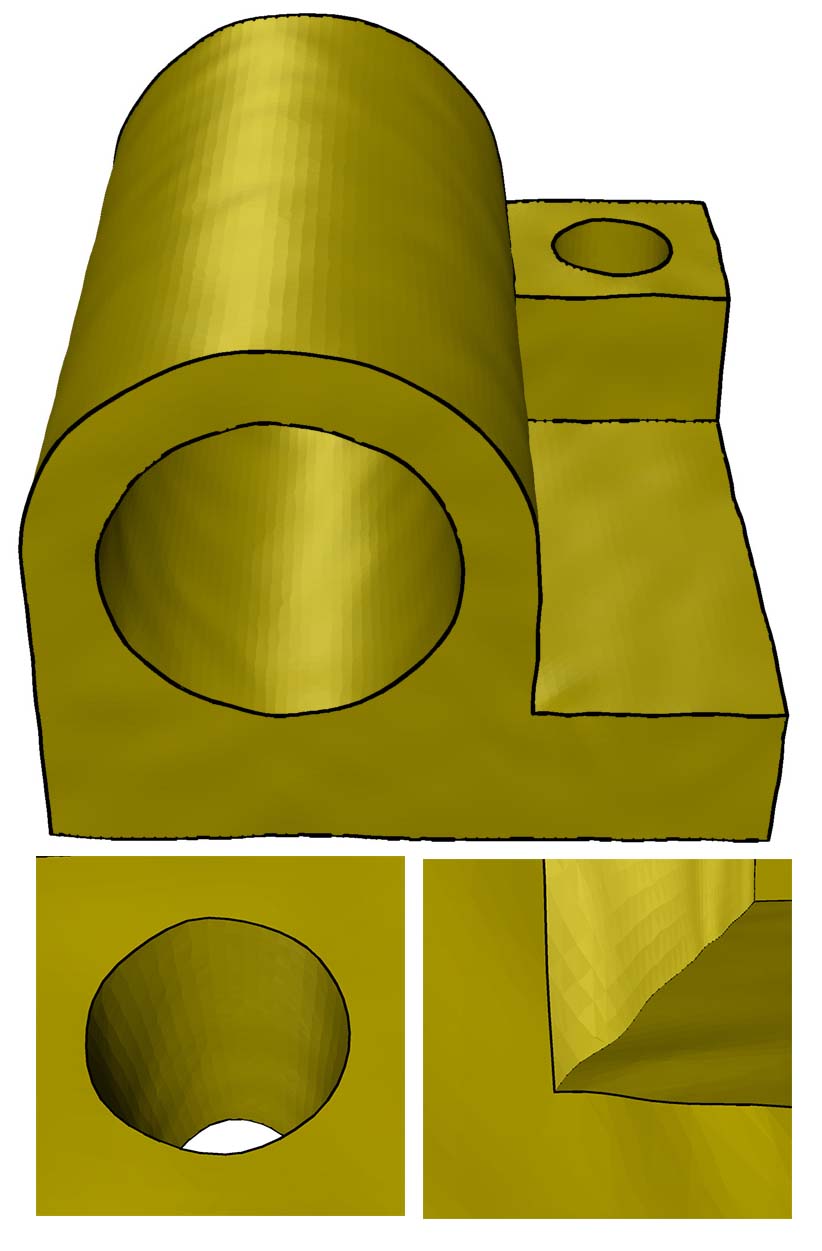} }}%
 	\subfloat[\cite{L0Mesh}]{{\includegraphics[width=2.2cm]{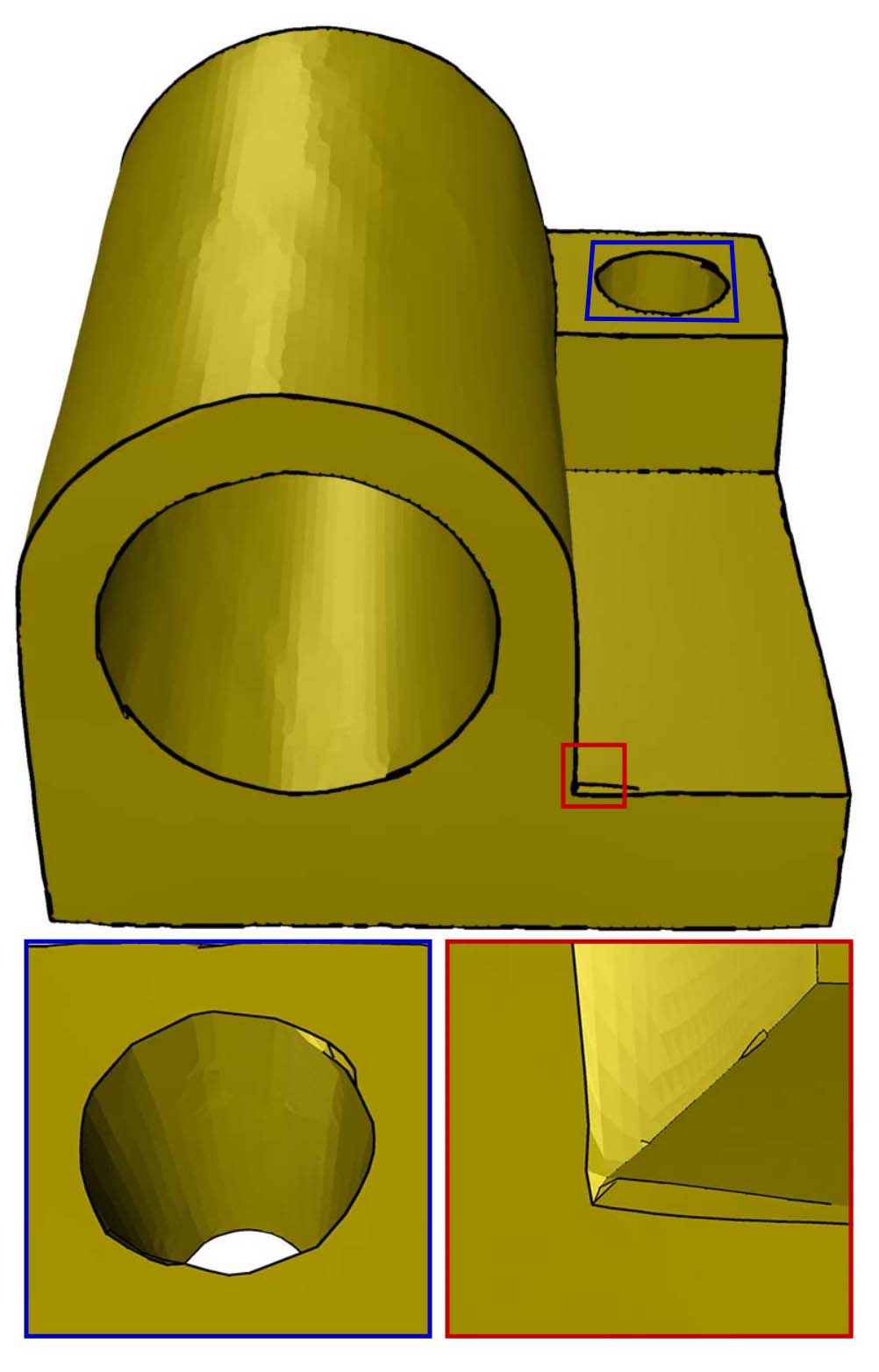} }}%
 	\subfloat[\cite{binormal}]{{\includegraphics[width=2.2cm]{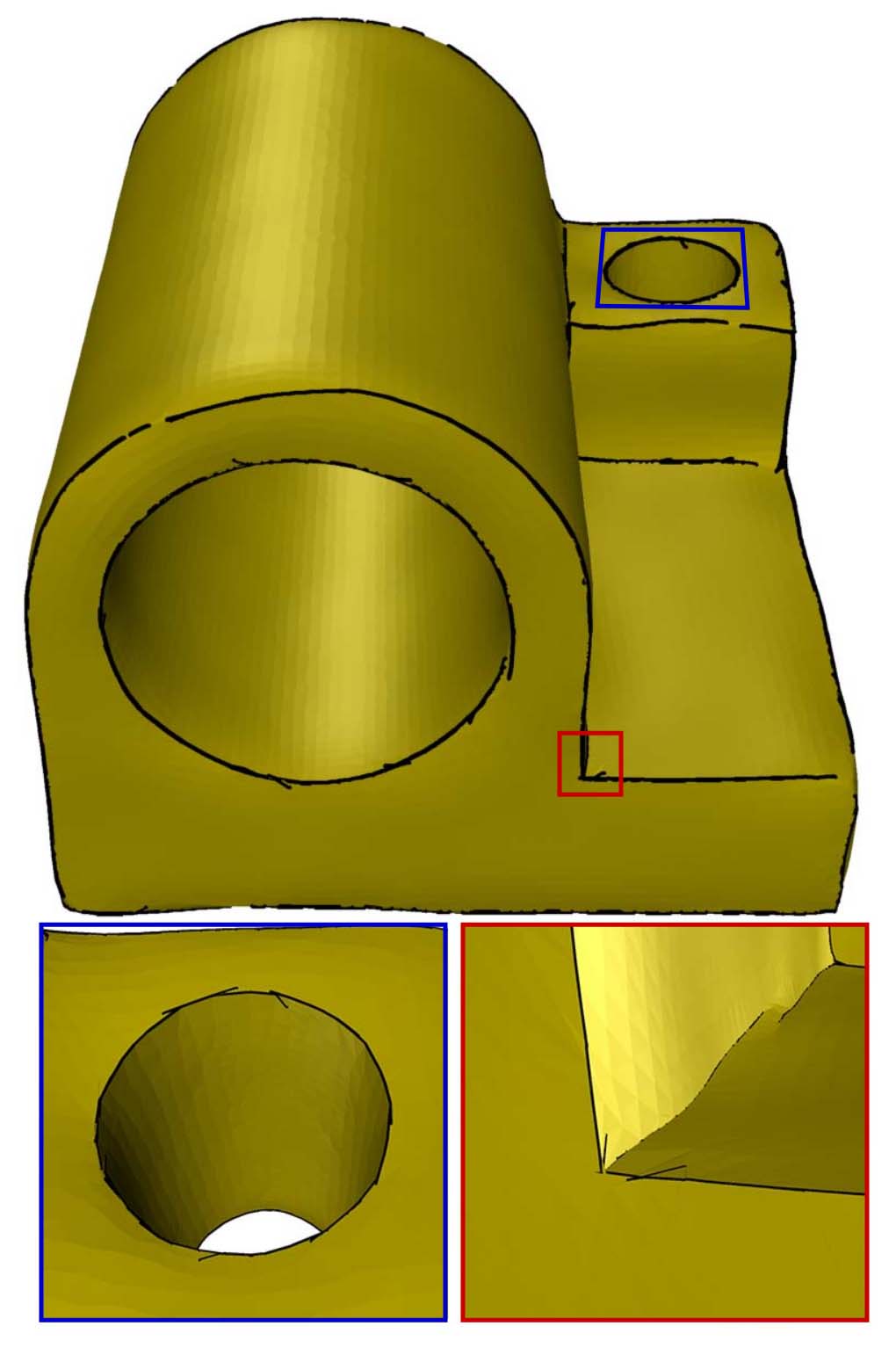} }}%
 	\subfloat[\cite{yadav17}]{{\includegraphics[width=2.2cm]{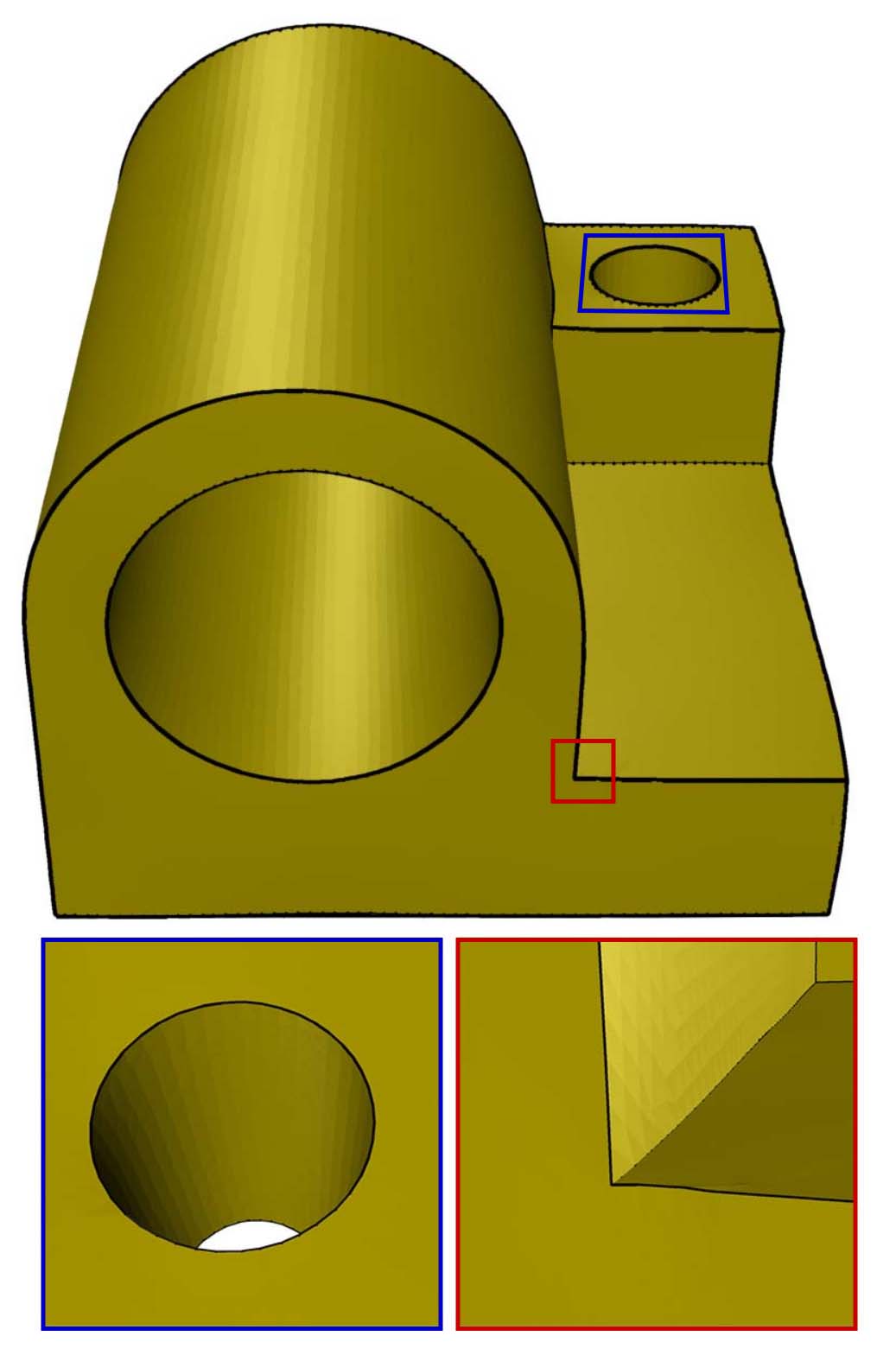} }}%
 		\subfloat[\cite{Centin}]{{\includegraphics[width=2.16cm]{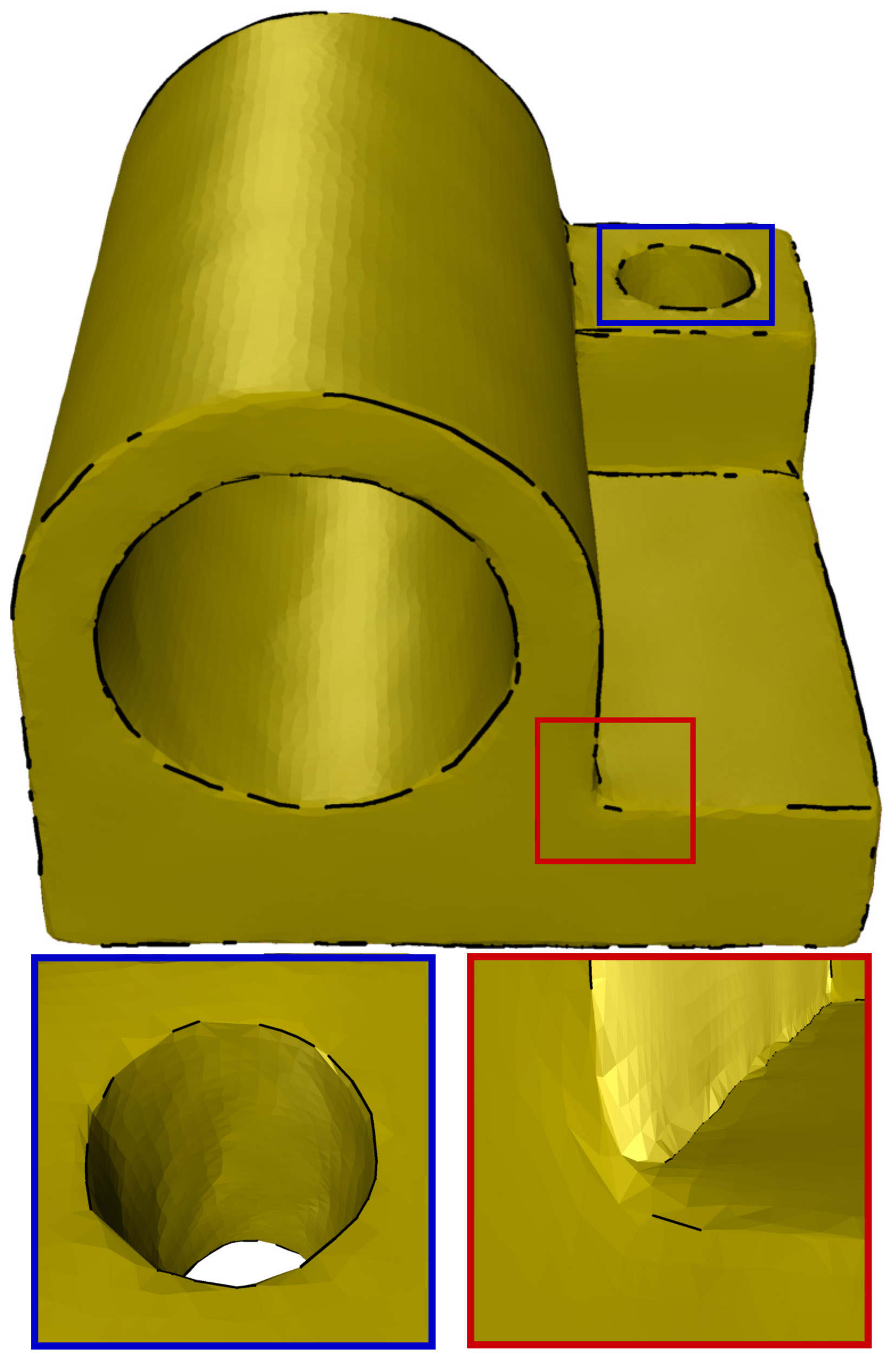} }}%
 	\subfloat[Ours]{{\includegraphics[width=2.12cm]{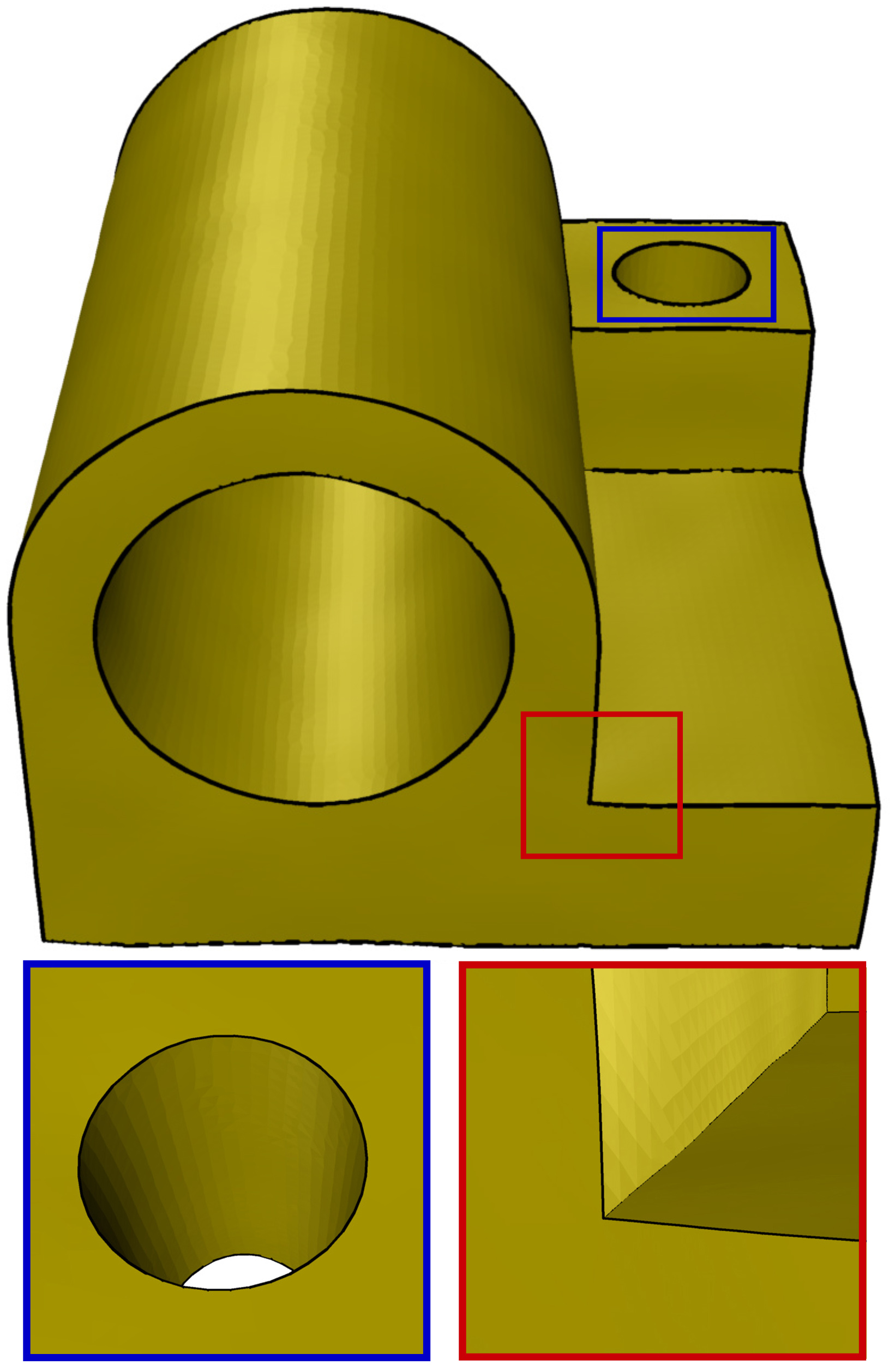} }}%
 	\centering
 	\caption{Non-uniform triangulated mesh surface corrupted by Gaussian noise ($\sigma_n = 0.35l_e$) in normal direction. The first row shows the results obtained by state-of-the-art methods and the proposed method. The second row shows the magnified view of the corner and the cylindrical hole of the corresponding geometry.  }
 	\label{fig:jSharp}
 \end{figure*}
 
 \begin{figure*}
 	\centering
 	\subfloat[Original]{\includegraphics[width=2.16cm]{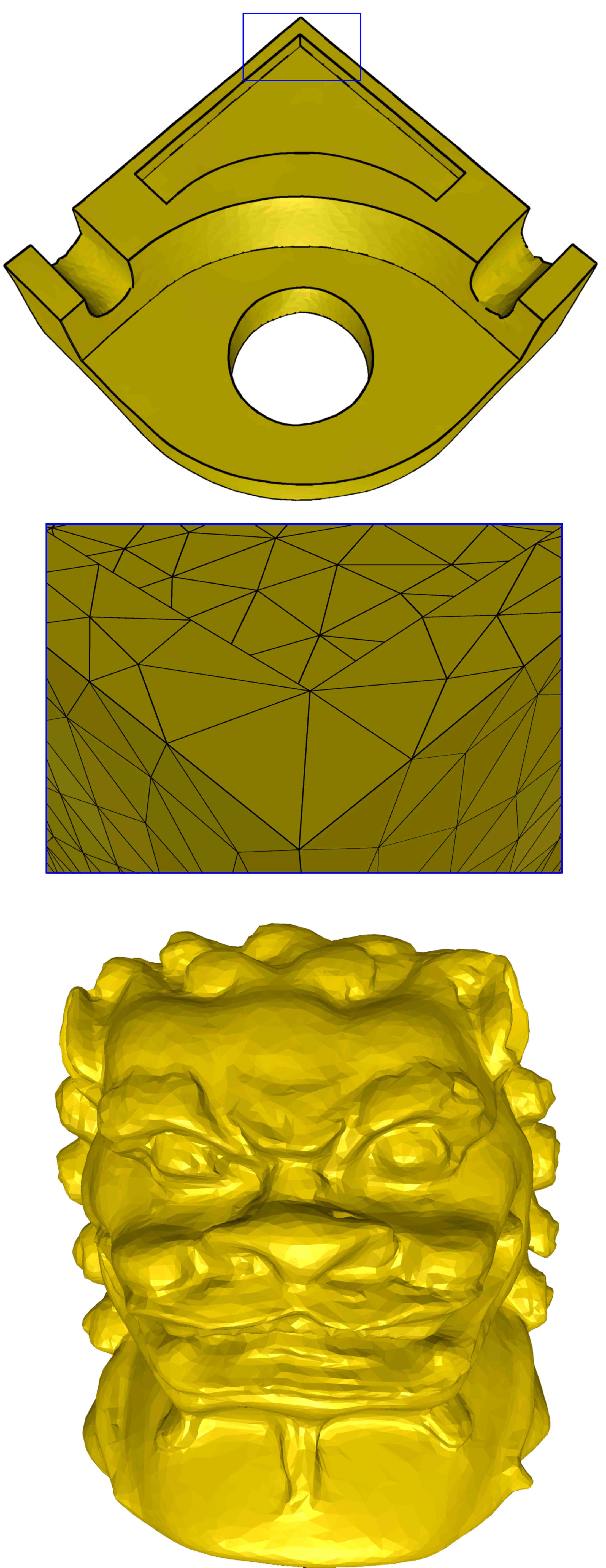}} 
 	\subfloat[Noisy]{\includegraphics[width=2.16cm]{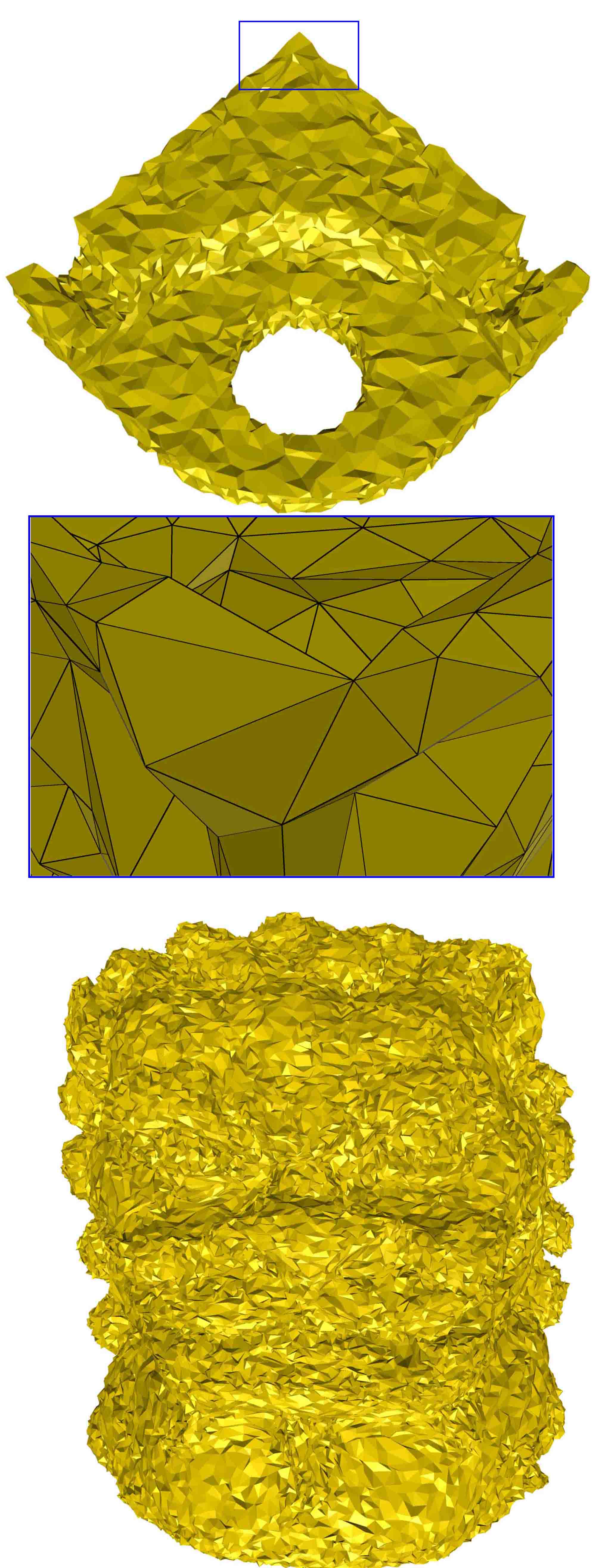}} 
 	\subfloat[\cite{aniso}]{\includegraphics[width=2.17cm]{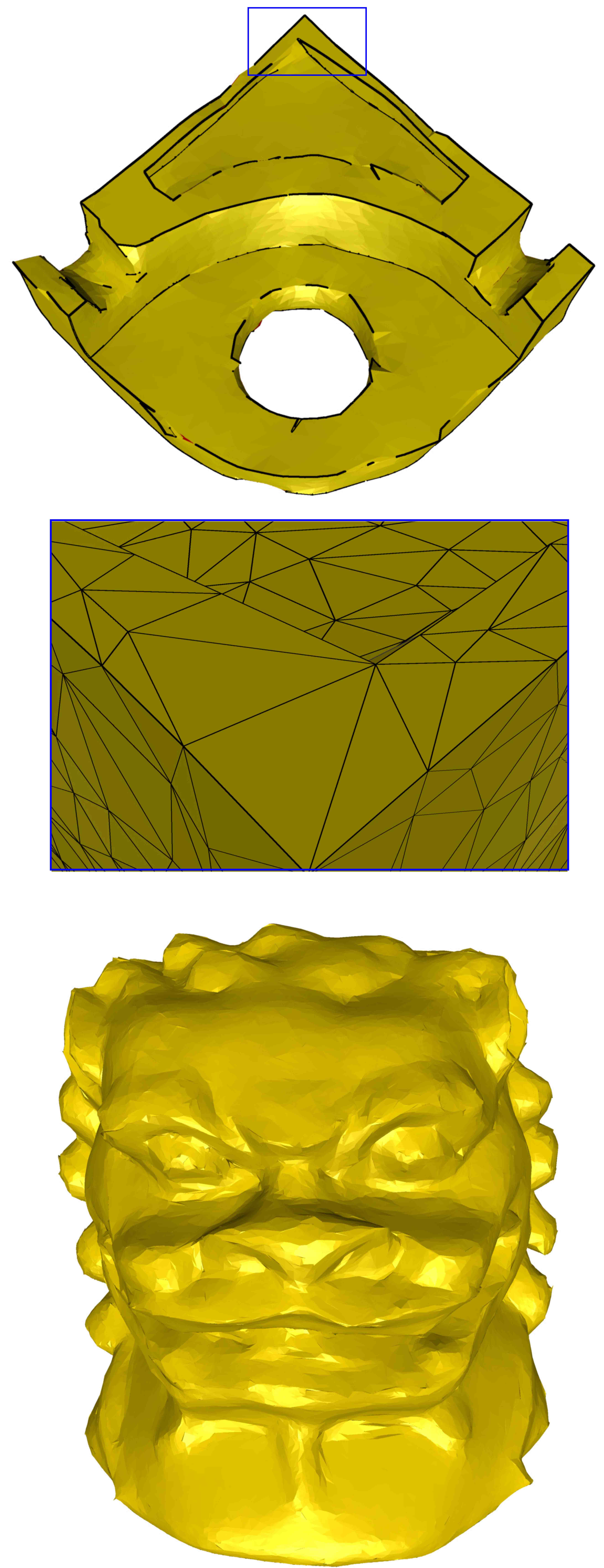}}
 	\subfloat[\cite{BilNorm}]{\includegraphics[width=2.17cm]{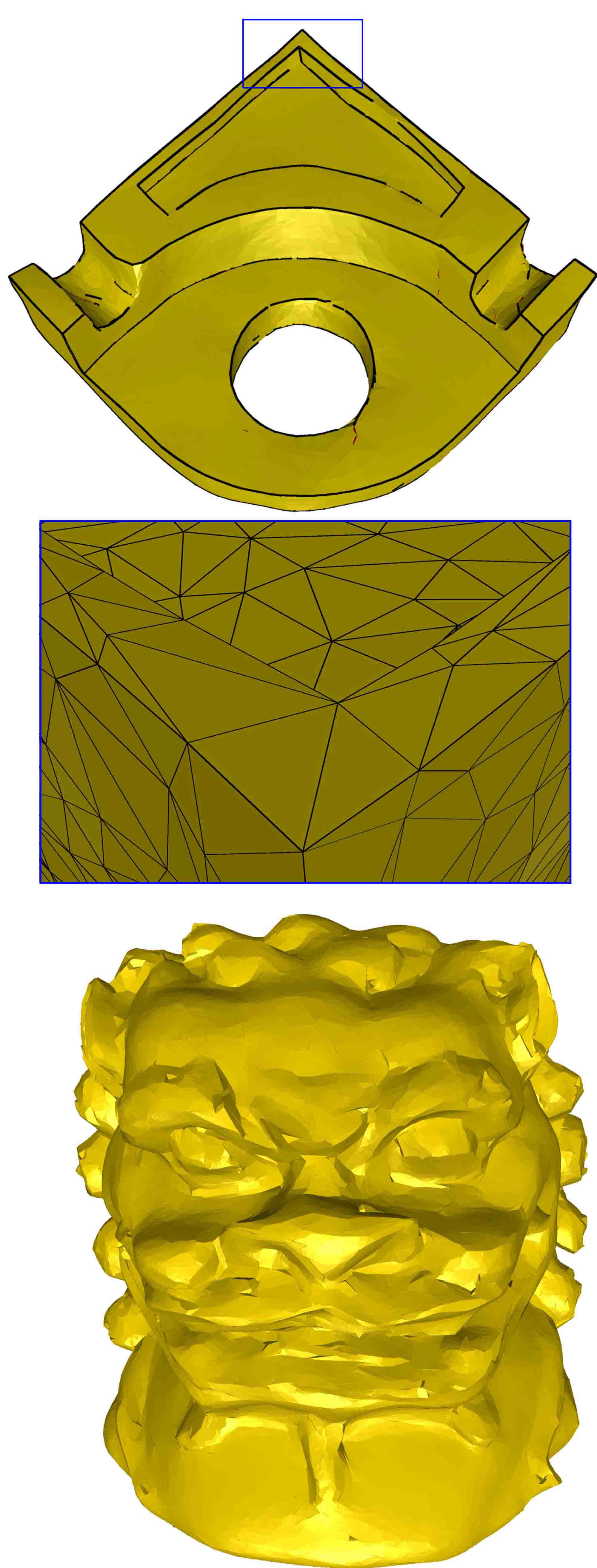}} 
 	\subfloat[\cite{L0Mesh}]{\includegraphics[width=2.2cm]{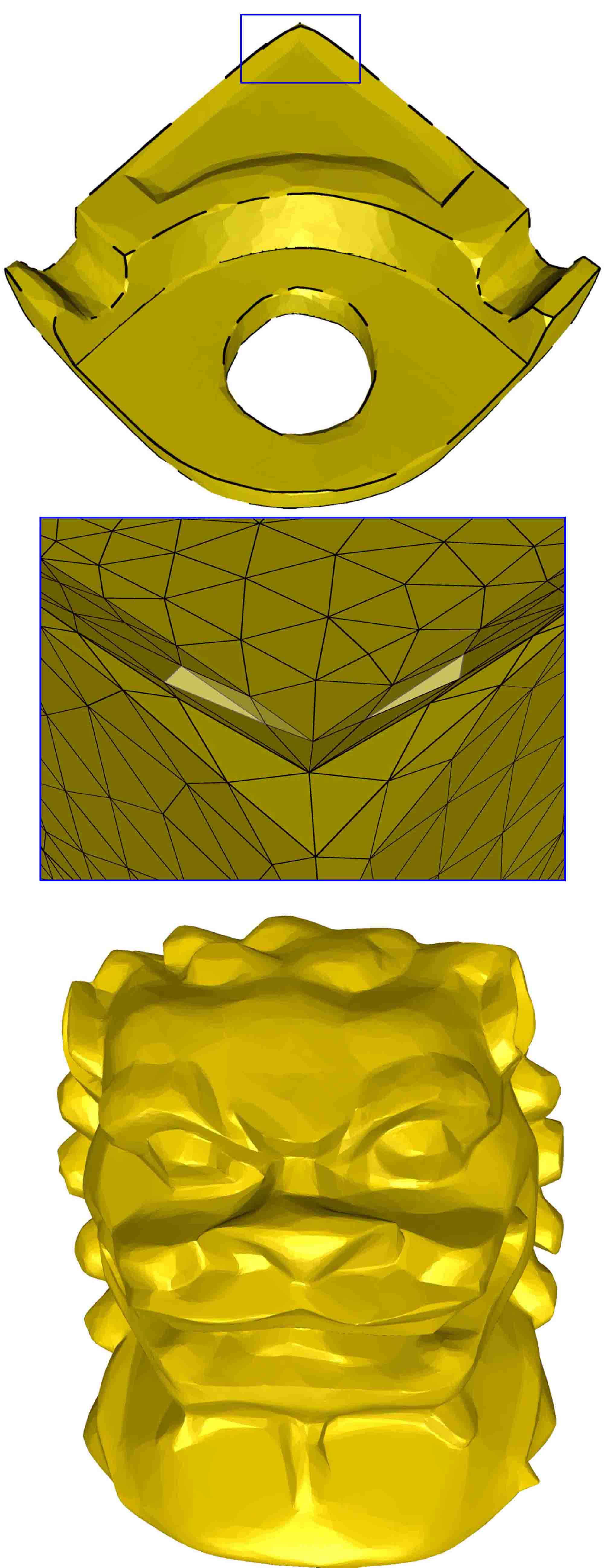}}
 	\subfloat[\cite{binormal}]{\includegraphics[width=2.2cm]{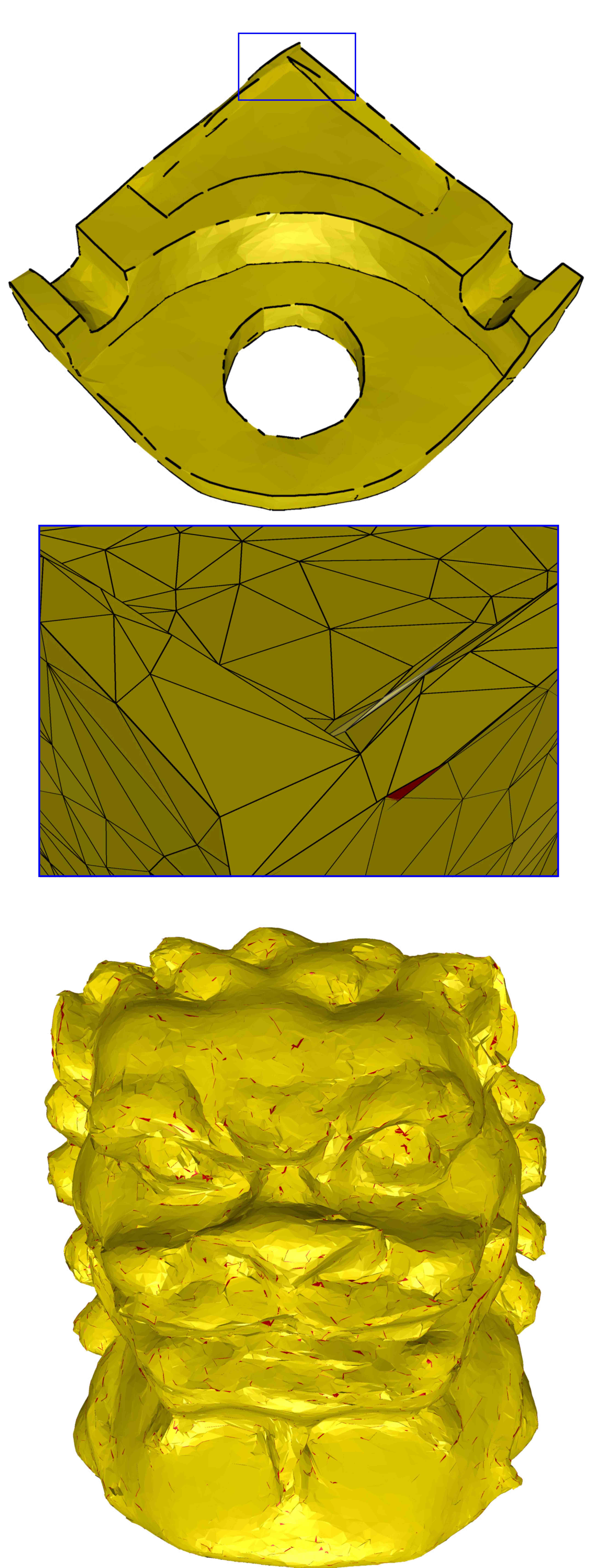}}
 	\subfloat[\cite{yadav17}]{\includegraphics[width=2.25cm]{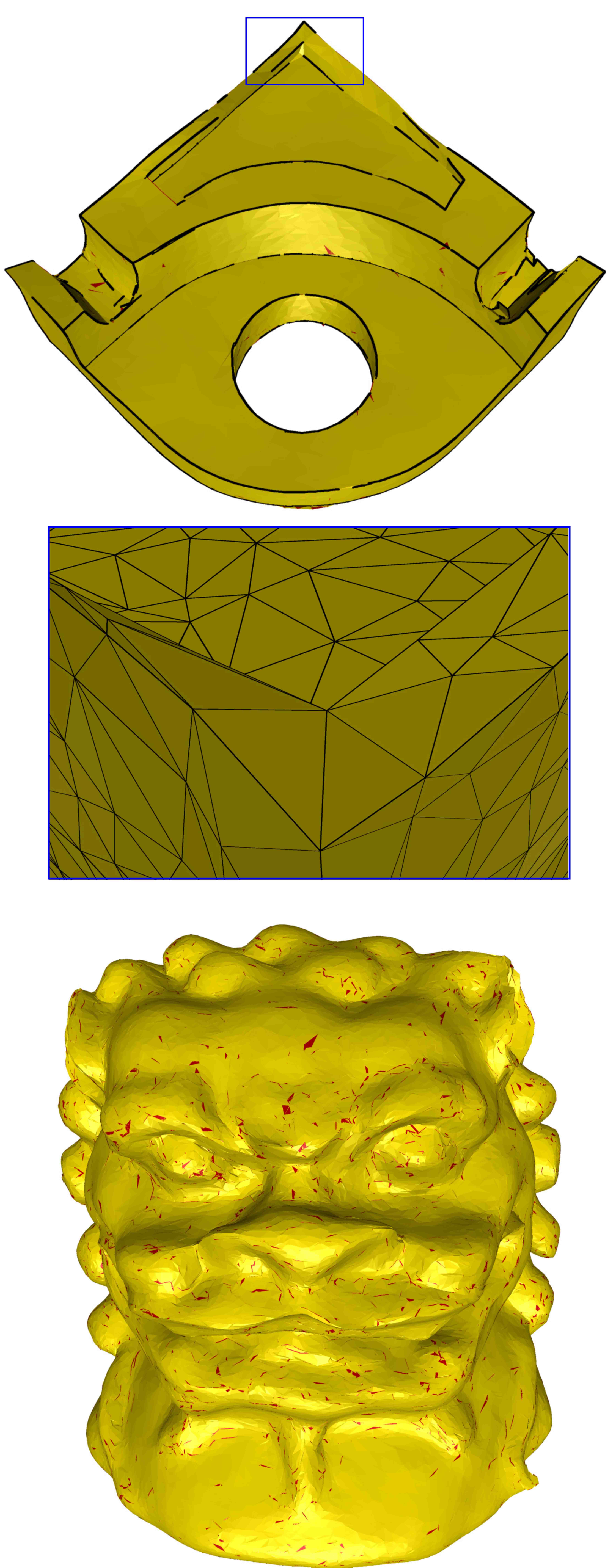}}
 	\subfloat[Ours]{\includegraphics[width=2.2cm]{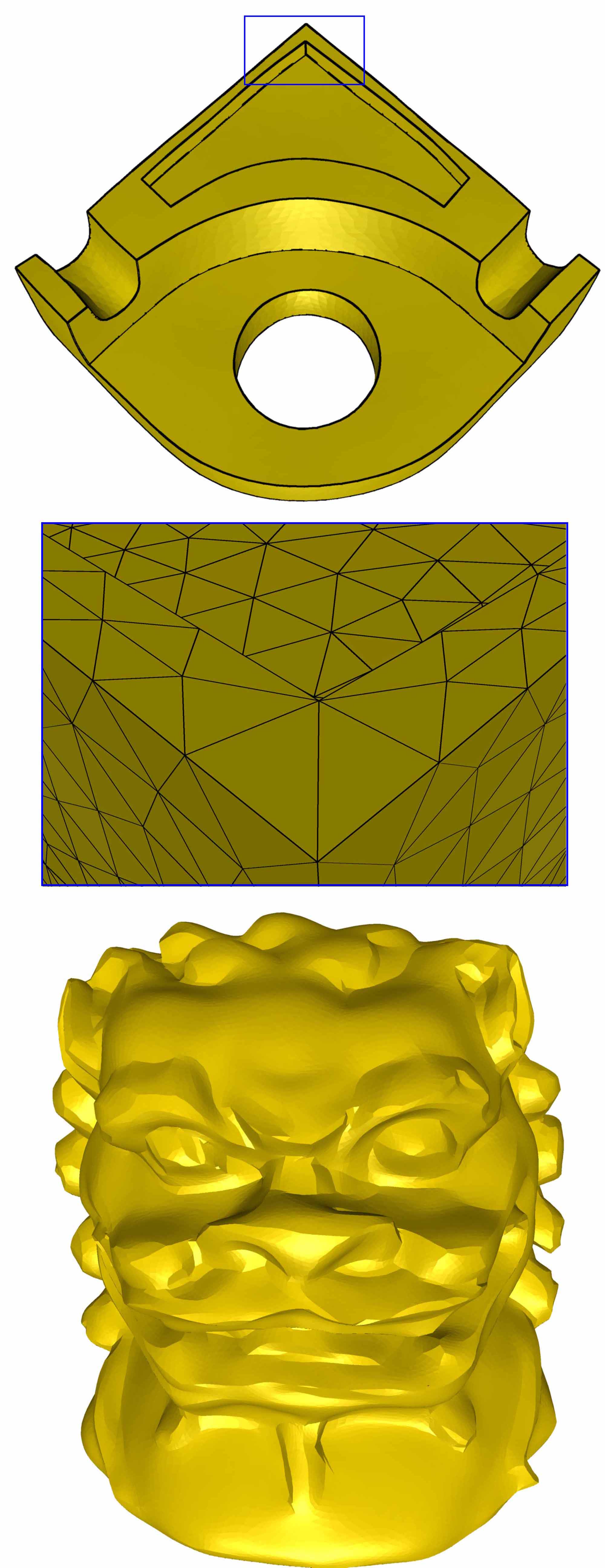}}
 	
 	\caption{Figure (a) shows the noisy models, which are corrupted by a uniform noise ($\sigma_n = 0.5l_e$) in random directions. Figure (b)-(h) show the results produced by state-of-the-art methods and the proposed method. The second row shows the magnified view of the mesh quality at the sharp corners. The sharp feature curves are computed at $\theta =65^\circ$ (for the Bearing model) and the red color faces are with wrong orientation (flipped normals).}
 	\label{fig:hn}
 \end{figure*}
 

  \begin{figure*}[h]%
  	\centering
  	\subfloat[Original]{{\includegraphics[width=2.2cm]{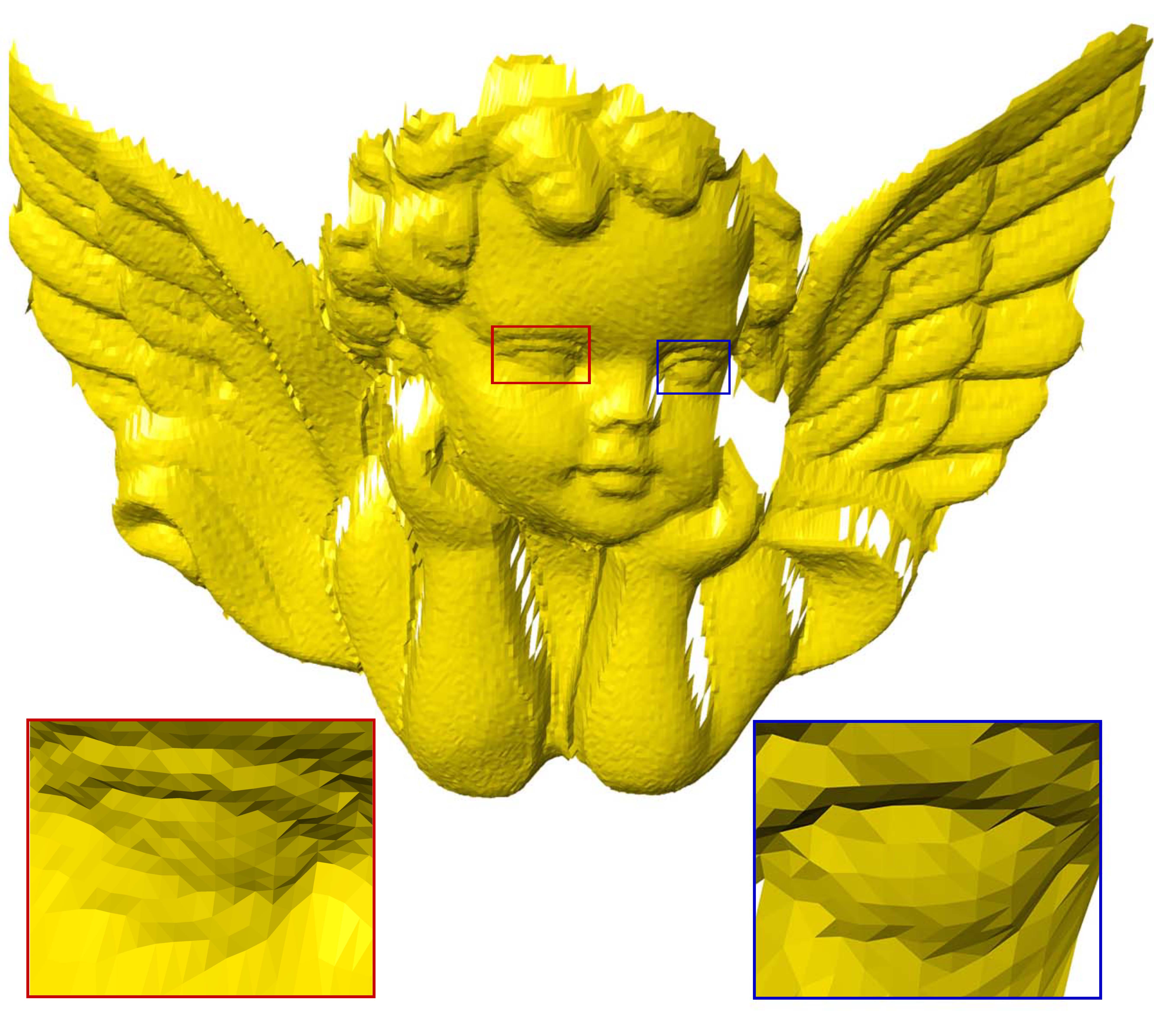} }}%
  	\subfloat[\cite{BilNorm}]{{\includegraphics[width=2.2cm]{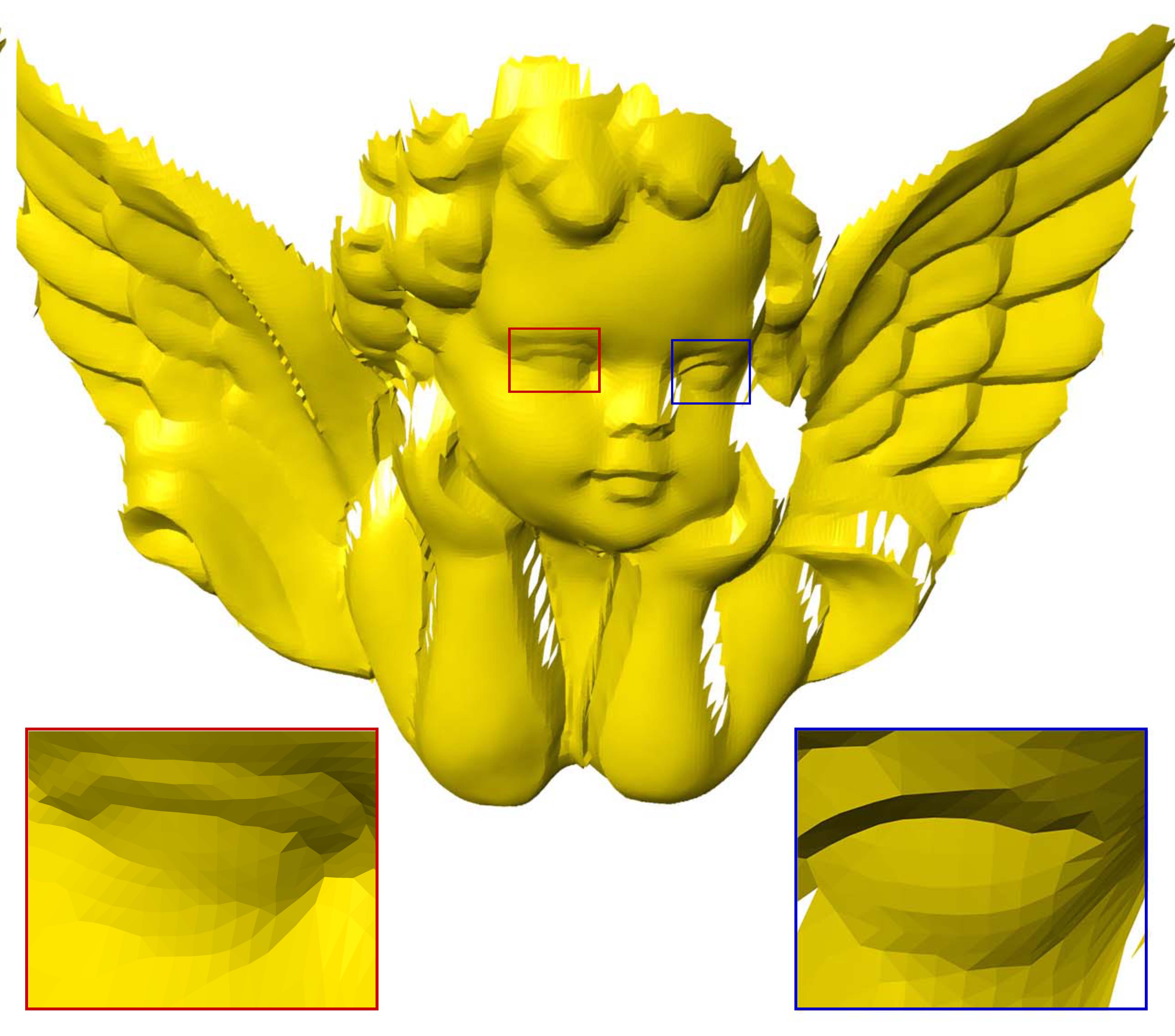} }}%
  	\subfloat[\cite{L0Mesh}]{{\includegraphics[width=2.2cm]{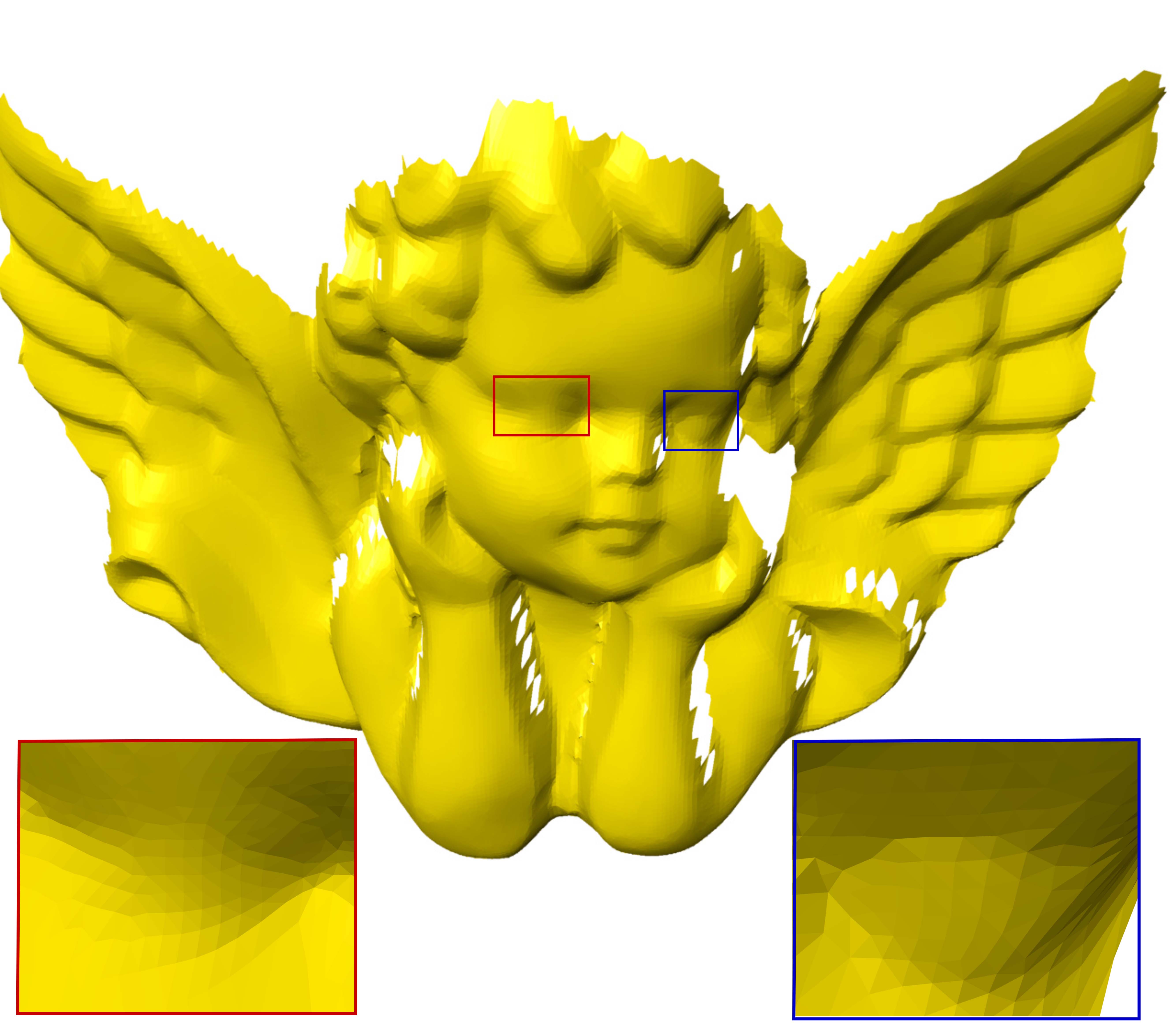} }}%
  	\subfloat[\cite{Guidedmesh}]{{\includegraphics[width=2.2cm]{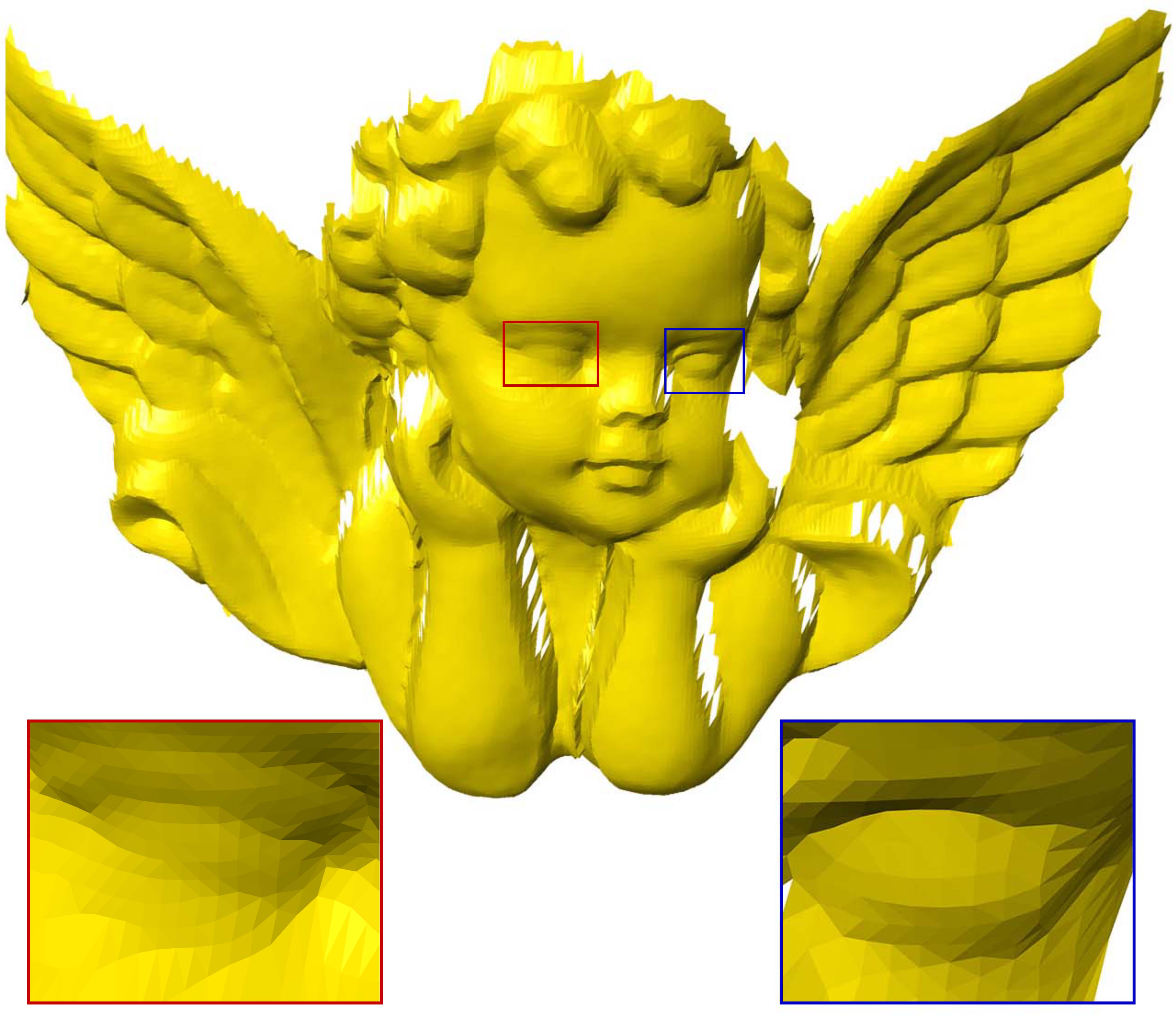} }}%
  	\subfloat[\cite{robust16}]{{\includegraphics[width=2.2cm]{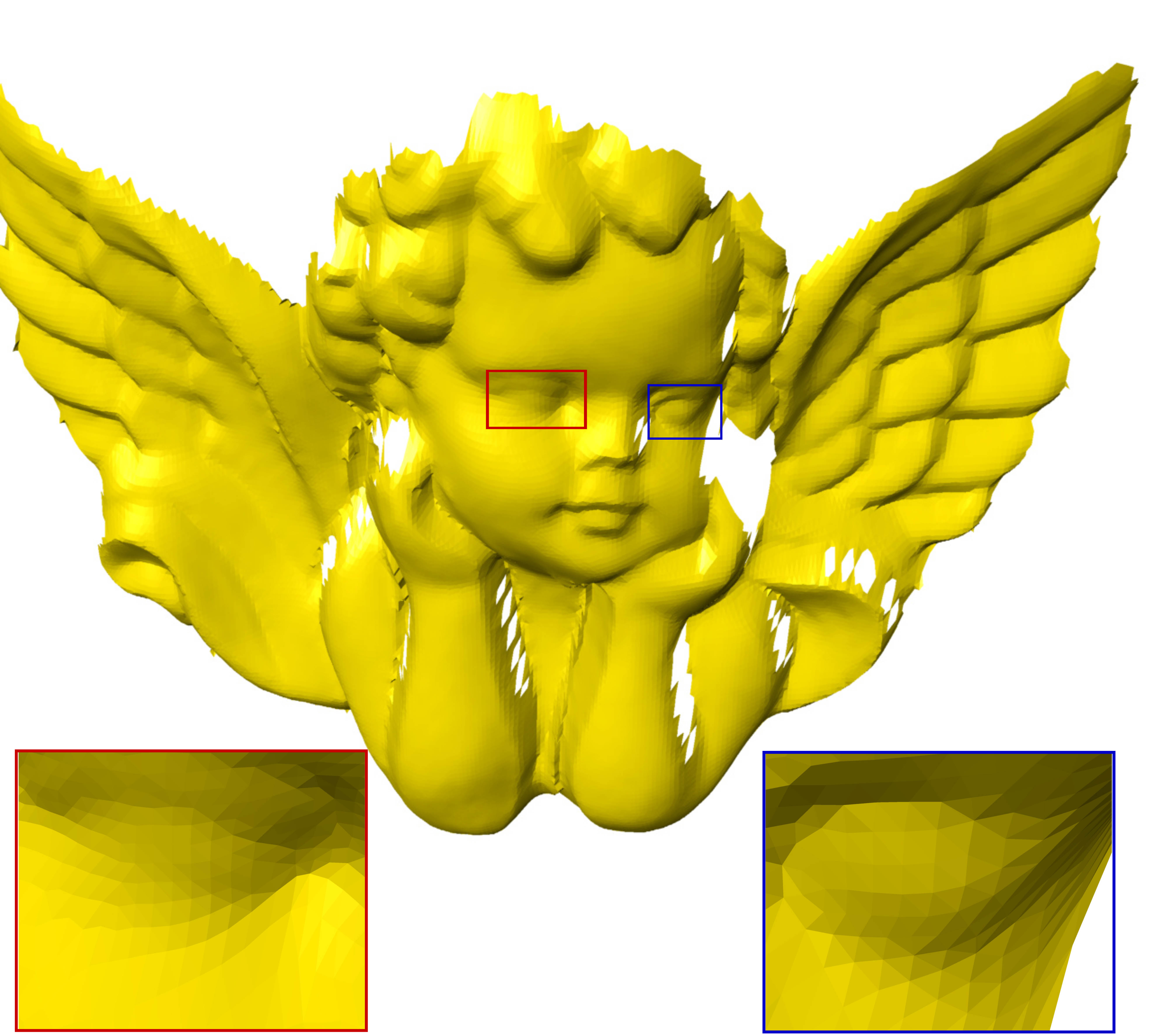} }}%
  	\subfloat[\cite{yadav17}]{{\includegraphics[width=2.2cm]{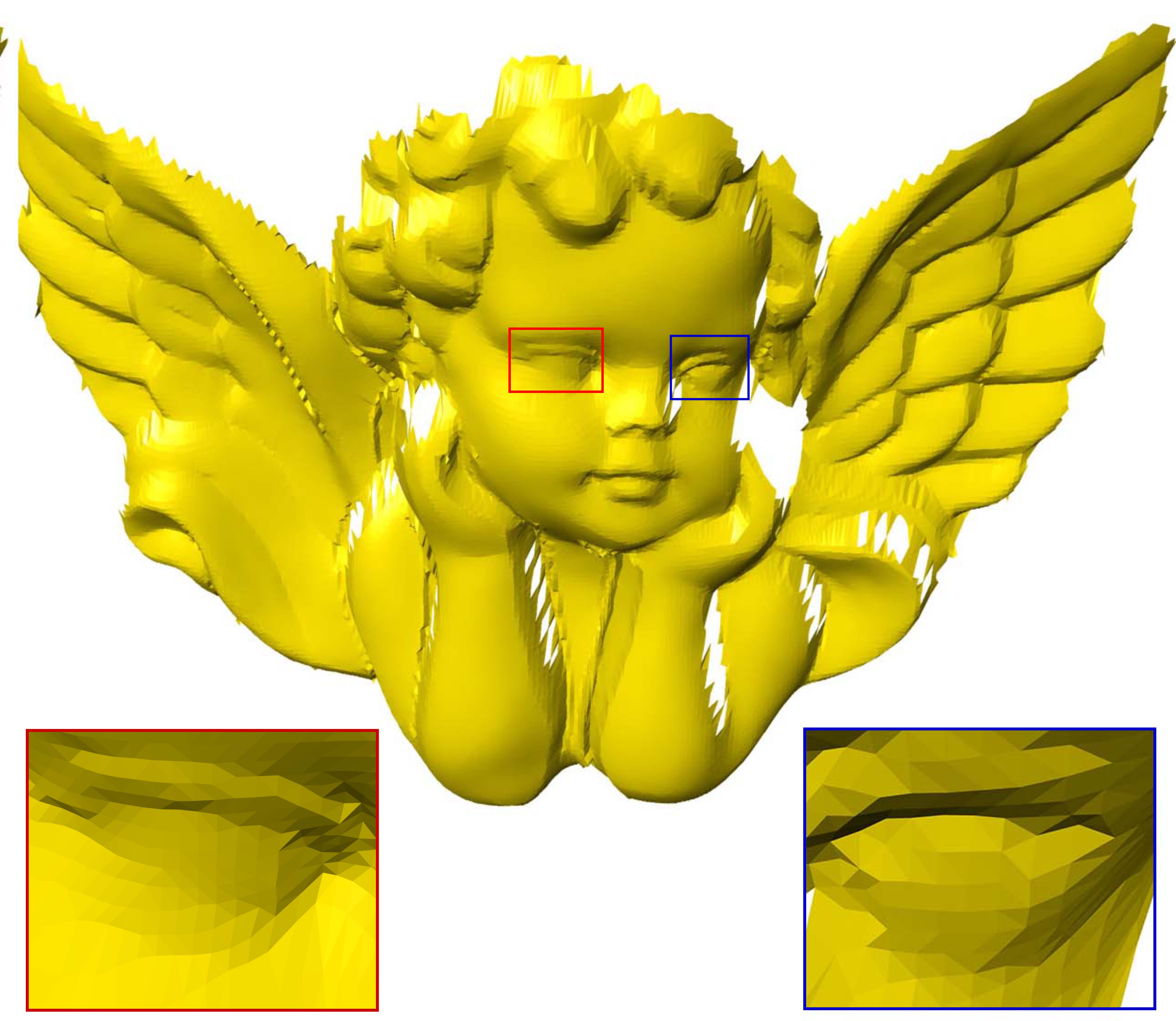} }}%
  	\subfloat[\cite{Centin}]{{\includegraphics[width=2.2cm]{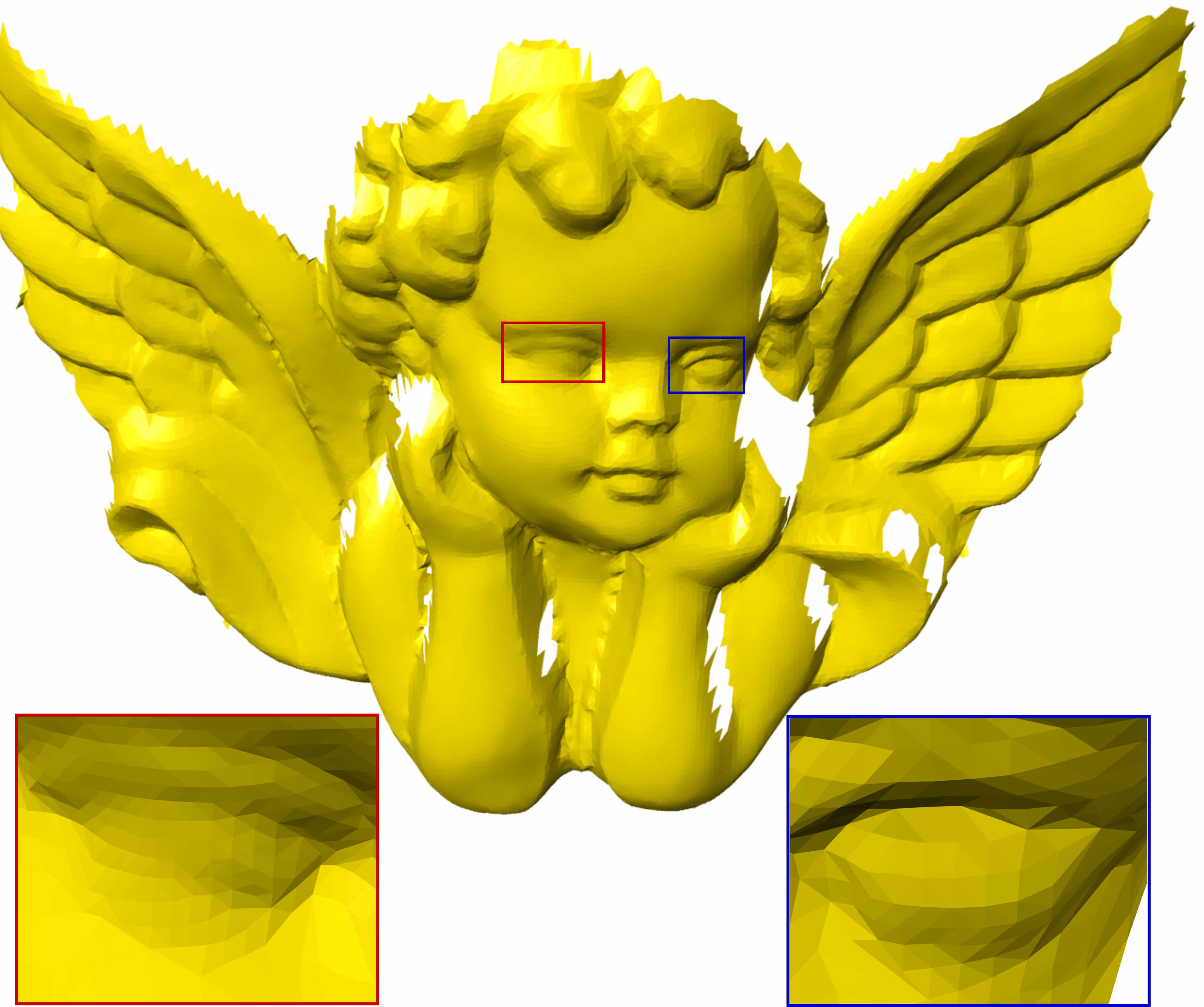} }}%
  	\subfloat[Ours]{{\includegraphics[width=2.2cm]{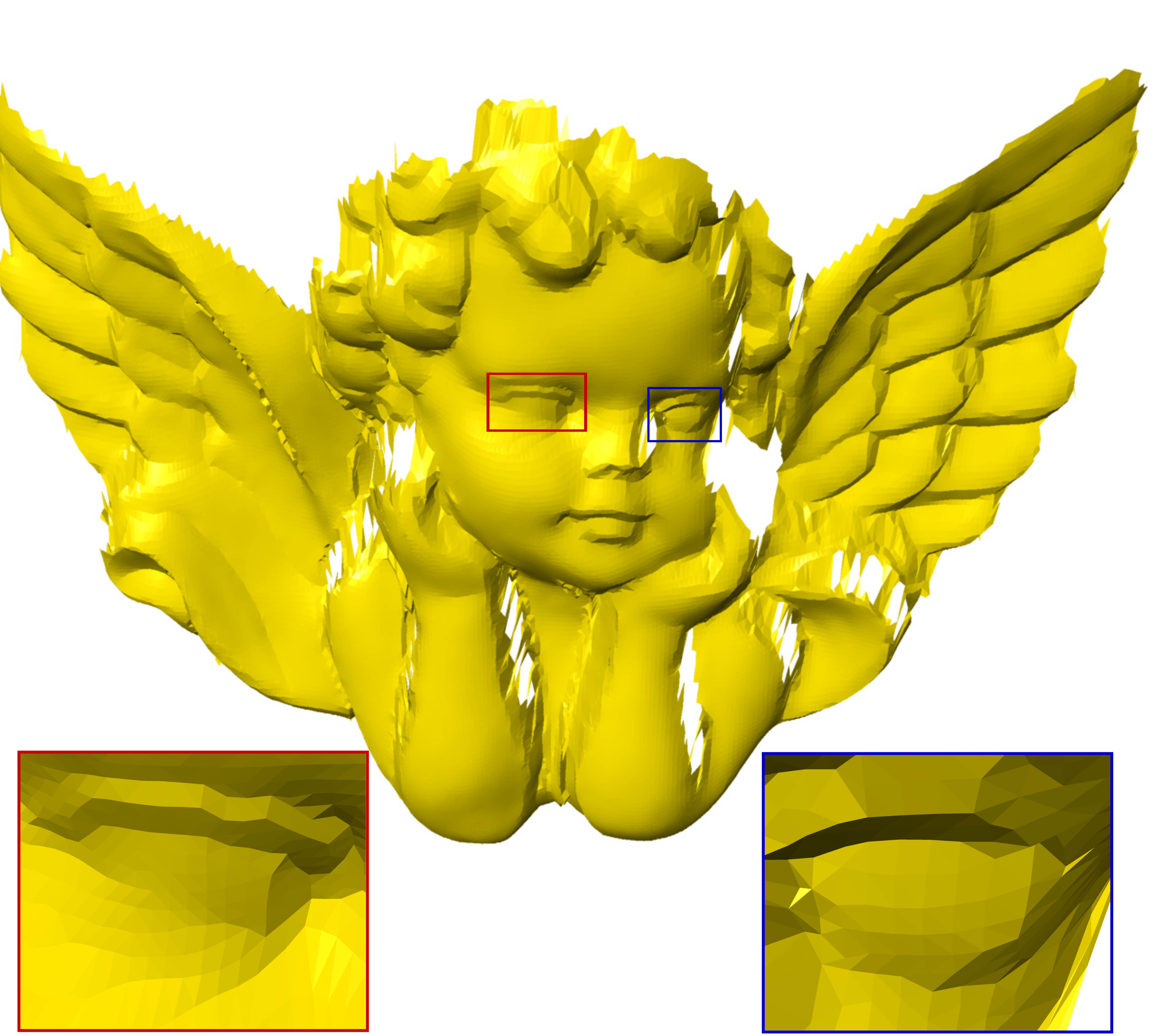} }}%
  	\centering
  	\caption{The Angel model, a triangulated mesh surface (real data) obtained by a 3D scanner. The figures shows the result produced by the proposed method, better preserving sharp features (both eyes) compared to state-of-the-art methods.  }
  	\label{fig:angel}
  \end{figure*}
  
  \begin{figure*}[h]%
  	\centering
  	\subfloat[Original]{{\includegraphics[width=2.2cm]{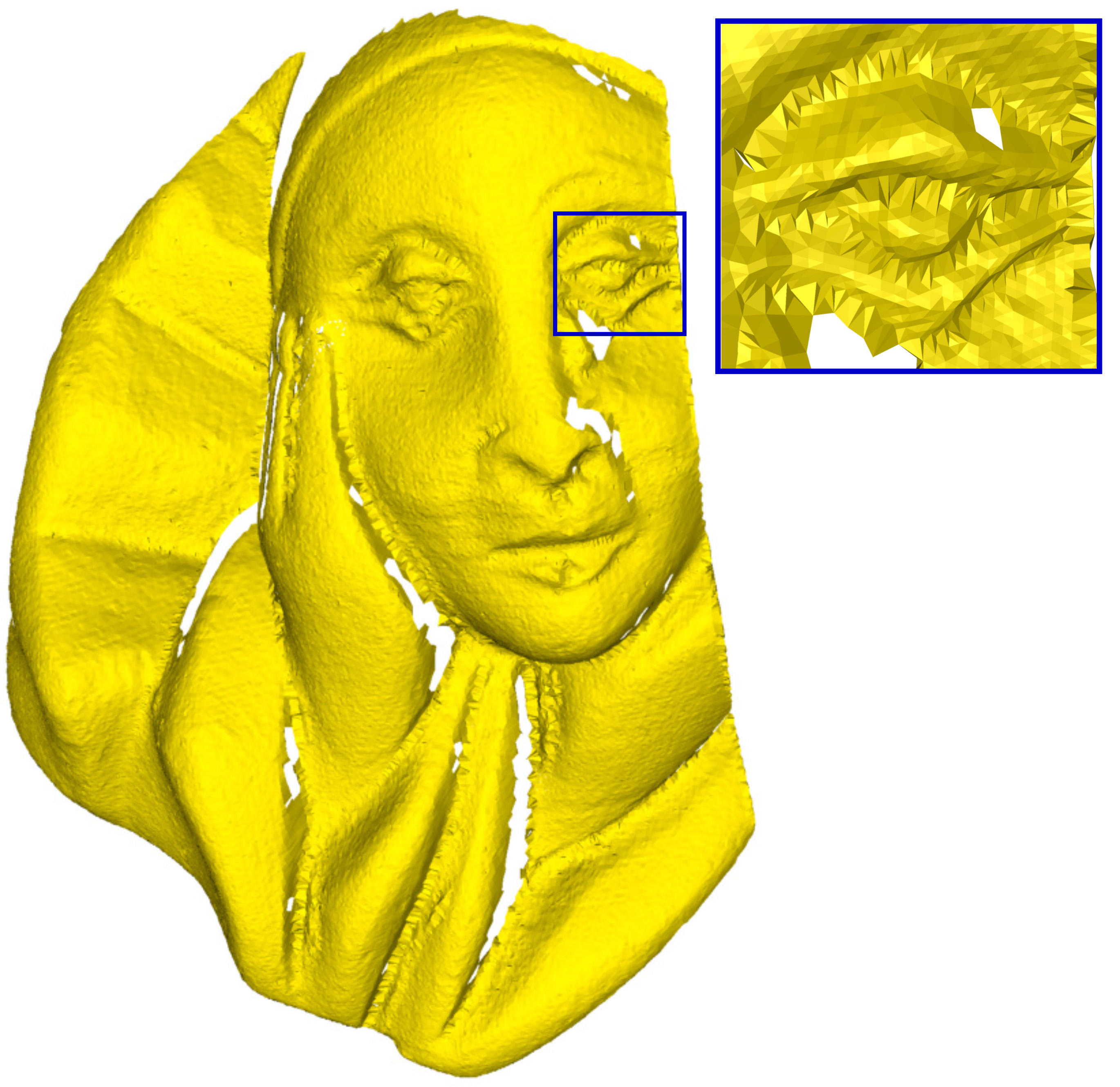} }}%
  	\subfloat[\cite{aniso}]{{\includegraphics[width=2.2cm]{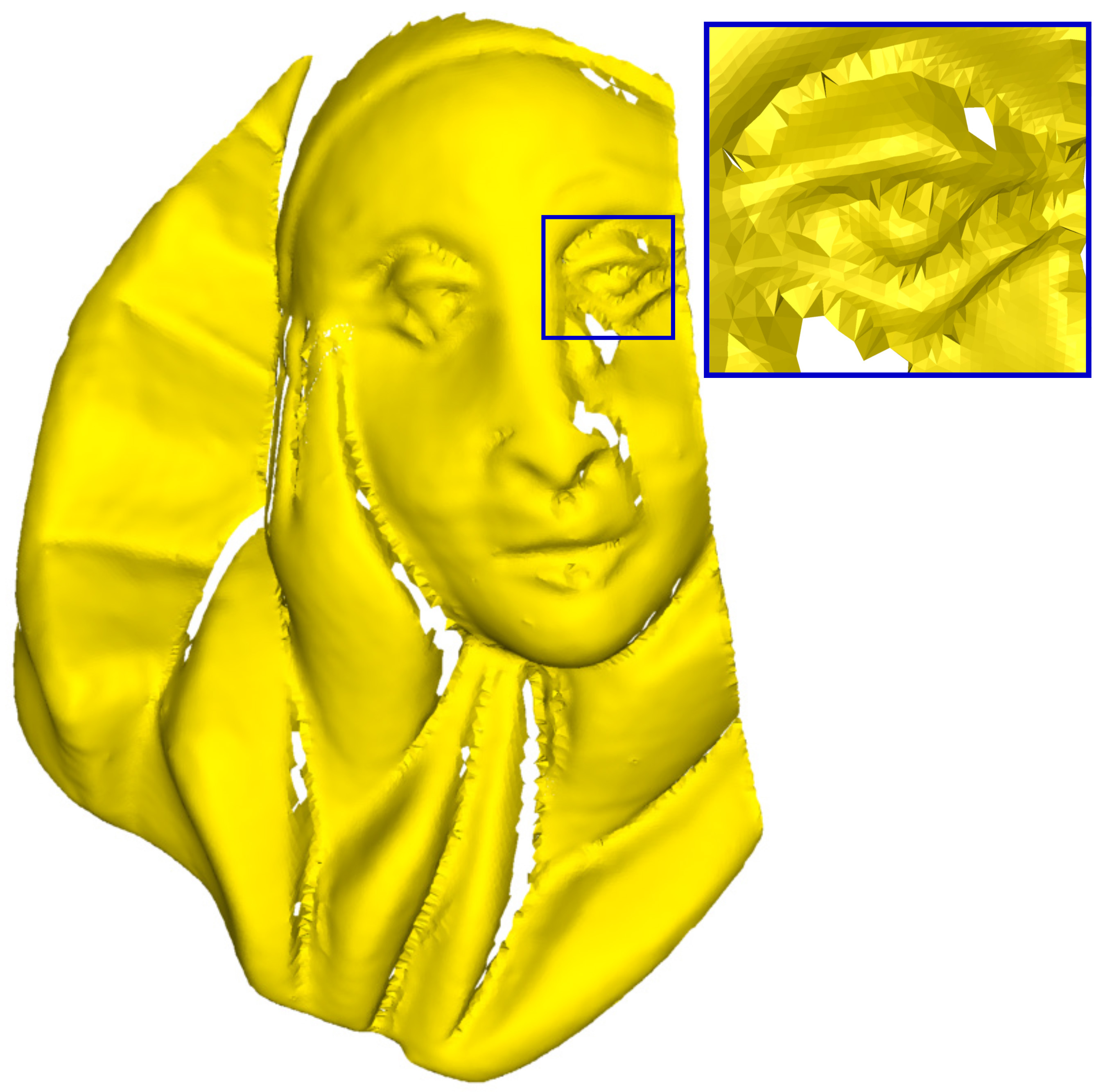} }}%
  	  	\subfloat[\cite{BilFleish}]{{\includegraphics[width=2.2cm]{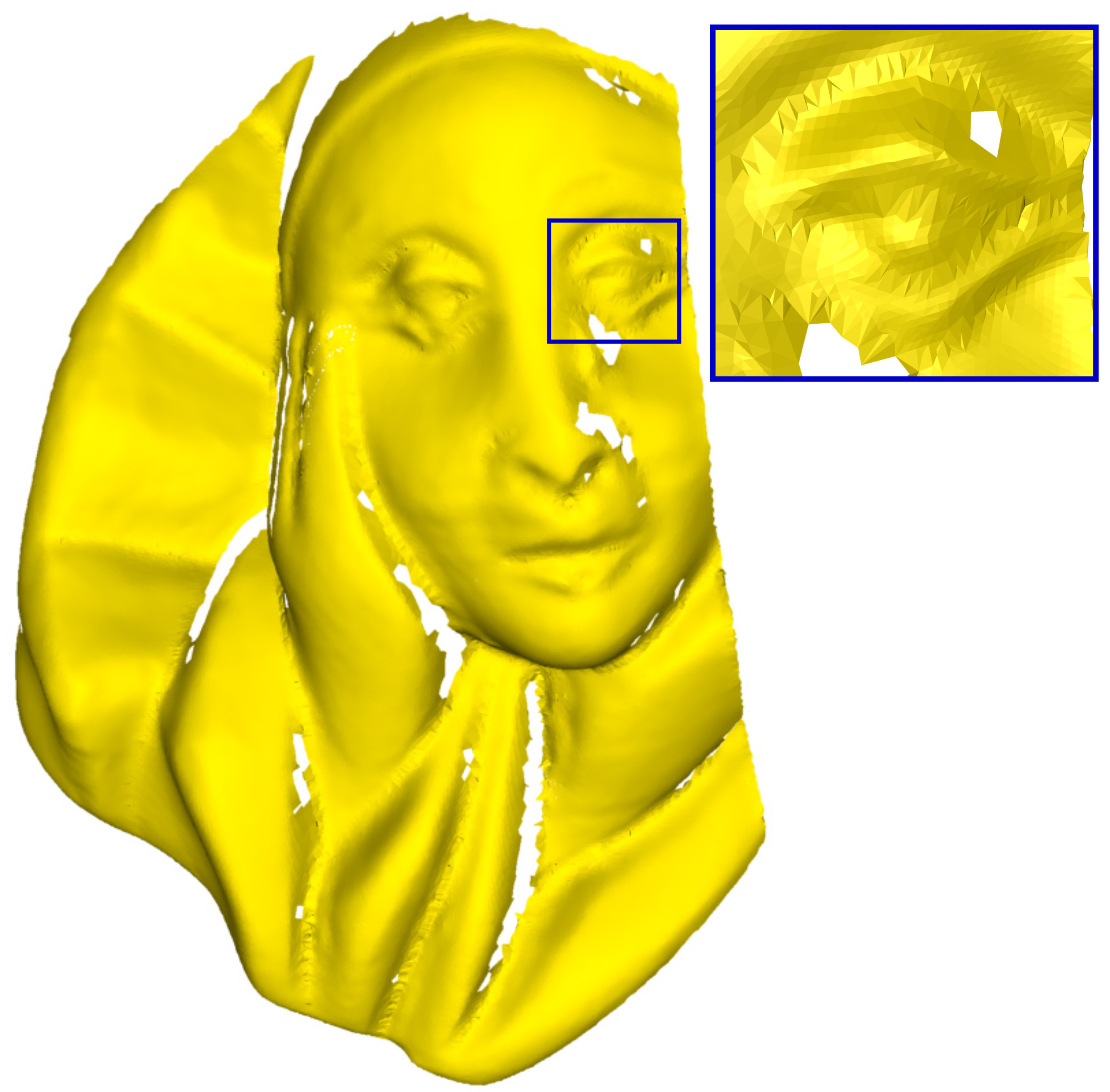} }}%
  	\subfloat[\cite{BilNorm}]{{\includegraphics[width=2.2cm]{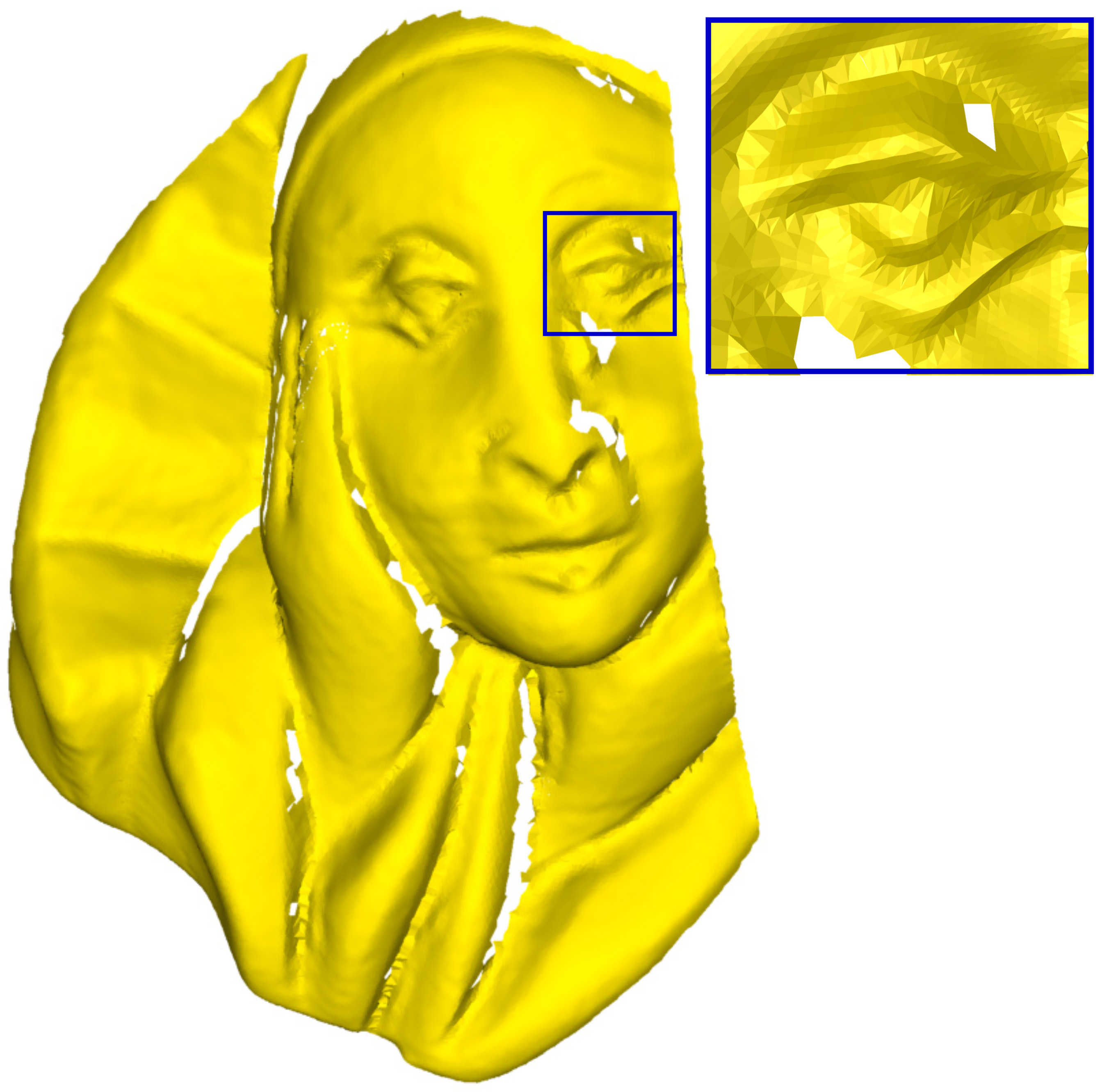} }}%
  	\subfloat[\cite{L0Mesh}]{{\includegraphics[width=2.2cm]{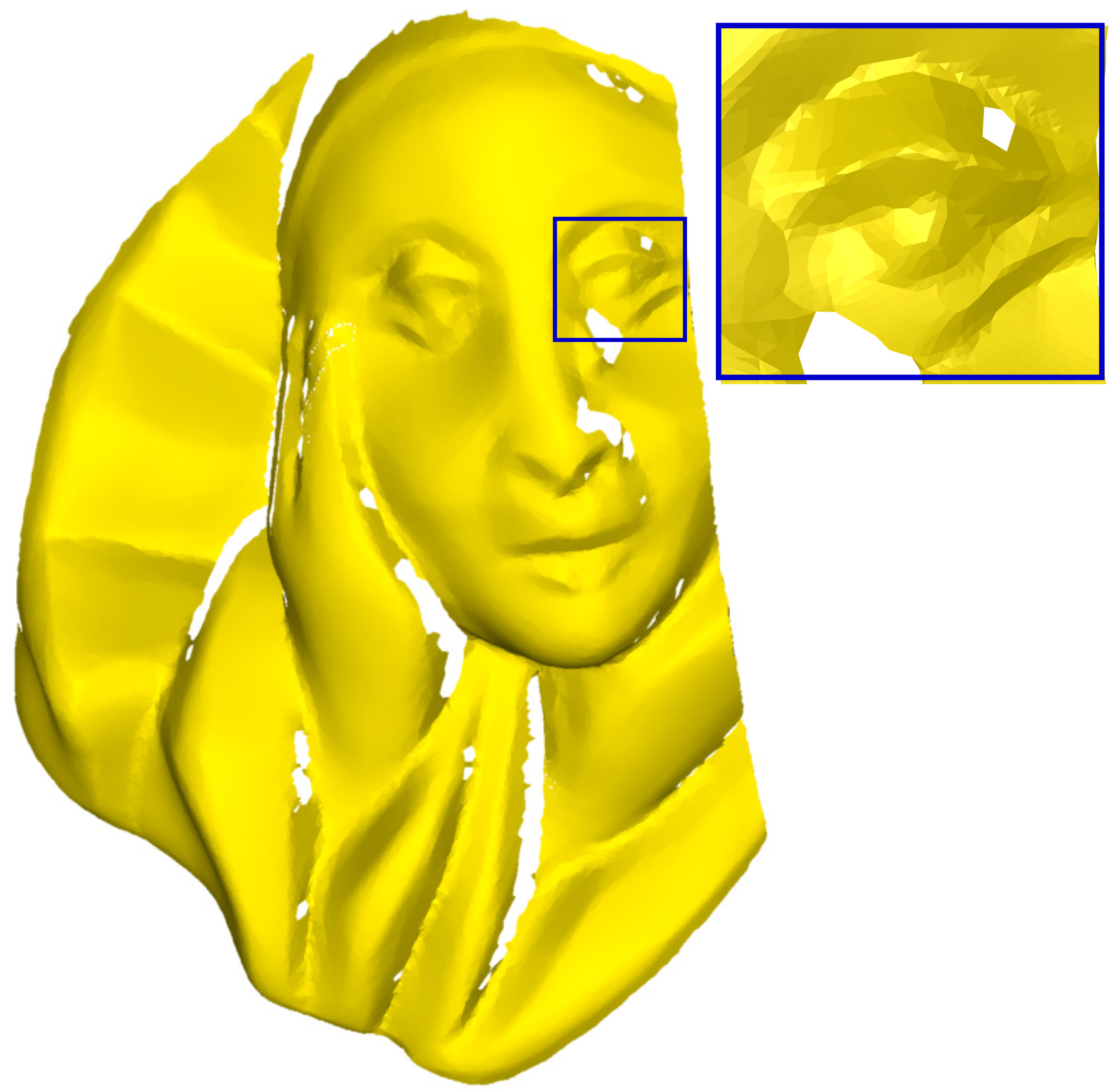} }}%
  	\subfloat[\cite{robust16}]{{\includegraphics[width=2.2cm]{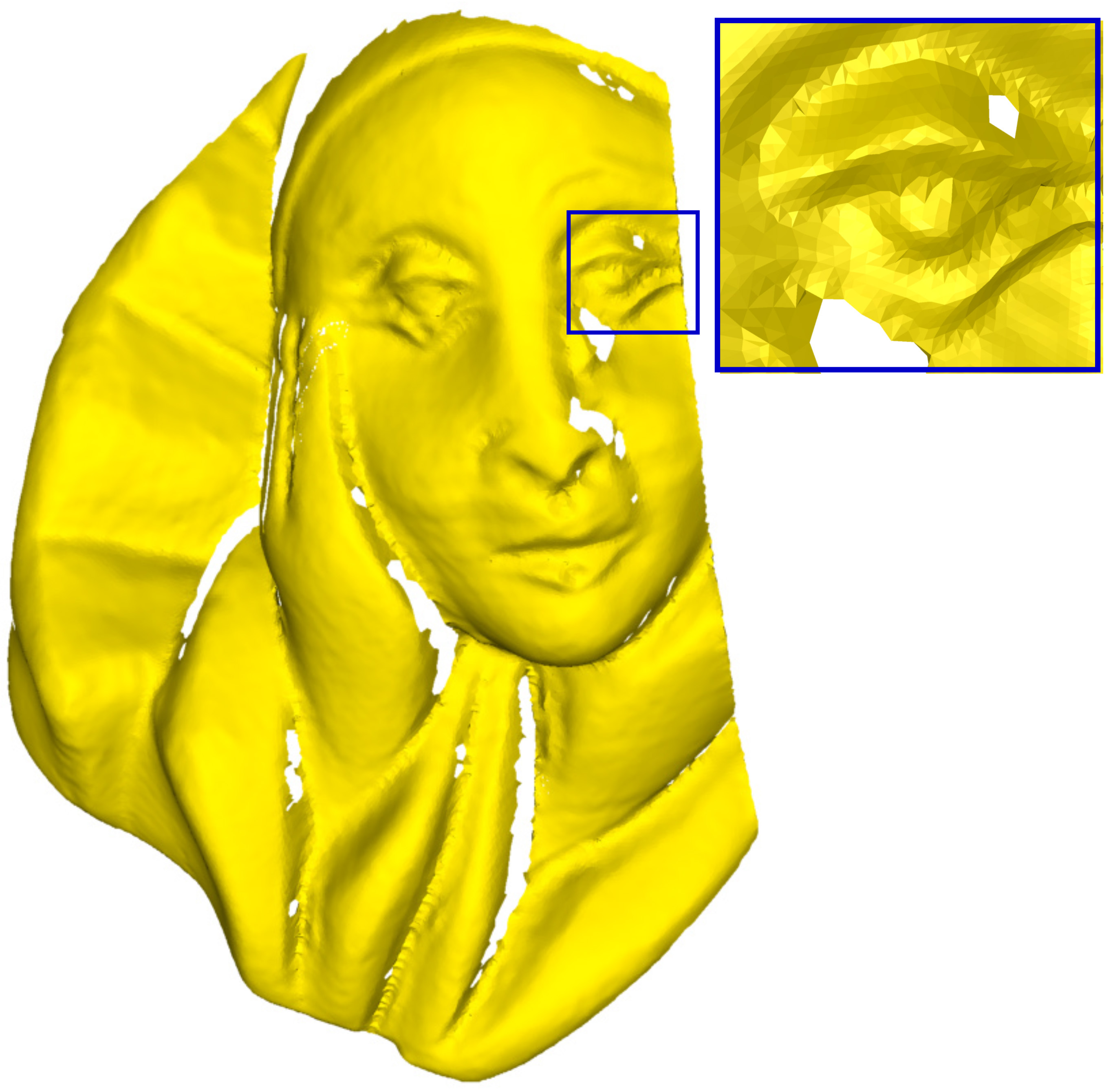} }}%
  	\subfloat[\cite{yadav17}]{{\includegraphics[width=2.2cm]{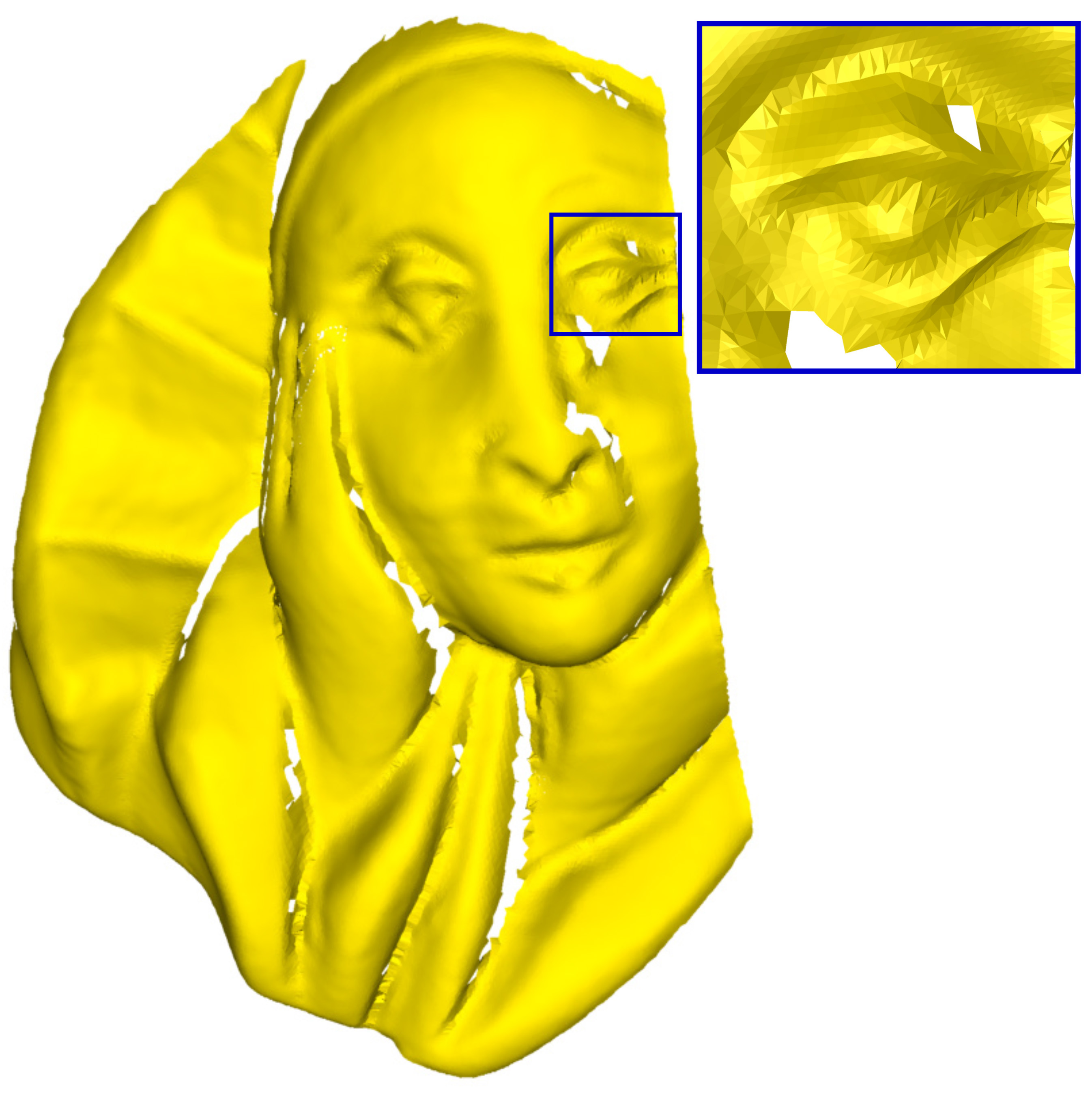} }}%
  	\subfloat[Ours]{{\includegraphics[width=2.2cm]{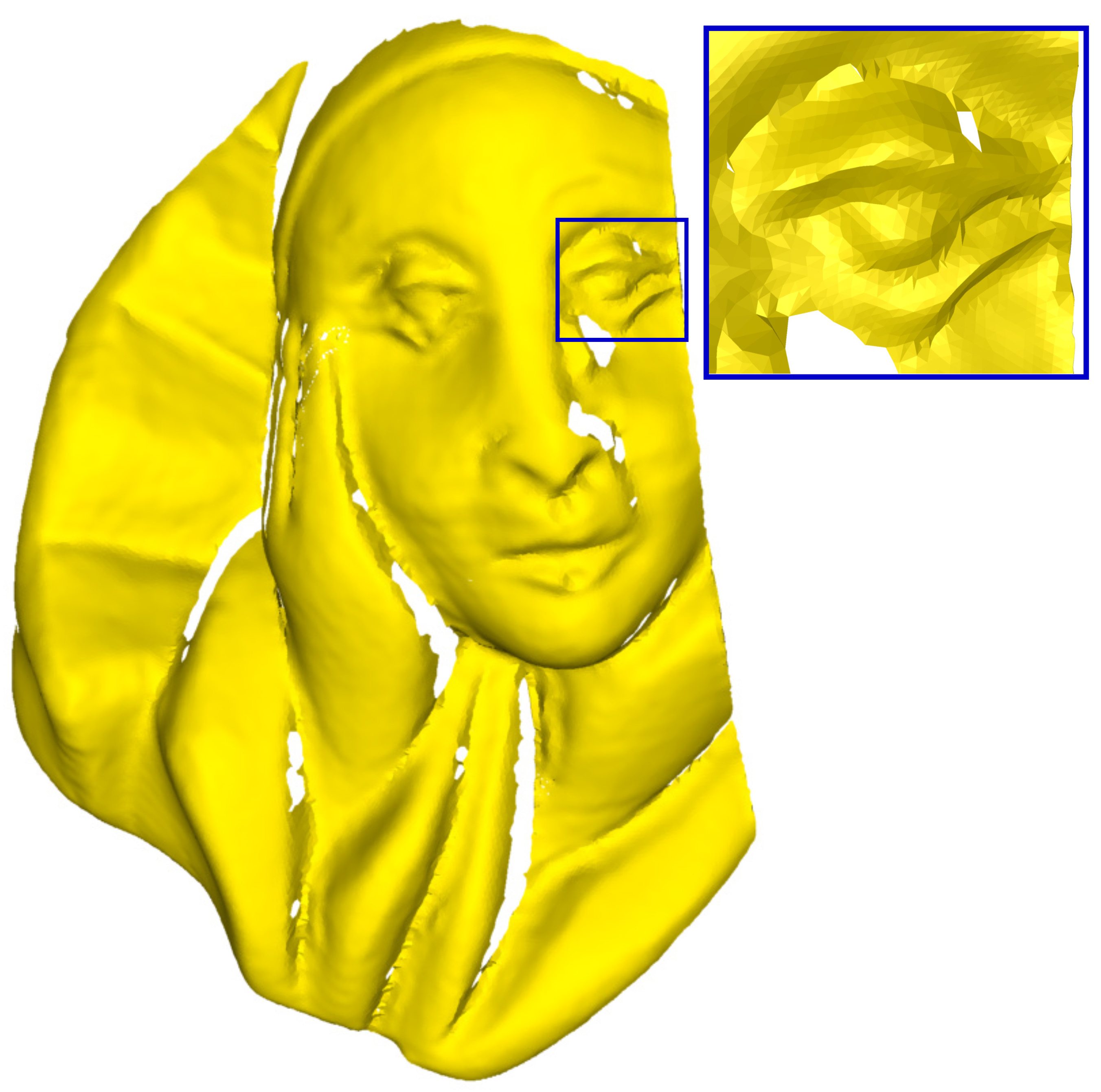} }}%
  	\centering
  	\caption{The Pierrot model (real data) is corrupted by a 3D scanner noise. The figures shows the results obtained by
  		state-of-the-art methods and the proposed method.  }
  	\label{fig:pierrot}
  \end{figure*}
  
  \begin{figure}
  	\centering
  	\subfloat[Noisy]{\includegraphics[width=4.5cm]{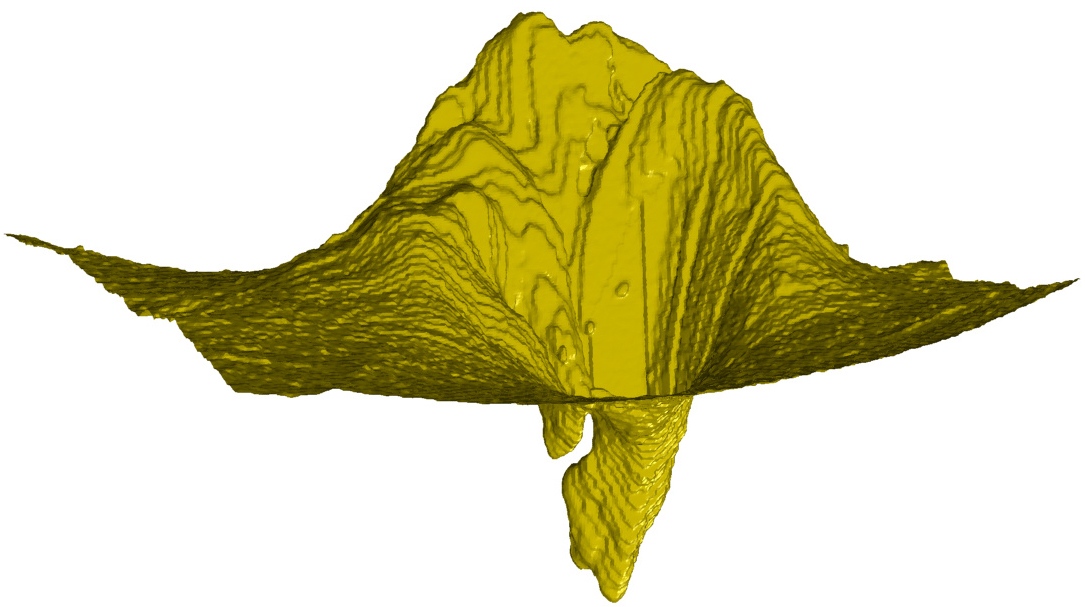}} 
  	\subfloat[Ours]{\includegraphics[width=4.5cm]{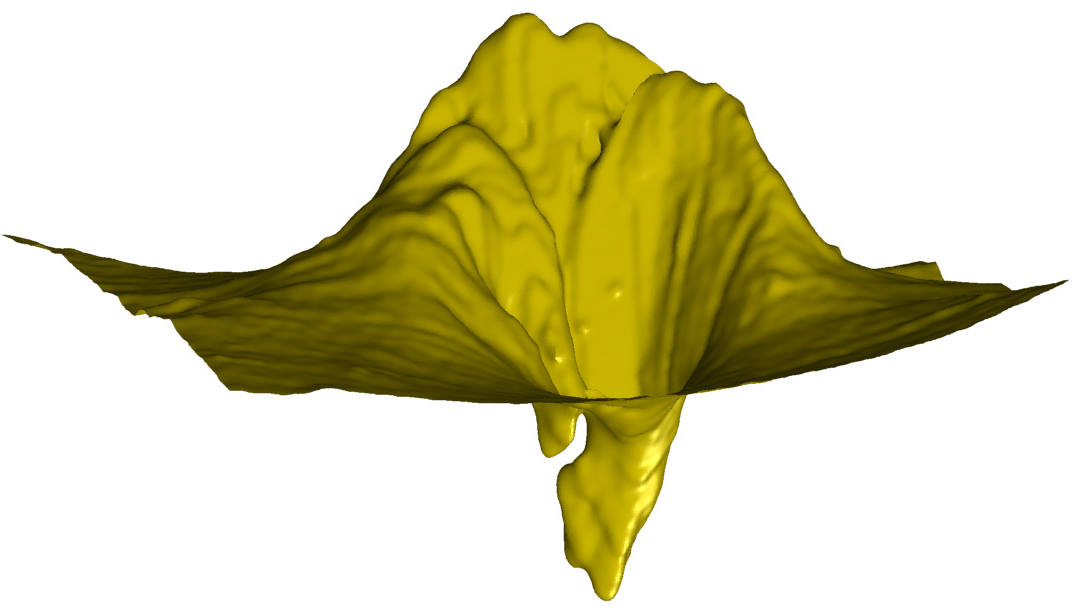}} 
  	
  	\caption{Figure (a) shows the noisy model of a human retina with marching cube artifacts and the proposed method removes these artefacts by choosing the appropriate value of $\lambda_I$.}
  	\label{fig:med}
  \end{figure}

The term $\mathbf{D}_{t_i}$ is computed w.r.t the corresponding vertex normal and the vertex normal at the vertex $\mathbf{v}_i$ is approximated using the processed face normals:
\begin{equation}
	\bar{\mathbf{n}}_{v_i}=\frac{1}{\sum\limits_{j\in F_v(i)}^{} a_j}\sum\limits_{j\in F_v(i)}^{} a_j\tilde{\mathbf{n}}_j,
\end{equation}
where $\tilde{\mathbf{n}}_j$ and $a_j$ represents the processed face normal and the area of the corresponding face, similar to Equation (\ref{equ:bil}). The term $F_v(i)$ represents the set of faces, which are connected to the vertex ${\mathbf{v}}_i$. Now by using this 1-ring vertex normal, the differential coordinate component in the tangent direction is computed as:
\begin{equation}
\mathbf{D}_{t_i}=\mathbf{D}_i-\langle \mathbf{D}_i,\bar{\mathbf{n}}_{v_i} \rangle \bar{\mathbf{n}}_{v_i}, 
\end{equation} 
where the term $\langle \mathbf{D}_i,\bar{\mathbf{n}}_{v_i} \rangle \bar{\mathbf{n}}_{v_i}$ represents the component of the differential coordinate in the vertex normal direction. The normal component of the differential coordinate will be more effective against high-intensity noise because mostly noise affects in surface normal direction. Diffusion only in normal direction leads to more shrinkage because the shape information of a geometry is embedded in the surface normal. Similarly, diffusion of tangential components leads to a better mesh quality, less shrinkage, and less shape deformation. However, it is less effective against high-intensity noise.
Figure ~\ref{fig:tanDir} shows the effect of diffusion only in tangent direction and another one by using Equation~(\ref{vertEnergy}). As it can be seen, the diffusion along the tangent and the normal directions produces a better result compared to the diffusion only in the tangent direction.

Minimization of energy function mentioned in equation~(\ref{vertEnergy}) can be done by the gradient descent method and the updated vertex position will be:
\begin{equation}
	\begin{split}
		\tilde{\mathbf{v}}_i = \mathbf{v}_i+ \frac{1}{3{ \lvert F_v(i)\rvert}} \sum\limits_{j=0}^{N_v(i)-1}\sum\limits_{(i,j)\in \partial F_k}^{}(\tilde{\mathbf{ n}}_k\cdot (\mathbf{v}_i -\mathbf{v}_j)) \\ +
		\lambda_I \mathbf{R}_i,
	\end{split}
	\label{finalEqu}
\end{equation}  
where $\lambda_I$ is the isotropic smoothing or mesh fidelity factor. We iterate the whole procedure mentioned in Equation (\ref{finalEqu}) several times to get the desired result. In each iteration, using the same value of $\lambda_I$ leads to feature blurring, shape deformation and volume shrinkage in the smoothed surface as noise is getting lower with the iterations. To overcome this problem, the effect of isotropic smoothing should get reduced after each iteration. We reduce  $\lambda_I$ in every iteration by $\lambda_I^{p+1}=0.6\lambda_I^p$, where $p$ represents the number iterations.

\section{Experiments, Results and Discussion }
We have performed our denoising scheme on various kinds
of CAD (Figure \ref{fig:tanDir}, \ref{fig:params}, \ref{fig:fandisk}, \ref{fig:jSharp} and \ref{fig:hn}) and CAGD (Figure \ref{fig:nicola} and \ref{fig:hn}) models. These models consist of different levels of features and are corrupted with different kind of synthetic noise (Gaussian and uniform) in different random directions.  We also included real scanned data (Figure \ref{fig:angel}, \ref{fig:pierrot} and \ref{fig:med}) models in our experiments. We provide the visual and quantitative comparison of our method to several state-of-the-art denoising methods in which we implemented \cite{yadav17}, \cite{BilFleish}, \cite{BilNorm}, \cite{aniso} and \cite{L0Mesh}
based on their published article and several results of \cite{Centin}, \cite{binormal},
\cite{Guidedmesh}, \cite{AdamsKd} and \cite{robust16} are provided by their authors.
\newline
\subsection{Parameters} In the proposed method, we discussed about five different parameters: kernel widths of the closeness and similarity functions ($\sigma_c$ and $\sigma_s$), geometric neighbour disk ($\Omega$), isotropic/mesh fidelity factor $\lambda_I$ and the number of iterations $p$. Throughout the algorithm, the radius of the neighbour disk is defined as $r = 2c_a$, where $c_a$ represents the average distance between the centroid of faces and the kernel width of the closeness function is half of the radius of the neighbour disk ($\sigma_c = c_a$). The average centroid distance-based radius will make the algorithm robust against irregular sampling as the denser region will have more faces compared to the rare region during the face normal processing. Effectively, the user has to tune only 3 parameters ($\sigma_s$, $\lambda_I$ and $p$) to get the desired results. In the quantitative comparison table, the proposed method parameters are mentioned in the following format $(\sigma_s, \lambda_I, p)$. Similarly, $(\tau, r, p)$ for \cite{yadav17}, where $\tau$ is the feature threshold, $r$ is the radius of the geometric neighborhood disk and $p$ is the number of iterations. The parameters $(\sigma_c,\sigma_s, p)$ for \cite{BilFleish}, $(\sigma_s, p)$ for \cite{BilNorm}, $(\lambda,s, p)$
for \cite{aniso} and $\alpha$ for \cite{L0Mesh}, where $\sigma_c$ and $\sigma_s$ are the standard deviation of the Gaussian function in the bilateral weighting. The variables $s$ and $\lambda$ represent the step size and the smoothing threshold. The term $\alpha$ controls the amount of smoothing in \cite{L0Mesh}.

Figure \ref{fig:params} shows the effect of the kernel width parameter $\sigma_s$ of the similarity function and isotropic factor $\lambda_I$. The black curve represents sharp features and is computed using the dihedral angle, $\theta = 65^\circ$ between neighboring faces. From Figure \ref{fig:params} (b-d), the isotropic factor is fixed, $\lambda_I = 0.02$. The small value of the $\sigma_s = 0.1$ produces a piecewise flat region and is not able produce a noise free surface with sharp features (Figure \ref{fig:params} (b)). For $\sigma_s = 1.5$, the sharp features are smoothed out and for $\sigma_s = 0.5$, the algorithm produces an optimal result. Similarly, from Figure \ref{fig:params} (e-g), the parameter $\sigma_s $ is fixed, $\sigma_s  = 0.8$. For $\lambda_I = 0$, we can see that the reconstructed surface has several edge flips (faces with wrongly oriented normals) and the algorithm is not able to preserve all desired features (Figure \ref{fig:params}(e)). However, the algorithm produces a high fidelity mesh with sharp features when we use $\lambda_I = 0.2$ (Figure \ref{fig:params}(f)) and blurs the sharp features with bigger isotropic factor (Figure \ref{fig:params} (g)). In general, the value of $\lambda_I$ depends on noise intensity on surfaces. The higher the intensity of noise the more normal flips will be produced during the denoising process, therefore, it is better to use a bigger value of $\lambda_I$ to avoid these normal flips. At the same time, higher values for $\lambda_I$ have to be carefully as these could produce edge blurring and volume shrinkage. 

\subsection{Visual Comparison}
Figure \ref{fig:fandisk}-\ref{fig:pierrot} represent a visual comparison of the proposed method with several state-of-the-art methods. Figure \ref{fig:fandisk} shows the Fandisk model, which is corrupted with a moderate intensity Gaussian noise in random direction. The sharp feature curve (black curve) shows that the proposed method preserves sharp features (especially at corners) effectively and does not produce false features at the cylindrical region compared to state-of-the-art methods. Method \cite{yadav17} and \cite{robust16} manage to produce features at sharp corners but do not recover the sharp features on the whole surface. Method \cite{L0Mesh} creates false features at the cylindrical region and is not able to preserve the corner features \cite{Centin}. The Nicola model (Figure \ref{fig:nicola}) has a non-uniform mesh and is corrupted by a Gaussian noise in random direction. The magnified view shows that    
our method preserves sharp features around the eye regions better than state-of-the-art methods and does not produce any false features around the nose area (piecewise flat areas). Figure \ref{fig:jSharp} shows the Joint sharp model with non-uniform mesh corrupted with a Gaussian noise ($\sigma_n = 0.35l_e$) in normal direction, where $l_e$ is the average edge length. As it shown, the proposed method reconstructs the cylindrical region and corners without creating any false features. Method \cite{yadav17} produces a quite similar result to ours while Method \cite{Guidedmesh} and \cite{L0Mesh} produce false features around the cylindrical region. The other state-of-the-art methods are not able to preserve sharp features. Figure \ref{fig:hn} provides a visual comparison between state-of-the-art methods and the proposed method where models are corrupted with high-intensity of noise ($\sigma_n = 0.5l_e$). As shown in Figure \ref{fig:hn}, our method is able to remove noise components properly and retains the sharp features without creating any edge flips. Method \cite{BilNorm} is able to preserve sharp features but produces edge flips and the mesh quality is not optimal. Method \cite{L0Mesh} is using an isotropic regularization factor which helps the algorithm to produce a good quality mesh on the planar areas but fails to produce similar triangles near the sharp edges (second row Figure \ref{fig:hn}(e)).

In general, real data is captured using the 3D scanners and have quite low intensity of noise. Thus, it is difficult to see a considerable difference between the results of the proposed method and state-of-the-art methods. Figure \ref{fig:angel} shows that our algorithm is capable of better preserving features around both eyes compared to state-of-the-art methods. Methods \cite{yadav17} and \cite{BilNorm} are able to retain the feature in the right eye but fail for the left eye. The other state-of-the-art methods are not able to preserve the fine level of features in both eyes. In case of the Pierrot model (Figure \ref{fig:pierrot}), our method removes noise effectively around the eyes and preserves features. Methods \cite{robust16} and \cite{yadav17} retain similar scale of the feature but are not able to remove noise properly.  Figure \ref{fig:med} shows the applicability of the proposed method to medical data. As shown in Figure \ref{fig:params}, the level of the feature can be controlled by using the parameter $\lambda_I$, which is effective in real data smoothing. For example, Figure \ref{fig:med}(a) shows a noisy model of human retina obtained by the optical coherence tomography \cite{yadavMed}. The surface has the marching cube artefacts which is removed by the proposed method along with noise components effectively.
{\footnotesize
	\begin{center}
		\captionof{table}{Quantitative Comparison}
		\label{tab:quant}
		\scalebox{0.85}{
			\begin{tabular}{ |c|c|c|c|c|l| }
				\hline
				Models & Methods & MSAE & $E_v$$\times 10^-3$& $Q$ & Parameters \\ \hline
				& \cite{BilFleish} & 7.45 & 172.7 &1.18 &(0.3, 0.3, 30)  \\
				& \cite{aniso} & 6.269 & 76.87 &1.00 &(0.05, 0.05, 50) \\
				Nicola & \cite{BilNorm} & \textbf{5.25} & \textbf{66.07}&1.02 &(0.3, 60)\\
				\small{${\lvert F \rvert = 29437 }$} & \cite{L0Mesh} & 6.50 & 99.58 & 0.91&(1.4)\\ 
				\small{${\lvert V \rvert = 14846 }$}& \cite{Guidedmesh} & 5.82 & 69.63 & 1.05&(Default)\\
				& \cite{yadav17} & {5.900} & {72.8} & 1.02 &(0.3, 0.15, 50)\\
				& \textbf{Ours} & {5.5} & {66.14} & \textbf{0.9} &(1.6, 0.06, 25)\\ \hline
				 & \cite{BilNorm} & 3.17 & 0.590  & 1.45 &(0.3, 50)\\
				Bearing	& \cite{L0Mesh} & 4.70 & 1.24 & 1.58 &(4.0)\\ 
				\small{${\lvert F \rvert = 6950 }$}	& \cite{binormal} & {3.95} & {0.783} & {2.62} &(Default)\\
				\small{${\lvert V \rvert =  3475}$}	& \cite{yadav17} & {4.43} & {0.698} & {1.66} &(0.3,0.08,80)\\
				& \textbf{Ours} & \textbf{2.32} & \textbf{0.588} & \textbf{0.74} &(0.8,0.2,100)\\
				\hline
				& \cite{BilFleish} & 3.777 & 0.530   & 0.83 &(0.3, 0.3, 40) \\
				& \cite{aniso} & 2.630 & 0.766  & 0.78 &(0.009, 0.05, 50)\\
				Joint & \cite{BilNorm} & 1.808 & 0.263 & 0.77 &(0.4, 100) \\
				\small{${\lvert F \rvert = 52226  }$} & \cite{L0Mesh} &1.768 & 0.500 & 0.75  &(1.4)\\ 
				\small{${\lvert V \rvert =  26111}$}& \cite{Guidedmesh} & 0.956 & 0.179 & 0.83 &(Default)\\
				& \cite{binormal} & 2.874 & 0.366 & 1.67 &(Default)\\
				& \cite{robust16} & {1.16} & {1.49}&\textbf{ 0.71} &(Default)\\
				& \cite{yadav17} & \textbf{0.829} & \textbf{0.171}& 0.91 &(0.3, 0.05, 60)\\
				& \textbf{Ours} & {1.032} & {0.198} & {0.73}&(1.0, 0.2, 150)\\ \hline
				& \cite{BilFleish} & 8.567 & 4.422 & 0.87 &(0.4, 0.4, 40)   \\
				Fandisk & \cite{aniso} & 5.856 & 4.910  & 0.79 &(0.07, 0.05, 30)\\
				\small{${\lvert F \rvert = 12946  }$}& \cite{BilNorm} & 2.727 & 1.877 & 0.76 &(0.4,70) \\
				\small{${\lvert V \rvert = 6475 }$}& \cite{L0Mesh} &4.788 & 5.415 & 0.766 &(1.4)\\ 
				& \cite{Guidedmesh} & {2.221} & \textbf{1.702}& 0.97 &(Default)\\ 
				& \cite{robust16} & {3.1} & {4.42}& {0.7} &(Default)\\ 
				& \cite{yadav17} & {2.692} & {1.964}& 0.89 &(0.3, 0.2, 50)\\ 
				& \textbf{Ours} & \textbf{2.201} & {1.844}& \textbf{0.7} &(0.55, 0.2, 100)\\\hline
				& \cite{BilNorm} & 14.88 & 19.64 & 2.72 &(1.0, 100) \\
				Chinese Lion & \cite{L0Mesh} & 8.93 & 25.34 & 1.51 &(1.4)\\
				\small{${\lvert F \rvert = 118276}$}& \cite{binormal} & {19.22} & {21.5} & {2.38} &(Default)\\
				\small{${\lvert V \rvert = 59140}$}& \cite{yadav17} & {21.39} & {21.89} & {2.16} &(0.3, 1.0,100)\\
				&  \textbf{Ours} & \textbf{8.795} & \textbf{18.35} &\textbf{0.85} &(0.4, 0.3, 100)\\ 
				& & & & &\\
				\hline
			\end{tabular}
		}
	\end{center}
}
\subsection{Quantitative Comparison}
For the quantitative comparison, we are using three different parameters. Two of them show differences between the ground truth and smooth model and the so-called $L^2$ vertex-based positional error and the orientation error respectively. The third parameter computes the mesh quality of a smoothed surface. 
The $L^2$ vertex-based positional error from the original ground truth model is represented by $E_v$ and defined as \cite{vertexUpdate}:
\begin{equation*}
E_v = \sqrt{\frac{1}{3\sum_{k\in F}^{} a_k} \sum_{i\in V}^{} \sum_{j\in F_v(i)}^{} a_j {\mathrm{dist}(\tilde{\mathbf{v}}_i, T)}^2},
\end{equation*}
where $F$ is the triangular element set and $V$ represents the set of vertices. The terms $a_k$ and $a_j$ are the corresponding face areas. The distance $\mathrm{dist}(\tilde{\mathbf{v}}_i, T)$ is the closest $L^2$-distance between the newly computed vertex $\tilde{\mathbf{v}}_i$ and the triangle $T$ of the reference model.

The MSAE computes the orientation error between the original model and the smooth model and is defined as:
\begin{equation*}
\text{MSAE} = E[\angle ({\mathbf{\tilde{n}, n}})],
\end{equation*}
where $\mathbf{\tilde{n}}$ is the newly computed face normal and $\mathbf{n}$ represents the face normal of the reference model. The term $E$ stands for the expectation value. 

Mesh quality of the surface mainly refers to the shape of the faces in the mesh. For a better mesh quality, the size and angles of the triangular face should not be too small or too large. To measure the quality of faces, we use the ratio of the circumradius-to-minimum edge length of the corresponding face \cite{meshQuality}: 
\begin{equation*}
Q=\frac{r}{e_{min}},
\end{equation*}  
where $r$ and $e_{min}$ are circumradius and minimum edge length of the corresponding triangle. Ideally, every face belonging to the mesh should be an equilateral triangle with a quality index of $Q=1/\sqrt{3}$.

Table \ref{tab:quant} shows the quantitative comparison of our method with state-of-the-art methods. For the models which are corrupted with high-intensity noise (Bearing, Chinese Lion and Sharp Sphere), the proposed method has lower values of $E_v$, MSAE and $Q$ compared to Methods \cite{BilNorm} and \cite{L0Mesh}. Method \cite{BilNorm} produces a quite similar value of $E_v$ but bigger values for MSAE and $Q$ because of the several edge flips. He et al.\cite{L0Mesh} produces a good quality mesh in planar regions but does not produce a similar quality of triangles near sharp features, which leads to a higher value mesh quality index. For the Nicola model, our method produces a quite similar values for the position and orientation error (slightly bigger) to Method\cite{BilNorm}, because our algorithm also follows the bilateral filtering technique and in addition, we use an isotropic factor which introduces a small amount of volume shrinkage. At the same time, it produces a good quality mesh.
\begin{center}
	\captionof{table}{Running Time (in seconds)}
	\scalebox{0.85}{	
		
		\label{tab:runn}
		\begin{tabular}{ |c|c|c|c|c|c|c|c|l| }
			
			\hline
			Models  & Fandisk & Bearing & Nicola & Joint & Lion \\ \hline
			Time (s) & 5.8 & 4.2 & 4.3 & 45.3 & 92 \\ 
			\hline
		\end{tabular}
	}	
\end{center}  

\subsection{Running Time}
As it is explained by \cite{yadav17}, the two-stage denoising methods have quite similar running time complexity. The proposed algorithm also follows the two-stage denoising scheme and has the complexity of $O(n_c\cdot n_f \cdot p)$, where $n_c$ is the number of faces within the neighbourhood, $n_f$ and $p$ are the numbers of faces and iterations respectively. The implementation of the algorithm is quite straightforward and simple. First of all, we compute the smooth face normals using Equation (\ref{equ:bil}) then rearrange vertices by following the Equation (\ref{finalEqu}). Our implementation is done in a single computation thread using Java. Table \ref{tab:runn} shows the running time of our algorithm for different models according to the parameter values mentioned in Table \ref{tab:quant}.      

\section{Conclusion}
In this paper, we presented a robust mesh denoising algorithm which produces a high-quality mesh along with sharp features. In the first step of the proposed method, a robust statistics framework is applied for noisy face normal smoothing where \textit{Tukey's bi-weight function} is used as the robust estimator. \textit{Tukey's bi-weight function} completely stops the diffusion across sharp edges and removes noise along the sharp features and from flat areas. In the second step, a vertex position is updated using edge-face normal orthogonality constraints along with differential coordinates. Differential coordinates helps the algorithm to produce a high-quality mesh, makes it robust against high-intensity of noise and avoids normal flips during the denoising process. In section 3, we have shown the robustness of the proposed algorithm not only against different kinds and levels of noise but also against the face normal flipping. The quantitative comparison table shows that our method produces better mesh quality compared to state-of-the-art methods. Similarly, for the positional and the orientation error, the proposed method is either better or similar to state-of-the-art methods.


\bibliographystyle{IEEEtran}
\bibliography{extrinsic}{}

\end{document}